\documentclass[a4paper,11pt]{article}
\usepackage{jcappub} 
\usepackage{lineno}
\usepackage{graphicx}
\usepackage{subfig}
\usepackage{amsmath}
\usepackage{mathtools}

\arxivnumber{2506.17926} 
\title{\boldmath Testing Cosmic Distance Duality Relation and Transparency with DESI DR2}

\author[1,2]{Xuwei Zhang}
\author[3,1]{Xiaofeng Yang$^*$}
\author[3,4]{Yunliang Ren}
\author[3,4]{Shuangnan Chen}
\author[3,5]{Yangjun Shi}
\author[1,2]{Cheng Cheng$^*$}
\author[1,2]{Ming Zhang$^*$}
\author[1,2]{Xiaolong He}

\affiliation[1]{\small State Key Laboratory of Radio Astronomy and Technology, Xinjiang Astronomical Observatory, Chinese Academy of Sciences, 150, Science 1-Street, Urumqi 830011, China}
\affiliation[2]{\small School of Astronomy and Space Science, University of Chinese Academy of Sciences, No.1 Yanqihu East Rd, Beijing 100049, China}
\affiliation[3]{\small School of Physics and Electronics, Henan University, Jinming Avenue, Kaifeng 475001, China}
\affiliation[4]{\small School of Physical Science and Technology, Xinjiang University, No. 666 Shengli Road, Urumqi 830046, China}
\affiliation[5]{\small School of Physics and Astronomy, China West Normal University, No.1 Shida Road, Nanchong 637002, China}

\emailAdd{xfyang@henu.edu.cn}
\emailAdd{chengcheng@xao.ac.cn}
\emailAdd{zhangm@xao.ac.cn}

\abstract{
The Cosmic Distance Duality Relation (CDDR) is a fundamental principle of standard cosmology, linking luminosity (LD) and angular diameter distances (ADD). This work investigates the validity of the CDDR and cosmic transparency by combining the latest Baryon Acoustic Oscillations (BAO) data from DESI DR2, Type Ia Supernovae from Pantheon+, and cosmic chronometers. To address the redshift mismatch between datasets, two distinct reconstruction techniques are employed: Gaussian Process Regression (GPR) and the Free-Knots Method (FKM). The analysis performs null tests on the CDDR under different cosmological priors, finding that the null hypothesis holds and the CDDR is valid within statistical uncertainties. Although mild deviations are observed from the local distance ladder prior, internal consistency calibration indicates that these discrepancies and the Hubble tension may share a common origin, possibly related to systematic effects or new physics. Using multiple phenomenological parameterizations, the deviation parameter is also found to be statistically consistent with zero (e.g., $\eta_1 = 0.023 \pm 0.027$ for the linear model under Planck priors). Furthermore, the study finds no statistically significant evidence for cosmic opacity. The average of the opacity derivative is compatible with zero ($\langle d\tau/dz \rangle = 0.0409 \pm 0.1024$ for GPR and $0.0730 \pm 0.1607$ for FKM). Based on these null results, stringent constraints are placed on the parameter space of Axion-Like Particles (ALPs) and Mini-Charged Particles (MCPs).\footnotetext{Corresponding authors.}}

\begin{document}
\maketitle
\flushbottom

\section{Introduction}\label{sec:intro}

The cosmic distance duality relation (CDDR), a basic test of standard cosmology, establishes a fundamental connection between the luminosity distance (LD, $D_L$) and the angular diameter distance (ADD, $D_A$) through the expression
\begin{equation}
    \eta(z) = \frac{D_L(z)}{D_A(z)(1+z)^2},\label{eq:ddr}
\end{equation}
where $z$ denotes the redshift \cite{1971grc..conf..104E}. This relation, first articulated by Etherington as the reciprocity theorem \cite{1933PMag...15..761E}, emerges from the geometrical reciprocity principle and relies on three key assumptions: 1) a metric theory of gravity, 2) conservation of photon number, and 3) photons traveling along unique null geodesics. Under these conditions, the CDDR predicts $\eta(z) = 1$ across all redshifts in a standard cosmological framework.

Testing the validity of CDDR serves as a powerful tool for probing potential deviations from the standard cosmological model (SM), the $\Lambda$CDM paradigm. Violations of this relation could indicate systematic errors in distance measurements, modifications of general relativity in which photons do not follow null geodesics \cite{Uzan_2004,Santana_2017} and nonminimal coupling (NMC) to the matter fields \cite{Azevedo_2021}, or exotic physical processes such as the absorption or scattering of photons by intergalactic dust \cite{Corasaniti_2006}, axion-photon conversion \cite{Tiwari_2017,buenabad2022constraintsaxionscosmicdistance}, and cosmological opacity \cite{More_2009,Remya_Nair_2012}. Apart from testing the CDDR itself, it can also be used to constrain the mass density profile of galaxy clusters \cite{cao2011distancedualityrelationtemperature,Cao_2016}. In cosmology, recent research shows that it can be used to explain the evolution of dark energy and the Hubble tension \cite{teixeira_implications_2025,alfano_cosmic_2025}. Investigating $\eta(z)$ thus provides a unique window into these unresolved issues and the underlying physics governing the universe.

To test CDDR, various observational probes have been employed, each with distinct advantages and limitations. Type Ia supernovae (SNIa), acting as standard candles, offer precise measurements of $D_L$ through their apparent magnitudes ($m_B$), as demonstrated by datasets like Pantheon+ \cite{Brout_2022}, DES \cite{Sanchez_2024} and Union \cite{rubin2025unionunitycosmology2000}. In addition to SNIa, other distance indicators such as Gamma Ray Bursts (GRBs) \cite{Wang_2022} and Quasars (QSOs) \cite{Zheng_2020} have been utilized, while forecast analyses have explored the future potential of Gravitational Wave (GW) standard sirens \cite{PhysRevD.99.063507,PhysRevD.99.083523,yang2019constraintscosmicdistanceduality}. ADD can also be obtained from the Sunyaev-Zel'dovich (SZ) effect in galaxy clusters (GC) \cite{PhysRevD.69.101305,Uzan_2004,DeBernardis:2006ii,Holanda_2010,alfano_cosmic_2025}, gas mass fraction measurements in GC \cite{Meng_2012,Bora_2021}, quasars (QSO) \cite{PhysRevD.99.063507,tonghua2023recentobservationstellinglight}, strong gravitational lensing (SGL) \cite{Rana_2017,PhysRevD.103.103513,Liao_2019,huang_opacity-free_2024,Qi_2025}, baryon acoustic oscillations (BAO) \cite{xu_model-independent_2022,favale2024quantification2dvs3d}, and gravitational-wave events \cite{deleo2025distinguishingdistancedualitybreaking}.

A key challenge in testing the CDDR is the mismatch in redshifts between LD and ADD data. To address this, some studies employ a binning strategy, selecting LD and ADD data pairs within a small redshift interval (e.g., $\Delta z = |z_\text{LD}-z_\text{ADD}| \leq 0.005$) \cite{Holanda_2010,Li_2011,yang_testing_2024}. Others utilize reconstruction techniques to bridge redshift gaps, such as linear or polynomial fitting \cite{Holanda_2016}, Gaussian process regression (GPR) \cite{PhysRevD.90.124064,bengaly_null_2022,wu_null_2023,gao_null_2025}, or artificial neural networks (ANN) \cite{tang_deep_2023,keil_probing_2025}. A widely used approach is to constrain parameterized forms of $\eta(z)$ \cite{Remya_Nair_2012,Liao_2016,wang_testing_2024}, while some works adopt template-free, statistical tests without explicit parameterizations \cite{ma_statistical_2018}.

In this work, we present an analysis of the CDDR using the latest Dark Energy Spectroscopic Instrument (DESI) Data Release 2 (DR2) combined with Pantheon+ SNIa and cosmic chronometer (CC) data (section~\ref{data}). To address the inherent redshift mismatch among these disparate datasets, we employ two distinct and complementary model-independent reconstruction techniques to align the distance indicators reliably (section~\ref{match}). Building upon this aligned framework, our approach employs multiple complementary methodologies: First, we directly measure the evolution of $\eta(z)$ under different priors, and conversely, we constrain the supernova absolute magnitude assuming the validity of the CDDR to investigate the impact of the Hubble tension (section~\ref{sec:reconstruction}). Second, we adopt parametric approaches that treat calibration parameters as free nuisance variables, thereby marginalizing over the absolute distance scale to provide robust, model-independent constraints on CDDR deviations (section~\ref{sec:parametrized}). Finally, we investigate cosmic transparency through differential opacity tests and constrain exotic physics models (section~\ref{sec:opacity}).

\section{Data}\label{data}

\subsection{Baryon Acoustic Oscillations from DESI Data Release 2}
Baryon Acoustic Oscillations (BAO), arising from sound waves in the early universe, serve as a standard ruler for measuring cosmological distances. BAO measurements constrain distance scales relative to the sound horizon at the drag epoch, $r_d$, which is defined as:
\begin{equation}
    r_d=r_s(z_d) = \int_{z_d}^\infty \frac{c_s(z')}{H(z')} dz',
\end{equation}
where $z_d$ is the redshift of the drag epoch and $c_s$ is the sound speed of the baryon-photon fluid. 

In this work, we utilize the high-precision BAO measurements from the Dark Energy Spectroscopic Instrument (DESI) Data Release 2 (DR2) \cite{desicollaboration2025desidr2resultsi,desicollaboration2025desidr2resultsii}. This dataset provides a robust distance ladder over the redshift range $0.295 \le z \le 2.33$. The primary observables summarized in table~\ref{tab:desi_dr2_full} include the transverse comoving distance $D_M(z)/r_d$, the Hubble distance $D_H(z)/r_d$, and the volume-averaged distance $D_V(z)/r_d$. These are related to the ADD $D_A(z)$ and the Hubble parameter $H(z)$ through the following relations:
\begin{align}
    \frac{D_M(z)}{r_d} = \frac{(1+z) D_A(z)}{r_d}, \quad \frac{D_H(z)}{r_d} = \frac{c}{H(z) r_d}.
\end{align}
For the BGS sample at $z_{\text{eff}} = 0.295$, only the isotropic measurement $D_V(z)/r_d = [z D_M^2(z) D_H(z)]^{1/3} / r_d$ is provided. 

For the anisotropic samples (LRG, ELG, QSO, and Ly$\alpha$), the transverse and radial measurements are obtained simultaneously from the same galaxy distribution. Consequently, they are statistically correlated, primarily due to the Alcock-Paczynski effect during the BAO feature fitting. To account for this, we define the covariance matrix for each redshift bin as:
\begin{equation}
\mathbf{C}_{\text{BAO}} = \begin{bmatrix}
\sigma_{D_M/r_d}^2 & r \cdot \sigma_{D_M/r_d} \sigma_{D_H/r_d} \\
r \cdot \sigma_{D_M/r_d} \sigma_{D_H/r_d} & \sigma_{D_H/r_d}^2
\end{bmatrix},
\end{equation}
where $\sigma_{D_M/r_d}$ and $\sigma_{D_H/r_d}$ are the standard deviations, and $r$ is the correlation coefficient between the two components. For the isotropic BGS measurement, the covariance reduces to a single variance term $\sigma^2_{D_V/r_d}$. The specific methodology for using these observables to test the CDDR is detailed in section~\ref{sec:reconstruction}.

\begin{table}[h]
\centering
\caption{DESI DR2 BAO measurements used in this analysis. The columns provide the isotropic ($D_V/r_d$), transverse ($D_M/r_d$), and radial ($D_H/r_d$) distance scales, along with the correlation coefficient $r_{M,H}$ between the transverse and radial components.}
\label{tab:desi_dr2_full}
\begin{tabular}{lccccc}
\hline
Sample & $z_{\text{eff}}$ & $D_V/r_d$ & $D_M/r_d$ & $D_H/r_d$ & $r_{M,H}$ \\
\hline
BGS       & 0.295 & $7.944 \pm 0.075$  & --                 & --                 & --      \\
LRG1      & 0.510 & $12.720 \pm 0.098$ & $13.587 \pm 0.169$ & $21.863 \pm 0.427$ & $-0.475$ \\
LRG2      & 0.706 & $16.048 \pm 0.110$ & $17.347 \pm 0.180$ & $19.458 \pm 0.332$ & $-0.423$ \\
LRG3+ELG1 & 0.934 & $19.720 \pm 0.091$ & $21.574 \pm 0.153$ & $17.641 \pm 0.193$ & $-0.425$ \\
ELG2      & 1.321 & $24.256 \pm 0.174$ & $27.605 \pm 0.320$ & $14.178 \pm 0.217$ & $-0.437$ \\
QSO       & 1.484 & $26.059 \pm 0.400$ & $30.519 \pm 0.758$ & $12.816 \pm 0.513$ & $-0.489$ \\
Ly$\alpha$ & 2.330 & $31.267 \pm 0.256$ & $38.988 \pm 0.531$ & $8.632 \pm 0.101$  & $-0.431$ \\
\hline
\end{tabular}
\end{table}

\subsection{Standard Candle Supernova Data}
Type Ia supernovae (SNIa) are widely regarded as standard candles and play a crucial role in testing the CDDR. In this work, we utilize the Pantheon+ dataset, which includes 1701 SNIa light curves from 1550 spectroscopically confirmed samples spanning the wide redshift range $0.001 < z < 2.26$ \cite{Scolnic_2022,Brout_2022}\footnote{\url{https://github.com/PantheonPlusSH0ES/DataRelease}} (DES-SN5YR \cite{Sanchez_2024} and Union3 \cite{rubin2025unionunitycosmology2000} are also available, but they lack high redshift data points). The observed distance modulus for each SNIa is defined as:
\begin{equation}
    \mu_\text{obs}(z) = m_{B,\text{obs}}(z) - (M_B - \alpha X_1 + \beta C),
\end{equation}
where $m_{B,\text{obs}}$ is the observed peak magnitude in the rest-frame $B$-band, $M_B$ is the absolute magnitude, $X_1$ is the light-curve stretch parameter, and $C$ is the color parameter. The nuisance parameters $\alpha$ and $\beta$ quantify the correlations between luminosity and the light-curve shape/color. 

In the Pantheon+ analysis, these light-curve corrections have already been applied to the provided magnitudes. Thus, the corrected observed magnitude, $m_{B} = m_{B,\text{obs}} + \alpha X_1 - \beta C$, can be directly used for cosmological analysis. The simplified distance modulus relates to the LD $D_L(z)$ as follows:
\begin{equation}
    \mu(z) = m_{B}(z) - M_B = 5\log_{10}(D_L(z)) + 25. \label{eq:DLmu}
\end{equation}
Equivalently, the LD can be explicitly expressed as a function of the apparent magnitude $m_B(z)$ and the absolute magnitude $M_B$:
\begin{equation}
    D_L(z, M_B) = 10^{\frac{m_B(z) - M_B - 25}{5}}. \label{eq:DL_mB}
\end{equation}
Here, $D_L$ is expressed in units of Mpc, and $M_B$ is the absolute magnitude which we treat as a nuisance parameter.

To ensure a rigorous statistical treatment, we utilize the full Pantheon+ covariance matrix $\mathbf{C}_{\text{SN}}$, which includes both statistical uncertainties and systematic errors from various sources such as calibration and bias corrections. This covariance matrix is incorporated into our Gaussian Process Regression (GPR) and Free-Knots Method (FKM) to accurately reconstruct the LD at the BAO effective redshifts.

\subsection{Cosmic Chronometers Data}
Cosmic chronometers (CC) provide a model-independent estimate of the Hubble parameter $H(z)$ by measuring the differential age evolution of passively evolving galaxies. In this analysis, CC measurements are essential for de-projecting the volume-averaged distance $D_V(z)/r_d$ of the BGS sample into its transverse and radial components without assuming a specific cosmological model. We adopt the compilation from \cite{Favale_2023}, which consists of 32 CC data points spanning the redshift range $0.07 < z < 1.965$ \cite{Jimenez_2003,Daniel_Stern_2010,Zhang_2014,PhysRevD.71.123001,Moresco_2012,Moresco_2015,Moresco_2016}. 

To ensure a rigorous error analysis, we follow the methodology of Moresco et al. \cite{moresco_setting_2020,moresco_unveiling_2022} to construct the full covariance matrix $\mathbf{C}_{\text{CC}}$ for our compiled 32 CC measurements. We generate this matrix utilizing the publicly available pipeline\footnote{\url{https://gitlab.com/mmoresco/CCcovariance}}. This matrix incorporates both statistical uncertainties and systematic effects:
\begin{equation}
    \mathbf{C}_{\text{CC}} = \mathbf{C}_{\text{stat}} + \mathbf{C}_{\text{sys}}.
\end{equation}
Specifically, the systematic covariance matrix is decomposed into physically motivated terms: $\mathbf{C}_{\text{sys}} = \mathbf{C}_{\text{met}} + \mathbf{C}_{\text{young}} + \mathbf{C}_{\text{model}}$. The components $\mathbf{C}_{\text{met}}$ and $\mathbf{C}_{\text{young}}$ account for uncertainties in estimating stellar metallicity and potential contamination from young stellar components, respectively. Since these effects depend on the properties of individual galaxies at specific redshifts, they are treated as purely diagonal (uncorrelated) matrices. In contrast, the modeling component $\mathbf{C}_{\text{model}}$ encapsulates systematic uncertainties arising from the assumed star formation history (SFH), initial mass function (IMF), stellar library, and stellar population synthesis (SPS) model. Because the same theoretical models are applied globally to all galaxies, these terms induce fully correlated off-diagonal errors across different redshift bins. By executing the aforementioned pipeline on our specific dataset, we construct the exact full covariance matrix and robustly account for these inter-bin correlations during our Hubble parameter reconstruction process.

\subsection{Other Supplementary Data}
In addition to the observational probes described above, we adopt several fundamental constants and priors for our analysis. The speed of light is taken as $c = 299,792.458$ km/s. To investigate how the current Hubble tension influences the tests of the CDDR and to ensure the logical independence of our results, we consider three distinct sets of priors for the SNIa absolute magnitude $M_B$ and the sound horizon $r_d$:

\begin{itemize}
    \item \emph{Local Distance Ladder (R22)}: We adopt $M_B^{\text{R22}} = -19.253 \pm 0.027$ mag, as calibrated by Cepheids in the local distance ladder \cite{Riess_2022}. This prior represents a high-$H_0$ perspective and is widely used as the local anchor for supernova data.
    \item \emph{CDDR-Independent Calibration (DB23)}: We adopt the prior obtained from the combination of SNIa and CC data, $M_B^{\text{DB23}} = -19.384 \pm 0.052$ mag \cite{Dinda_2023}. Crucially, this derivation relies solely on low-redshift geometric probes and does not invoke BAO or CMB constraints which typically assume the validity of the CDDR ($\eta=1$). Consequently, this prior provides a statistically independent and unbiased anchor suitable for testing CDDR.
    \item \emph{Early Universe (P18)}: For the sound horizon at the drag epoch, we adopt the value from the Planck 2018 final release, $r_d^{\text{P18}} = 147.09 \pm 0.26$ Mpc \cite{planck2018}. This prior, derived from the Cosmic Microwave Background (CMB) under the assumption of $\Lambda$CDM, represents the early-universe perspective.
\end{itemize}

These priors allow us to systematically assess how the choice of cosmological calibration ranging from local measurements to early-universe constraints affects the reconstructed $\eta(z)$ evolution in section~\ref{sec:reconstruction}.

\section{Methods for Redshift Matching}\label{match}

As discussed in section~\ref{sec:intro}, testing the CDDR requires matching the redshifts of the two main probes: the Pantheon+ SNIa dataset (with many discrete points) and the DESI DR2 BAO dataset (reported at specific effective redshifts). To compute the CDDR estimator $\eta(z)$ at each BAO redshift, the LD $D_L(z)$ and the Hubble parameter $H(z)$ must be inferred at the same $z_{\text{eff}}$.

To ensure the results are not biased by a specific cosmological model, we employ two distinct and complementary reconstruction methodologies: Gaussian Process Regression (GPR) and the Free-Knots Method (FKM). The comparison between these two methods allows us to quantify the systematic uncertainties stemming from the redshift-matching process itself.

\subsection{Gaussian Process Regression (GPR)}

Gaussian Process Regression provides a non-parametric Bayesian framework to reconstruct continuous functions from discrete, noisy observations without the need for a pre-defined functional form \cite{Rasmussen2004}. In our analysis, we model the evolution of cosmological quantities as a Gaussian process, denoted as $f(z) \sim \mathcal{GP}(\mu(z), k(z, z'))$.

A Gaussian process is defined such that any finite collection of function values follows a joint multivariate Gaussian distribution. Given the observed data vector $\mathbf{y}$ at redshifts $\mathbf{z}$ and their corresponding full covariance matrix $\mathbf{C}$ (incorporating both statistical and systematic uncertainties), the joint distribution of the observed data and the predicted function values $\mathbf{f}_*$ at the target BAO redshifts $\mathbf{z}_*$ is given by:
\begin{equation}
\begin{bmatrix} \mathbf{y} \\ \mathbf{f}_* \end{bmatrix} \sim \mathcal{N} \left( 0, \begin{bmatrix} \mathbf{K}(\mathbf{z}, \mathbf{z}) + \mathbf{C} & \mathbf{K}(\mathbf{z}, \mathbf{z}_*) \\ \mathbf{K}(\mathbf{z}_*, \mathbf{z}) & \mathbf{K}(\mathbf{z}_*, \mathbf{z}_*) \end{bmatrix} \right),
\end{equation}
where $\mathbf{K}$ represents the kernel matrix that encodes the correlation between different points in the redshift space. 

The choice of the kernel function $k(z, z')$ is crucial, as it determines both the smoothness of the reconstructed function and the characteristic correlation scale. In this work, we evaluate several commonly used kernels by training them on the CC dataset and comparing their goodness-of-fit using the reduced chi-square statistic and the log-marginal likelihood (see appendix~\ref{appendix:a}). Previous studies have shown that kernels from the Matérn family generally outperform the widely used Radial Basis Function (RBF) kernel in cosmological applications involving CC, SNIa, and GRB data \cite{Zhang_2023}. Motivated by these results, we consider the general Matérn class of covariance functions, parametrized by a smoothness parameter $\nu$, which allows for a controlled interpolation between rough and overly smooth reconstructions. The Matérn kernel is defined as:
\begin{equation}
k_\nu(z, \tilde{z}) = \sigma_f^2 \, \frac{2^{1-\nu}}{\Gamma(\nu)}
\left( \frac{\sqrt{2\nu}\,|z-\tilde{z}|}{\ell} \right)^{\nu}
K_\nu\!\left( \frac{\sqrt{2\nu}\,|z-\tilde{z}|}{\ell} \right),
\end{equation}
where $\sigma_f^2$ denotes the signal variance and $\ell$ the characteristic correlation length. Based on the kernel comparison analysis presented in appendix~\ref{appendix:a}, we find that the Matérn kernel with $\nu = 3/2$ provides the most consistent overall performance across the datasets considered, and we therefore adopt it as our fiducial choice. We perform full Bayesian marginalization over the hyperparameters using an MCMC sampler based on the log-marginal likelihood (we use the \texttt{emcee}\cite{emcee} for visualization):
\begin{equation}
\ln p(\mathbf{y}|\mathbf{z}, \ell, \sigma_f) = -\frac{1}{2} \mathbf{y}^T [\mathbf{K} + \mathbf{C}]^{-1} \mathbf{y} - \frac{1}{2} \ln |\mathbf{K} + \mathbf{C}| - \frac{n}{2} \ln(2\pi).
\end{equation}

The implementation details differ between the two probes to optimize reconstruction performance. For the CC, the reconstruction of $H(z)$ is carried out in linear redshift space. Conversely, for the Pantheon+ dataset, we reconstruct the apparent magnitude $m_B(z)$ in logarithmic redshift space. This logarithmic transformation linearizes the magnitude-redshift relation. The resulting reconstructed functions evaluated at the BAO effective redshifts are displayed in figure~\ref{fig:recons}.

\subsection{Free-Knots Method (FKM)}

As a statistical cross-check, we also implement the Free-Knots Method (FKM), a control-point based reconstruction technique. Unlike GPR, which relies on a global kernel, FKM approximates the target function by optimizing its values at a set of predefined nodes (knots). The knot positions are set to coincide with the BAO effective redshifts, with an additional low-redshift anchor to fix the global normalization. To maintain a smooth and physically consistent evolution, we employ the Monotonic Piecewise Cubic Hermite Interpolating Polynomial (PCHIP) for interpolation, which prevents unphysical oscillations often associated with high-order polynomials.

\begin{figure}[htb]
  \centering
  \subfloat 
  { 
  \includegraphics[width=0.49\textwidth]{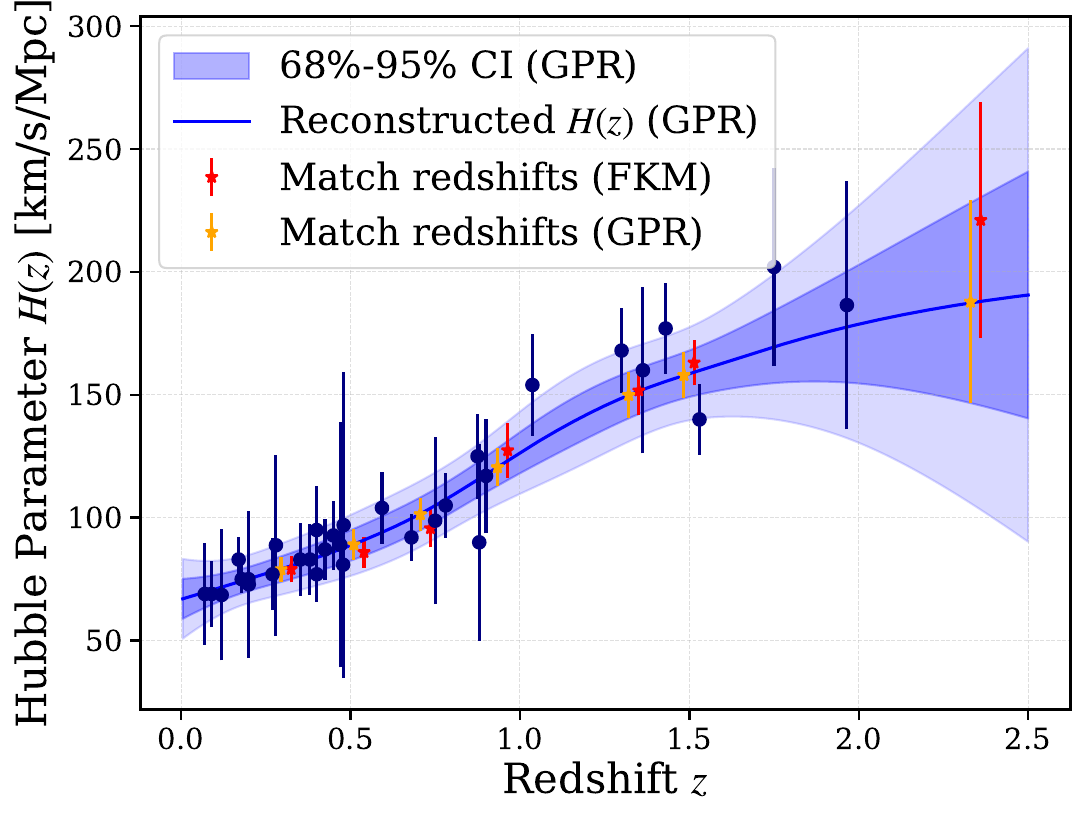}
  }
  \subfloat
  {
  \includegraphics[width=0.49\textwidth]{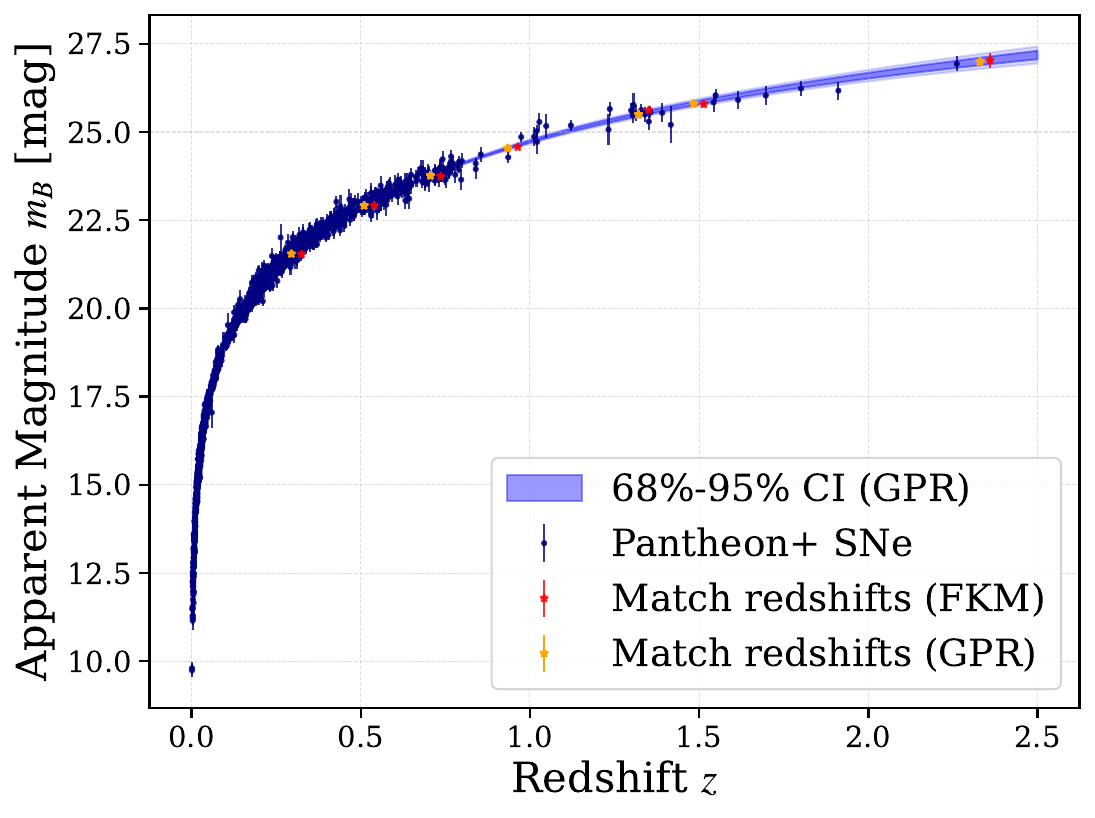}
  }
  \caption{Reconstruction of the expansion and distance indicators. In both panels, the blue points represent the original observational data (CC for the left panel and Pantheon+ for the right panel). The orange and red points denote the reconstructed values at the target BAO redshifts derived using the GPR and FKM methodologies, respectively. The shaded areas indicate the 68\%  and 95\% confidence intervals associated with the GPR reconstructions.}
  \label{fig:recons} 
\end{figure}

The reconstruction is performed by optimizing the knot values via MCMC sampling to minimize the $\chi^2$ statistics for both datasets. For the CC data, the objective function is defined as:
\begin{equation}
\chi^2_H = \left[\mathbf{H}^{\rm model}(z_{\rm CC})-\mathbf{H}^{\rm obs}(z_{\rm CC})\right]^T \mathbf{C}_{\rm CC}^{-1} \left[\mathbf{H}^{\rm model}(z_{\rm CC})-\mathbf{H}^{\rm obs}(z_{\rm CC})\right],
\end{equation}
where $\mathbf{C}_{\rm CC}$ is the full covariance matrix including systematic errors. Similarly, for the SNIa data, we reconstruct the corrected apparent magnitude $m_B(z)$ in logarithmic redshift space by minimizing:
\begin{equation}
\chi^2_{\rm SN} = \left[\mathbf{m}_B^{\rm model}(z_{\rm SN})-\mathbf{m}_B^{\rm obs}(z_{\rm SN})\right]^T \mathbf{C}_{\rm SN}^{-1} \left[\mathbf{m}_B^{\rm model}(z_{\rm SN})-\mathbf{m}_B^{\rm obs}(z_{\rm SN})\right].
\end{equation}
The optimal knot values obtained from these chains provide the reconstructed $H(z)$ and $m_B(z)$ values at the specific BAO redshifts required for the subsequent CDDR test (also seen red points in figure~\ref{fig:recons}).

\section{The Test of Cosmic Distance Duality Relation}
Having successfully reconstructed the expansion history $H(z)$ and the supernova magnitude-redshift relation $m_B(z)$ at the corresponding BAO effective redshifts, we can combine these independent distance indicators to test the reciprocity theorem. By substituting the matched LD and ADD into Eq.~(\ref{eq:ddr}), we derive the observational constraints on the duality parameter $\eta(z)$ and evaluate its consistency with the standard expectation of a transparent universe.

\subsection{Measurement Directly with Different $M_B$ Priors}\label{sec:reconstruction}
To perform a model-independent measurement of the CDDR, we first obtain the LD $D_L(z)$ using the reconstructed SNIa apparent magnitudes via eq.~(\ref{eq:DL_mB}). For the ADD $D_A(z)$, we derive it by combining the transverse ($D_M/r_d$) and radial ($D_H/r_d$) BAO measurements with the reconstructed Hubble parameter $H(z)$ from CC:
\begin{equation}
D_A(z) = \frac{c}{(1+z)H(z)} \frac{D_M(z)/r_d}{D_H(z)/r_d}.
\end{equation}
Note that the $r_d$ terms in the ratio of transverse and radial BAO observables cancel out, making this distance measurement purely geometric and independent of early-universe physics. The isotropic BGS sample is excluded from this specific test as it does not allow for this $r_d$-cancellation.

The covariance matrix $\mathbf{C}_{D_A}$ is obtained via Jacobian error propagation, assuming independence between BAO observables and $H(z)$ data:
\begin{equation}
\mathbf{C}_{D_A} = \mathbf{J}_{BAO} \mathbf{C}_{BAO} \mathbf{J}_{BAO}^T + \mathbf{J}_{H} \mathbf{C}_{H} \mathbf{J}_{H}^T,
\end{equation}
where Jacobian matrices are defined by diagonal elements:
\begin{equation}
\mathbf{J}_{BAO} = \left[ \text{diag}\left(\frac{D_A}{D_M/r_d}\right), \; \text{diag}\left(-\frac{D_A}{D_H/r_d}\right) \right], \quad \mathbf{J}_{H} = \text{diag}\left(-\frac{D_A}{H(z)}\right).
\end{equation}

Assuming the null hypothesis: $\eta(z) = 1$, the measured $D_A$ is converted to a model-equivalent LD $D_L(z) = (1+z)^2 D_A(z)$ with error propagation:
\begin{equation}
\mathbf{C}_{D_L} = \mathbf{S} \mathbf{C}_{D_A} \mathbf{S}^T, \quad \text{with} \quad \mathbf{S} = \text{diag}\left((1+z)^2\right).
\end{equation}

\begin{figure}[ht]
    \centering
    {\includegraphics[width=0.5\textwidth]{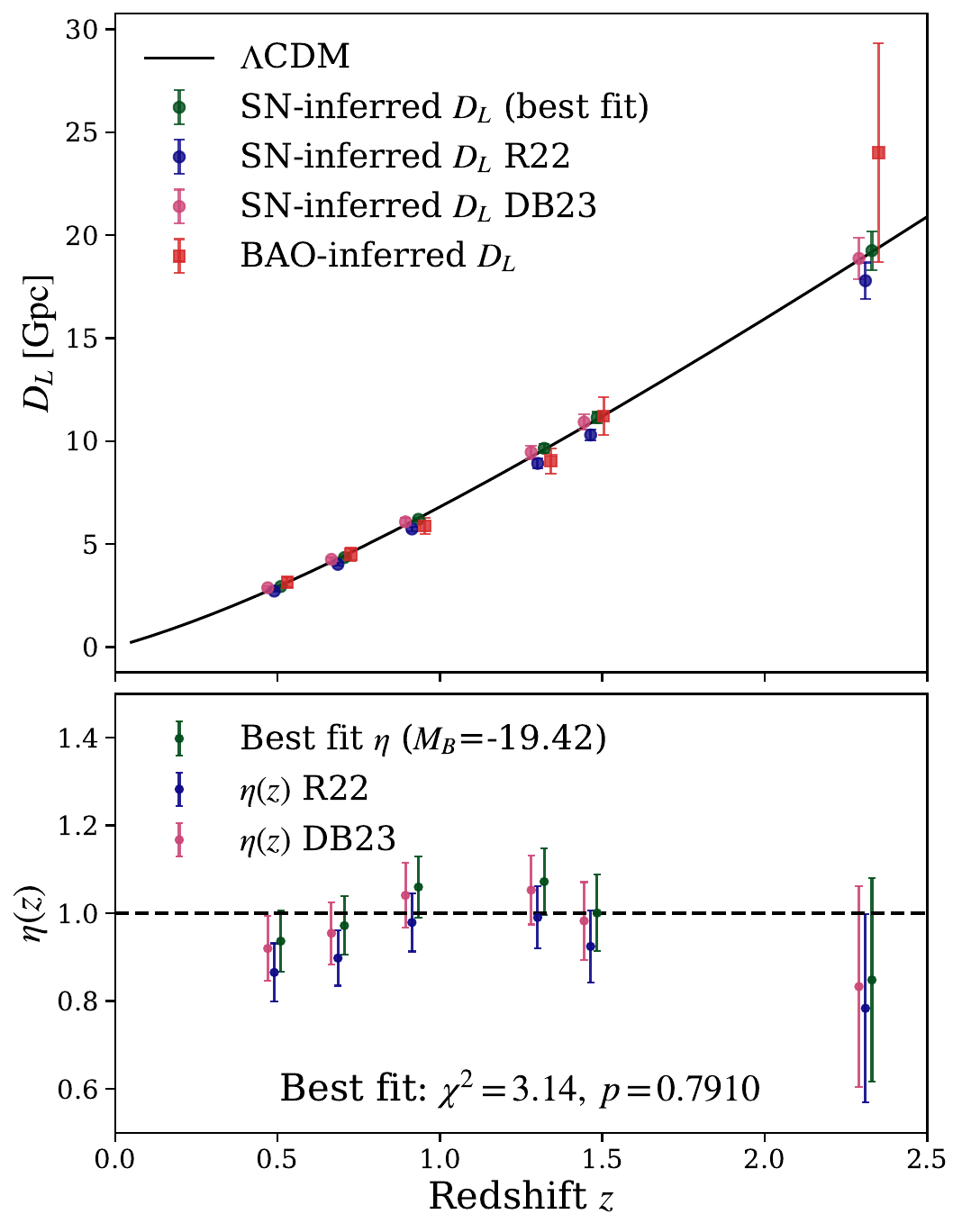}}\hfill
    {\includegraphics[width=0.5\textwidth]{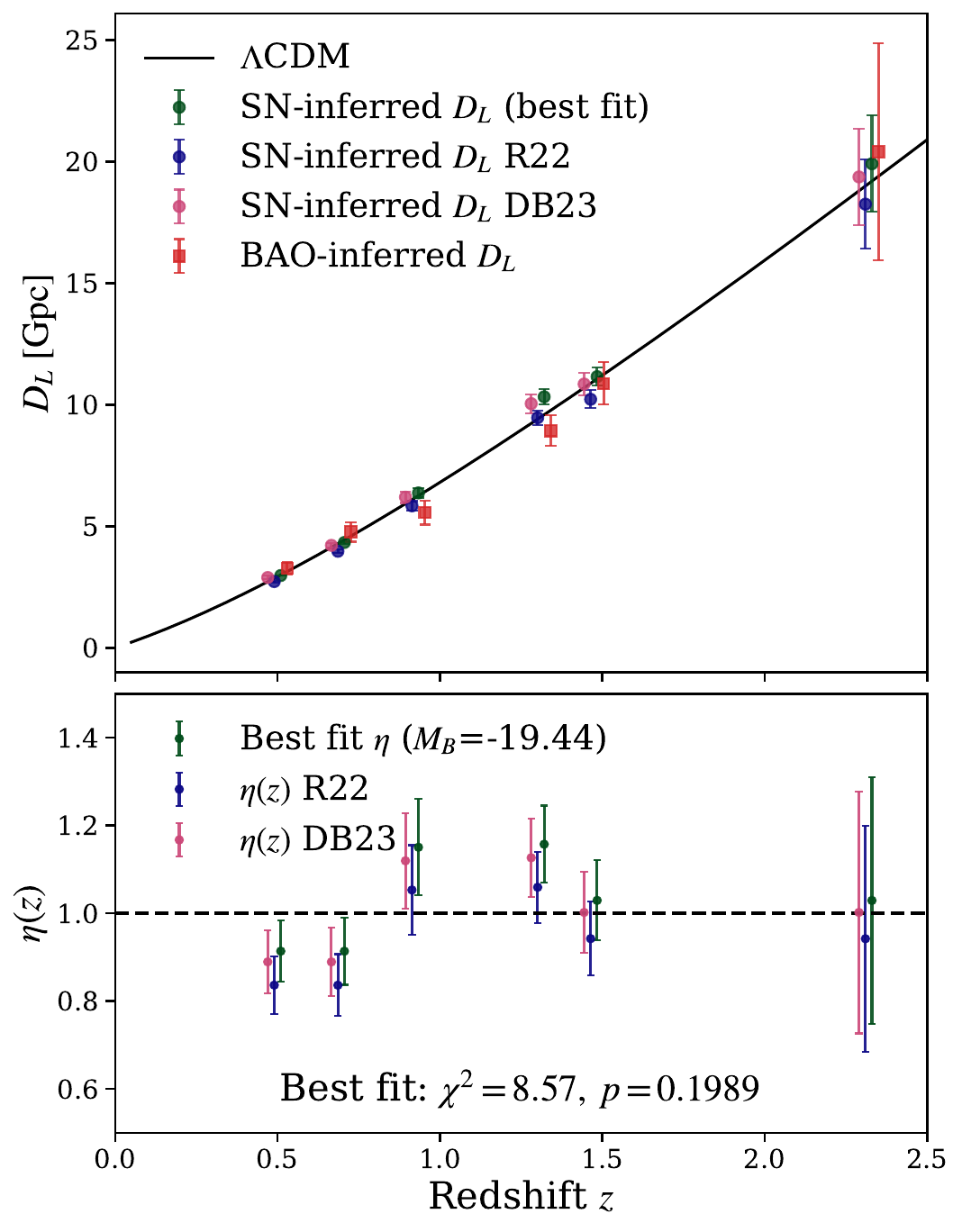}}
    \caption{Direct and model-independent measurement of $\eta(z)$ at DESI BAO effective redshifts. The left and right panels display results obtained using the GPR and FKM redshift-matching techniques, respectively. The data points illustrate the sensitivity of $\eta(z)$ to different supernova absolute magnitude calibrations: the local distance ladder prior (R22), the CDDR-independent calibration (DB23), and the internal best-fit value. The horizontal dashed line represents the null hypothesis: $\eta(z)=1$ (i.e., a transparent universe with metric duality). Error bars denote $1\sigma$ uncertainties.}
    \label{fig:measurement}
\end{figure}

As summarized in table~\ref{tab:measurement_results} and illustrated in figure~\ref{fig:measurement}, we evaluate the statistical consistency of our measurements with the null hypothesis $\eta(z) = 1$. The resulting $p$-values and significance levels indicate that the observations are broadly consistent with the null hypothesis across the matching methods and calibration scenarios considered. In all cases, the deviations from the reciprocity theorem remain within the $2\sigma$ confidence level, suggesting that the CDDR holds under current observational precision. We also note that the $p$-values tend to increase as the $M_B$ prior shifts towards the lower $H_0$ regime (DB23 or the Best-fit scenario), indicating a slightly improved alignment with the null hypothesis.

As shown in figure~\ref{fig:measurement}, the reconstructed $\eta(z)$ values fluctuate around the unity line without displaying a smooth redshift-dependent trend. Even in the case with the relatively higher significance (FKM under the R22 prior, where $p=0.0721$ and significance is $1.80\sigma$), the scatter remains random. This suggests that such discrepancies may be driven by statistical noise or a potential underestimation of observational uncertainties for individual data points. Thus, the discrepancies induced by different calibration priors and the Hubble tension may share a common origin.

\begin{figure}[h]
    \centering
    {\includegraphics[width=0.49\textwidth]{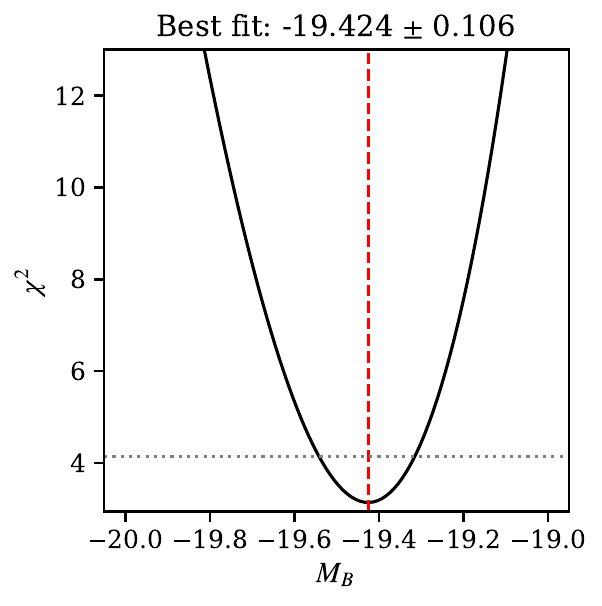}}\hfill
    {\includegraphics[width=0.49\textwidth]{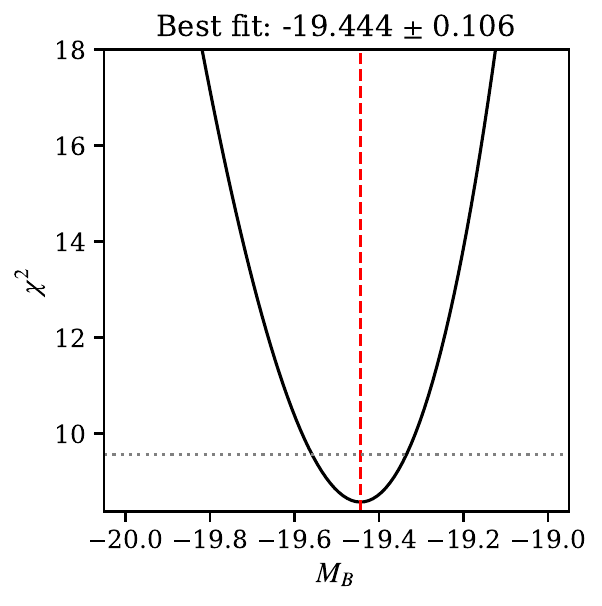}}
    \caption{Constraints on the SNIa absolute magnitude $M_B$ assuming the validity of the CDDR ($\eta=1$). The left and right panels display the $\chi^2$ profiles derived from the GPR and FKM methods, respectively. The horizontal dashed line represents the $1\sigma$ constraints. The red vertical dashed line indicates the central value.}
    \label{fig:bestfit_plots}
\end{figure}

The ``Best-fit" scenario in table~\ref{tab:measurement_results} provides the absolute minimum $\chi^2$ by treating $M_B$ as a free parameter. Assuming the CDDR is valid, we find the best-fit values to be $M_B = -19.424 \pm 0.106$ mag for GPR and $M_B = -19.444 \pm 0.106$ mag for FKM (see figure~\ref{fig:bestfit_plots}). These internally calibrated values are consistent with the DB23 prior ($-19.384 \pm 0.052$ mag) but show a more noticeable discrepancy with the R22 prior ($-19.253 \pm 0.027$ mag). This indicates that the DESI and Pantheon+ datasets are compatible with the reciprocity theorem when anchored to a Planck-like absolute distance scale.

\begin{table}[h]
\centering
\caption{Statistical consistency of measured $\eta(z)$ with null hypothesis: $\eta(z)=1$. The $\chi^2$ statistics and $p$-values are calculated using the full covariance matrix. The significance levels quantify the deviation from the null hypothesis.}
\label{tab:measurement_results}
\begin{tabular}{ccccc}
\hline
Matching Method & Calibration Prior & $\chi^2$ & $p$-value & Significance \\
\hline
GPR & R22      & 5.52  & 0.4785 & $0.71\sigma$ \\
GPR & DB23     & 3.25  & 0.7769 & $0.28\sigma$ \\
GPR & Best-fit & 3.14  & 0.7910 & $0.26\sigma$ \\
\hline
FKM & R22      & 11.57 & 0.0721 & $1.80\sigma$ \\
FKM & DB23     & 8.85  & 0.1820 & $1.33\sigma$ \\
FKM & Best-fit & 8.57  & 0.1989 & $1.28\sigma$ \\
\hline
\end{tabular}
\end{table}

\subsection{Parameterized Constraint of CDDR}\label{sec:parametrized}   
The direct measurements suggest that the choice of priors for $M_B$ and $r_d$ primarily induces a constant vertical offset in $\eta(z)$, rather than a distinct redshift-dependent evolution. This observation implies that fixing these calibration parameters to values in tension could lead to a misleading detection of CDDR violation that is merely a normalization artifact. Therefore, fitting for $\eta(z)$ evolution without handling this degeneracy might risk overinterpretation. To rigorously disentangle the constant calibration offset from genuine physical deviations, we proceed with a parametrized analysis where $M_B$ and $r_d$ are treated as free nuisance parameters. By marginalizing over the absolute scale, we can robustly test whether the data supports any intrinsic redshift evolution of $\eta(z)$, or if the deviation is fully absorbed by the calibration parameters.

The key advantage of this parametric approach lies in its ability to marginalize over the degenerate nuisance parameters $M_B$ and $r_d$, either numerically or analytically, thereby eliminating their explicit prior dependence. While this comes at the cost of introducing phenomenological models for $\eta_i$, it allows us to directly test the CDDR null hypothesis by examining whether the deviation parameters $\eta_i$ are consistent with zero.
Following extensive literature on CDDR tests, we consider four widely used phenomenological parameterizations that encompass different possible redshift dependencies of CDDR violations (see e.g., \cite{Avgoustidis_2009, Holanda_2010, Liang_2013, yang_testing_2024}):
\begin{align}
\text{(Linear)} \quad & \eta(z) = 1 + \eta_1 z, \\
\text{(Fractional)} \quad & \eta(z) = 1 + \frac{\eta_2 z}{1+z}, \\
\text{(Logarithmic)} \quad & \eta(z) = 1 + \eta_3 \ln(1+z), \\
\text{(Power law)} \quad & \eta(z) = (1+z)^{\epsilon}.
\end{align}
The first model corresponds to a first-order Taylor expansion, though it may diverge at high $z$. The second model behaves well at high $z$ and evolves more slowly. The third is a logarithmic form. The fourth is a power-law parametrization, which is commonly applied in cosmic transparency tests\cite{Liang_2013}. The parameters here are respectively denoted as $\eta_i$ (where $i = 1, 2, 3$). Particularly, in the transparency test, we use the conventional symbol $\epsilon = \eta_4$.

To perform a model-independent test, we treat the SNIa absolute magnitude $M_B$ and the sound horizon scale $r_d$ as free nuisance parameters. This approach allows the relative evolution of distance scales to be tested without being biased by the absolute distance scale tension.

In our joint Bayesian analysis, the observed CDDR parameter $\vec{\eta}_{\text{obs}}$ is calculated at each BAO effective redshift, denoted here as $z_j$ (to distinguish from the model index $i$). The LD term is directly derived from the reconstructed SNIa apparent magnitudes $m_B(z_j)$ following eq.~(\ref{eq:DL_mB}). For the ADD, our approach differs from the direct measurement in section \ref{sec:reconstruction}. Depending on the signal-to-noise ratio and survey volume, BAO measurements are classified as either ``anisotropic'' or ``isotropic'' based on their clustering signal extraction \cite{desicollaboration2025desidr2resultsii}. For anisotropic points, high signal-to-noise allows the robust separation of transverse ($D_M/r_d$) and radial ($D_H/r_d$) components via the Alcock-Paczynski effect. Since $r_d$ is treated as a free parameter in our Bayesian inference, we can directly utilize $D_M/r_d$ to derive $D_A$. This advantageously avoids reliance on external Hubble data, thereby reducing systematic uncertainties. In contrast, for isotropic points (samples of nearby bright galaxies), limited survey volume only permits a 1D spherically averaged measurement, yielding a single volume-averaged distance ($D_V/r_d$). For these points, we must still incorporate external $H(z)$ data from CC to disentangle the angular distance. Consequently, the ADD is modeled as:
\begin{align}
    D_A^{\text{BAO}}(z_j, r_d) &= \frac{(D_M/r_d)_j \cdot r_d}{1+z_j}, \quad \text{for anisotropic points}, \\
    D_A^{\text{BAO}}(z_j, r_d) &= \frac{1}{1+z_j} \sqrt{\frac{(D_V/r_d)_j^3 \cdot r_d^3 \cdot H(z_j)}{z_j \cdot c}}, \quad \text{for isotropic points}.
\end{align}

Combining these distance definitions with eq.~(\ref{eq:ddr}), we construct the observed $\eta$ vector. The joint log-likelihood is then defined in the $\eta$-space:
\begin{equation}
    \ln L(\theta) = -\frac{1}{2} \left[ \Delta\vec{\eta}^T \mathbf{C}(\theta)^{-1} \Delta\vec{\eta} + \ln |\mathbf{C}(\theta)| + N \ln(2\pi) \right],
\end{equation}
where $\theta = \{\eta_i, M_B, r_d\}$ is the parameter vector (where $\eta_i$ refers to the deviation parameter of the $i$-th model), $\Delta\vec{\eta}$ is the residual vector with elements $\Delta\eta_j = \eta_{\text{obs}}(z_j) - \eta_{\text{th}}(z_j)$, and $N$ is the number of data points.

Crucially, our analysis accounts for the fact that the uncertainties in $\eta$-space are parameter-dependent due to the non-linear transformations and the marginalized distance scales. We construct the total covariance matrix $\mathbf{C}(\theta)$ by projecting the raw observational covariance matrices through the Jacobian matrices:
\begin{equation}
    \mathbf{C}(\theta) = \mathbf{J}_{m} \mathbf{C}_{m_B} \mathbf{J}_{m}^T + \mathbf{J}_{b} \mathbf{C}_{\text{BAO}} \mathbf{J}_{b}^T + \mathbf{J}_{h} \mathbf{C}_{H} \mathbf{J}_{h}^T.
\end{equation}
The elements of the diagonal Jacobian matrices $\mathbf{J}$ are derived from the partial derivatives of the model value $\eta(z_j)$ with respect to the observables:
\begin{align}
    \text{SNIa:} \quad & w_{m, j} = \frac{\partial \eta(z_j)}{\partial m_{B, j}} = \eta(z_j) \frac{\ln 10}{5}, \\
    \text{BAO ($D_M$):} \quad & w_{b, j} = \frac{\partial \eta(z_j)}{\partial (D_M/r_d)_j} = -\frac{\eta(z_j)}{(D_M/r_d)_j}, \\
    \text{BAO ($D_V$):} \quad & w_{b, j} = \frac{\partial \eta(z_j)}{\partial (D_V/r_d)_j} = -\frac{3}{2} \frac{\eta(z_j)}{(D_V/r_d)_j}, \\
    \text{CC ($D_V$ only):} \quad & w_{h, j} = \frac{\partial \eta(z_j)}{\partial H(z_j)} = -\frac{1}{2} \frac{\eta(z_j)}{H(z_j)}.
\end{align}
By updating $\mathbf{C}(\theta)$ at each step of the MCMC sampling, we ensure that the non-linear error propagation and the statistical correlations (especially from the GPR-reconstructed SNIa and CC data) are self-consistently incorporated into the final constraints. The inclusion of the log-determinant term $\ln |\mathbf{C}(\theta)|$ ensures that the parameter estimation remains unbiased against the variations in the error volume.

\begin{table}[h]
\centering
\caption{Joint constraints on the CDDR deviation parameter $\eta_i$ and nuisance parameters $M_B$, $r_d$. We compare results from GPR and FKM across four phenomenological models and three calibration priors. The ``Noprior'' prior corresponds to flat priors on nuisance parameters. Uncertainties represent the asymmetric $1\sigma$ limits.}
\label{tab:combined_results}
\resizebox{\textwidth}{!}{%
\begin{tabular}{llcccc}
\hline
Model & Method & Prior & $\eta_i$ & $M_B$ [mag] & $r_d$ [Mpc] \\ 
\hline
\textbf{Linear} & GPR & P18 & $0.0227^{+0.0272}_{-0.0265}$ & $-19.4740^{+0.0445}_{-0.0450}$ & $147.0939^{+0.2583}_{-0.2553}$ \\
$\eta(z) = 1 + \eta_1 z$ & & R22 & $0.0130^{+0.0248}_{-0.0245}$ & $-19.4017^{+0.0491}_{-0.0503}$ & $143.5005^{+3.8588}_{-3.8381}$ \\
 & & Noprior & $0.0312^{+0.0290}_{-0.0282}$ & $-19.6463^{+0.1920}_{-0.1738}$ & $157.9996^{+11.3495}_{-11.8680}$ \\
\cline{2-6}
 & FKM & P18 & $0.0368^{+0.0317}_{-0.0297}$ & $-19.4964^{+0.0480}_{-0.0501}$ & $147.0992^{+0.2603}_{-0.2622}$ \\
 & & R22 & $0.0245^{+0.0274}_{-0.0268}$ & $-19.4021^{+0.0500}_{-0.0494}$ & $142.2937^{+3.9322}_{-3.9587}$ \\
 & & Noprior & $0.0453^{+0.0329}_{-0.0319}$ & $-19.6614^{+0.1925}_{-0.1788}$ & $157.5117^{+11.5526}_{-11.6428}$ \\
\hline
\textbf{Fractional} & GPR & P18 & $0.0986^{+0.1086}_{-0.1010}$ & $-19.5235^{+0.0878}_{-0.0912}$ & $147.0975^{+0.2570}_{-0.2585}$ \\
$\eta(z) = 1 + \frac{\eta_2 z}{1+z}$ & & R22 & $0.0229^{+0.0803}_{-0.0759}$ & $-19.3992^{+0.0486}_{-0.0501}$ & $143.3103^{+4.8493}_{-4.7882}$ \\
 & & Noprior & $0.1504^{+0.1237}_{-0.1143}$ & $-19.7676^{+0.2378}_{-0.2087}$ & $161.1123^{+11.0885}_{-12.6281}$ \\
\cline{2-6}
 & FKM & P18 & $0.1363^{+0.1153}_{-0.1079}$ & $-19.5561^{+0.0926}_{-0.0946}$ & $147.0971^{+0.2608}_{-0.2588}$ \\
 & & R22 & $0.0509^{+0.0859}_{-0.0817}$ & $-19.4011^{+0.0504}_{-0.0520}$ & $141.7995^{+4.9950}_{-4.9680}$ \\
 & & Noprior & $0.1947^{+0.1279}_{-0.1227}$ & $-19.7992^{+0.2387}_{-0.2047}$ & $160.6817^{+11.3302}_{-12.6074}$ \\
\hline
\textbf{Logarithmic} & GPR & P18 & $0.0513^{+0.0567}_{-0.0539}$ & $-19.4974^{+0.0627}_{-0.0646}$ & $147.0973^{+0.2624}_{-0.2549}$ \\
$\eta(z) = 1 + \eta_3 \ln(1+z)$ & & R22 & $0.0206^{+0.0471}_{-0.0439}$ & $-19.4019^{+0.0505}_{-0.0507}$ & $143.2621^{+4.3013}_{-4.3106}$ \\
 & & Noprior & $0.0735^{+0.0613}_{-0.0585}$ & $-19.6987^{+0.2076}_{-0.1879}$ & $159.2595^{+11.6140}_{-11.9017}$ \\
\cline{2-6}
 & FKM & P18 & $0.0740^{+0.0628}_{-0.0593}$ & $-19.5227^{+0.0667}_{-0.0694}$ & $147.0911^{+0.2588}_{-0.2618}$ \\
 & & R22 & $0.0407^{+0.0515}_{-0.0490}$ & $-19.4024^{+0.0499}_{-0.0508}$ & $141.6999^{+4.4632}_{-4.3058}$ \\
 & & Noprior & $0.0986^{+0.0691}_{-0.0634}$ & $-19.7218^{+0.2035}_{-0.1869}$ & $158.8310^{+11.3081}_{-11.5728}$ \\
\hline
\textbf{Power-law} & GPR & P18 & $0.0473^{+0.0508}_{-0.0530}$ & $-19.4933^{+0.0626}_{-0.0613}$ & $147.0919^{+0.2591}_{-0.2535}$ \\
$\eta(z) = (1+z)^{\epsilon}$ & & R22 & $0.0188^{+0.0464}_{-0.0447}$ & $-19.4010^{+0.0501}_{-0.0507}$ & $143.3554^{+4.3926}_{-4.3696}$ \\
 & & Noprior & $0.0652^{+0.0520}_{-0.0538}$ & $-19.6819^{+0.2117}_{-0.1841}$ & $158.5539^{+11.5080}_{-12.1315}$ \\
\cline{2-6}
 & FKM & P18 & $0.0698^{+0.0553}_{-0.0561}$ & $-19.5200^{+0.0643}_{-0.0642}$ & $147.0927^{+0.2565}_{-0.2541}$ \\
 & & R22 & $0.0368^{+0.0482}_{-0.0481}$ & $-19.4029 \pm 0.0496$ & $142.0959^{+4.3664}_{-4.4114}$ \\
 & & Noprior & $0.0869 \pm 0.0581$ & $-19.7037^{+0.2054}_{-0.1886}$ & $158.4236^{+11.5874}_{-11.8153}$ \\
\hline
\end{tabular}}
\end{table}

\begin{figure}[htbp]
    \centering
    \subfloat[Linear]{\includegraphics[width=0.5\textwidth]{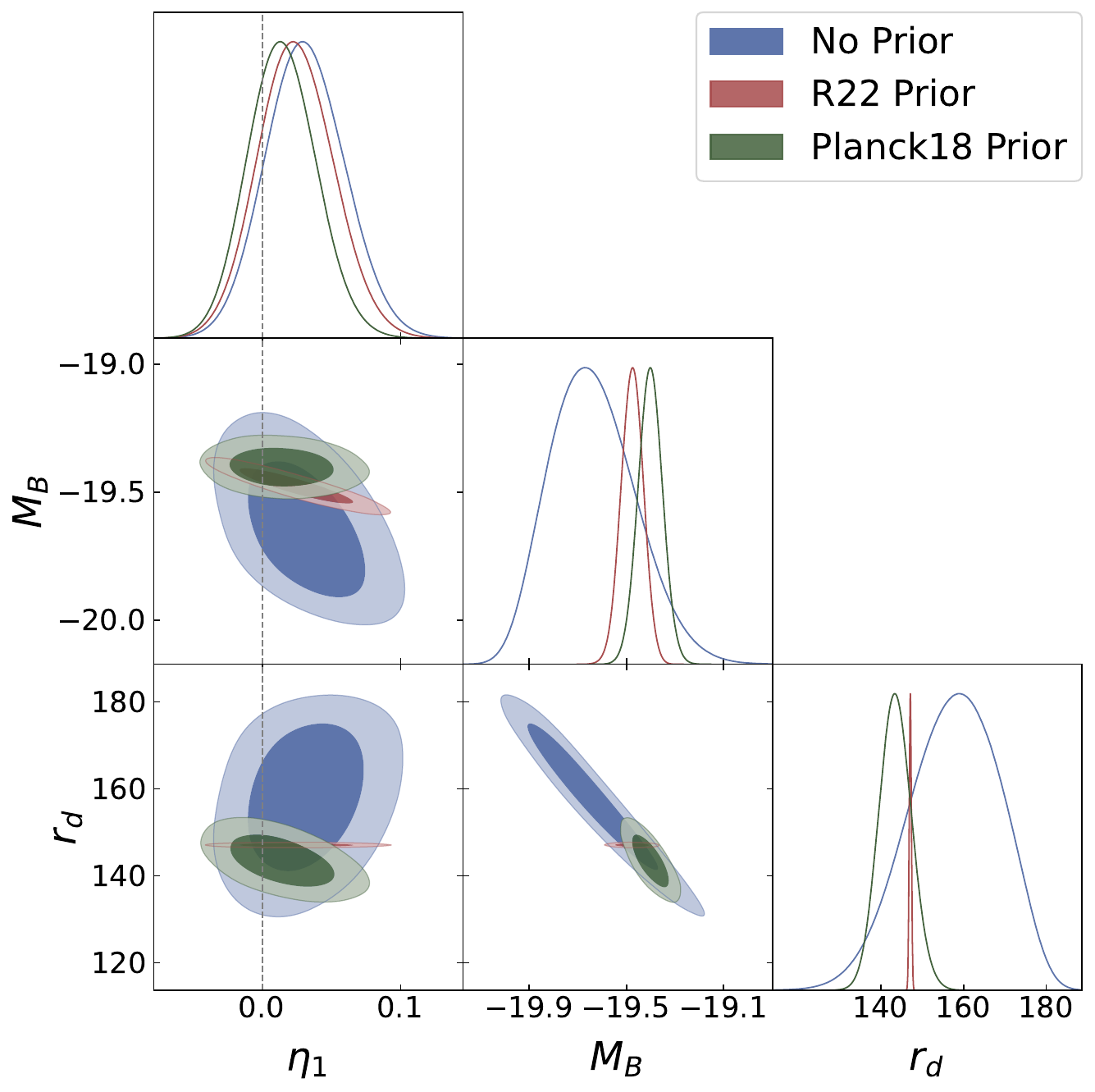}}\hfill
    \subfloat[Fractional]{\includegraphics[width=0.5\textwidth]{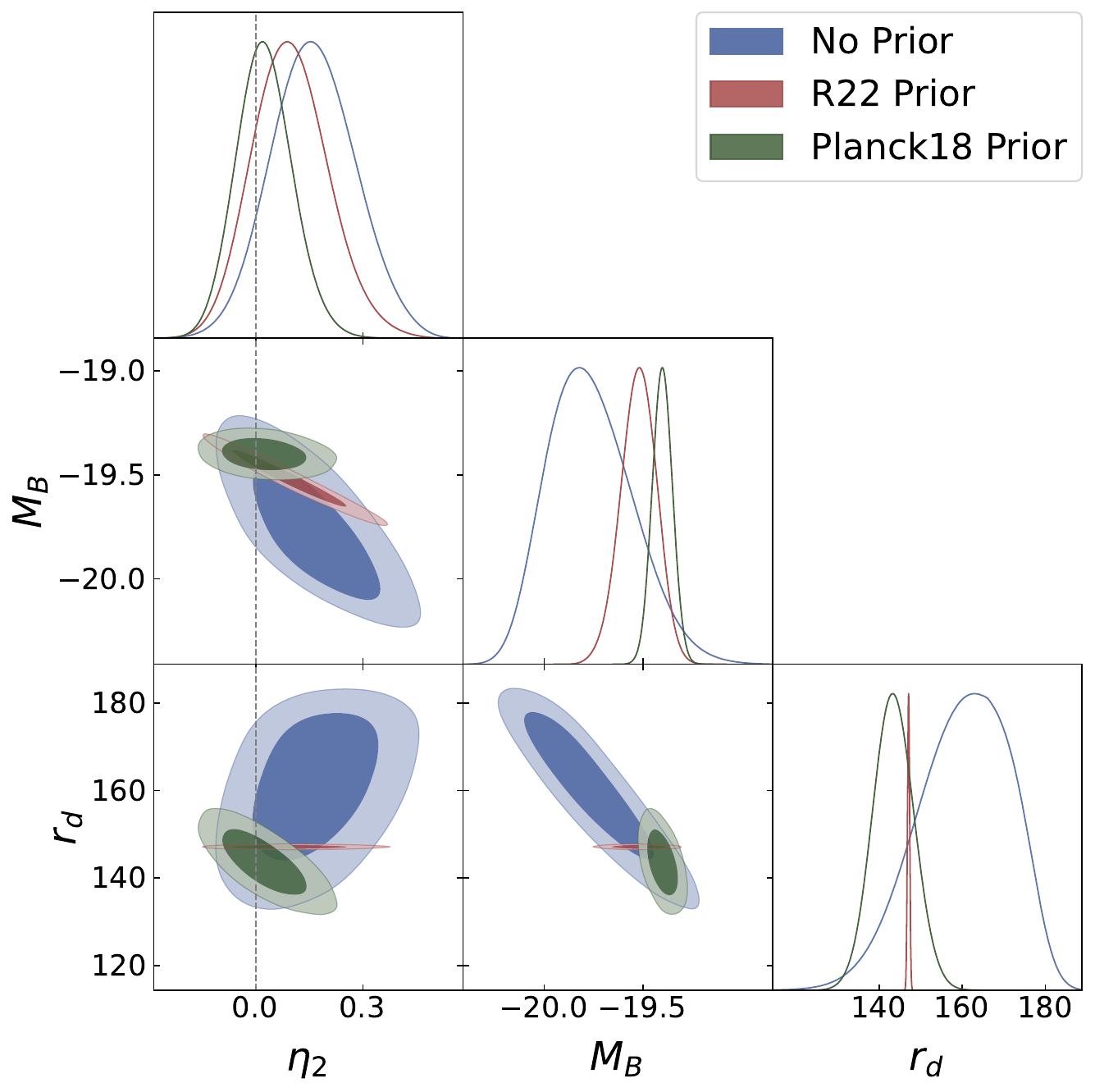}}\\
    \subfloat[Logarithmic]{\includegraphics[width=0.5\textwidth]{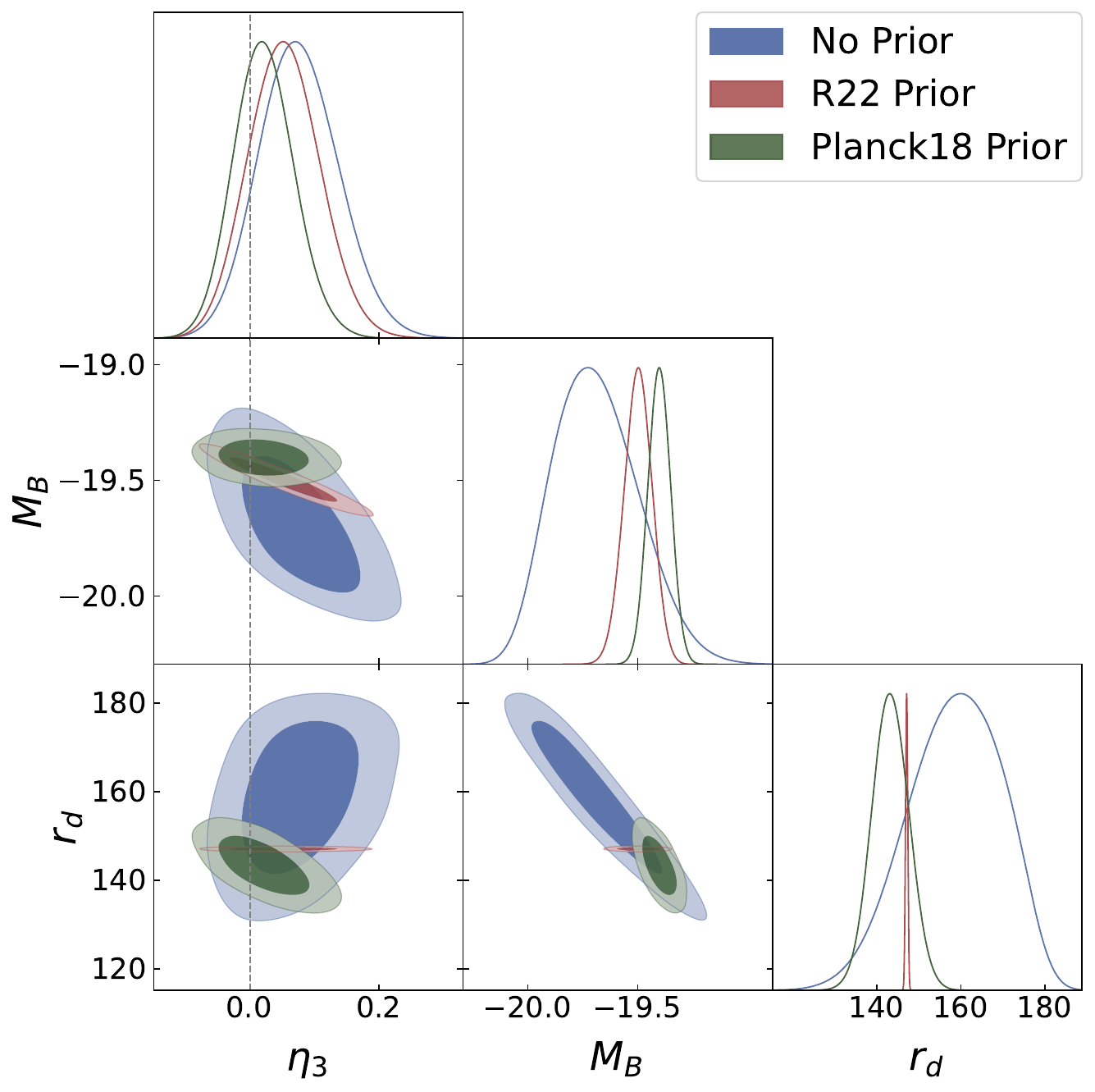}}\hfill
    \subfloat[Power-law]{\includegraphics[width=0.5\textwidth]{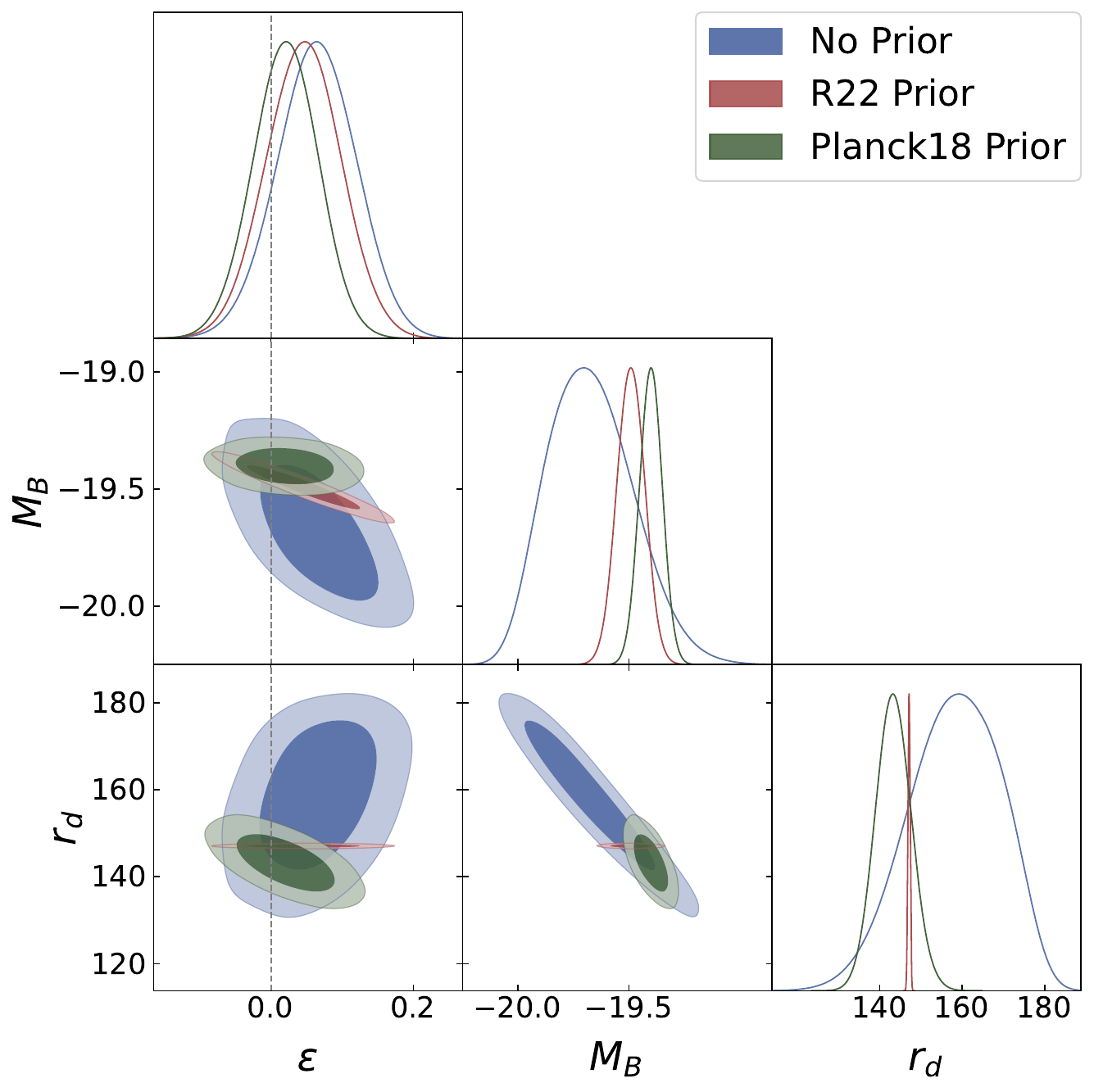}}
    \caption{Joint posterior distributions and 1D marginalized constraints for the CDDR deviation parameters $\eta_i$ and nuisance parameters ($M_B, r_d$) derived using the GPR method. The four panels correspond to the (a) Linear, (b) Fractional, (c) Logarithmic, and (d) Power-law parameterizations. The contours represent the 68\% and 95\% confidence levels under three different calibration scenarios: Planck 2018 (P18), local distance ladder (R22), and flat priors (Noprior). The vertical dashed lines in the 1D distributions indicate the best-fit values.}
    \label{fig:constraint_comparison}
\end{figure}

The results of our joint Bayesian analysis for the four phenomenological models are summarized in table~\ref{tab:combined_results} and visually represented in figure~\ref{fig:constraint_comparison} (We use package \texttt{GetDist} \cite{Lewis:2019xzd} for visualization). It is worth noting that for the P18 and R22 calibration scenarios, Gaussian priors are applied to the corresponding nuisance parameters ($r_d$ and $M_B$), whereas flat priors are assumed for the ``Noprior'' case. Using both the GPR and FKM reconstructions, we observe that the deviation parameters $\eta_i$ exhibit a slight positive preference compared to the null hypothesis, yet they remain statistically consistent with zero within $1\sigma$ or $2\sigma$ confidence levels across all parameterizations. For instance, in the linear model with the P18 prior, we find $\eta_1 = 0.0227^{+0.0272}_{-0.0265}$ (for GPR) and $\eta_1 = 0.0368^{+0.0317}_{-0.0297}$ (for FKM), which are both compatible with the standard expectation of $\eta_i=0$.

Notably, the GPR constraints yield uncertainty estimates ($\sigma(\eta_i) \approx 0.027$ for the linear P18 case) that are comparable to those obtained from the FKM analysis ($\sigma(\eta_i) \approx 0.031$). This indicates that when full marginalization over kernel hyperparameters is performed, GPR provides a conservative estimation of reconstruction errors similar to the knot-based method. The ``Noprior'' case, which assumes no external constraints on $M_B$ and $r_d$, generally yields slightly larger uncertainties and higher central values for $\eta_i$. Interestingly, the degradation in the precision of the deviation parameter $\eta_i$ is relatively small. This phenomenon arises because the parameters $\eta_i$ primarily characterize the redshift-dependent evolution of the CDDR, whereas the nuisance parameters $M_B$ and $r_d$ act as degenerate overall scaling factors that govern the absolute distance scale. While the absence of priors drastically increases the marginalized uncertainties on $M_B$ and $r_d$ themselves (as evident in table~\ref{tab:combined_results}), the extensive redshift coverage of the combined SNIa, BAO, and CC datasets provides stringent constraints on the relative distance evolution. This strong relative constraining power self-calibrates the shape parameters, preventing the uncertainties on $\eta_i$ from degrading significantly even without external calibration anchors. Conversely, the R22 prior pulls the deviation parameter closer to zero but necessitates significant shifts in the nuisance parameters to accommodate the local $H_0$ calibration. Despite these variations, the CDDR itself remains robust across different priors and reconstruction methods.

\section{The Test of Cosmology Transparency}\label{sec:opacity}
The previous sections have shown that, while the CDDR is broadly consistent with current observational data, weak deviations can emerge depending on the adopted priors and reconstruction techniques. This naturally raises the question of whether such deviations originate from genuine physical effects or are instead induced by methodological or systematic uncertainties.

In section~\ref{sec:parametrized}, we explored a phenomenological power-law parametrization of CDDR deviations, which is commonly employed in the literature as an effective description of cosmic opacity. The results obtained there did not reveal any statistically significant departure from a transparent universe. However, such parametrized approaches inevitably rely on specific functional assumptions about the redshift dependence of the opacity.

Among the physically motivated mechanisms that could lead to violations of the CDDR, photon number non-conservation, commonly referred to as cosmic opacity provides a particularly well-studied and testable possibility. Any process that absorbs, scatters, or otherwise removes photons along the line of sight would modify the observed LD and thereby induce a breakdown of the CDDR. In this section, we therefore revisit the cosmic transparency hypothesis using a fully model-independent, differential approach and then constrain the physics which affects the process.

\subsection{The Differential Calculation of Transparency }
This section therefore tests the cosmic transparency hypothesis directly. Following the methodology presented in Ref.~\cite{More_2009}, we quantify potential violations of the CDDR that could arise from photon absorption. 
The presence of any photon-absorbing medium would lead to a deviation from the CDDR. Assuming the existence of a dust medium which can absorb or scatter photons, the observed flux from a source at redshift $z$ would be diminished by a factor of $e^{-\tau(z)}$, where $\tau(z)$ denotes the optical depth between the source and the observer. Equivalently, this effect can be expressed in terms of the LD:
\begin{equation}
    D_{L,\text{obs}}^2=D_{L,\text{model}}^2e^{\tau(z)},
\end{equation}
where $D_{L,\text{obs}}=10^{\frac{m_{B,\text{obs}}-M_B-25}{5}}$ and $D_{L,\text{model}}=D_{A}(1+z)^2$ (assuming the standard duality). If the CDDR holds true, we can derive the relation for the differential optical depth between two adjacent redshift bins $z_{j-1}$ and $z_j$ as given in \cite{More_2009,Remya_Nair_2012}:
\begin{equation}
    \Delta  \tau_{j} = \frac{\ln 10}{2.5}\left[ \Delta m_{B,j} - 2.5\log_{10}\left(\frac{z_{j-1}(1+z_{j})^2 H(z_{j})}{z_{j}(1+z_{j-1})^2H(z_{j-1})}\left(\frac{(D_V/r_d)_{j}}{(D_V/r_d)_{j-1}}\right)^3\right)\right],
\end{equation}
where $\Delta m_{B,j}=m_{B}(z_j)-m_{B}(z_{j-1})$. This quantity is model-independent since it involves only the differential measurements between two adjacent redshifts.

\begin{table}[h]
\centering
\caption{Reconstructed differential optical depth $\Delta \tau$ and corresponding $1\sigma$ uncertainties $\sigma_{\Delta_j}$ between adjacent DESI DR2 redshift bins $[z_{i-1}, z_i]$. The evaluations are performed using both the GPR and the FKM. The column $z_{\text{mid}}$ denotes the mid-value of each redshift pair.}
\label{tab:delta_tau}
\begin{tabular}{ccccc}
\hline
Redshift Pair $[z_{i-1}, z_i]$ & $z_{\rm mid}$ & $\Delta \tau_{\rm GPR}$ & $\Delta \tau_{\rm FKM}$ \\ \hline
$[0.295, 0.510]$ & $0.403$ & $-0.034 \pm 0.057$ & $0.024 \pm 0.068$ \\
$[0.510, 0.706]$ & $0.608$ & $0.036 \pm 0.050$ & $0.029 \pm 0.077$ \\
$[0.706, 0.934]$ & $0.820$ & $-0.050 \pm 0.064$ & $-0.105 \pm 0.130$ \\
$[0.934, 1.321]$ & $1.128$ & $0.026 \pm 0.066$ & $0.149 \pm 0.129$ \\
$[1.321, 1.484]$ & $1.403$ & $-0.001 \pm 0.066$ & $-0.152 \pm 0.097$ \\
$[1.484, 2.330]$ & $1.907$ & $0.240 \pm 0.231$ & $0.159 \pm 0.314$ \\ \hline
\end{tabular}
\end{table}

\begin{figure}[h]
    \centering
    \includegraphics[width=0.5\linewidth]{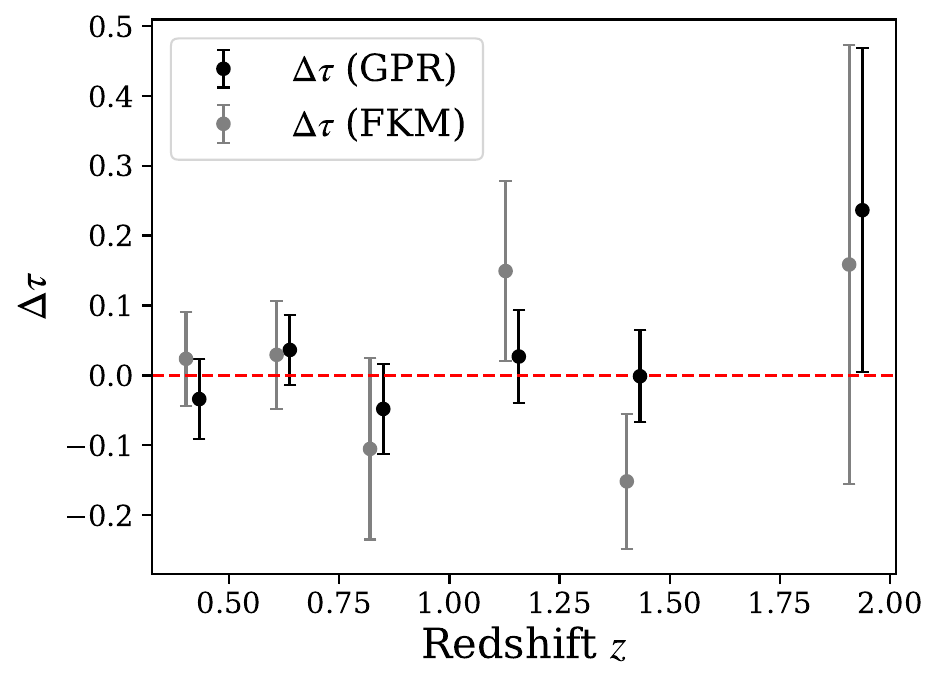}
    \caption{Calculated $\Delta \tau$ values and their $1\sigma$ uncertainties at $z_\text{mid}$ (mid value of two adjacent redshifts). The black points are calculated from the GPR, and the grey one is calculated from FKM. The red dashed horizontal line at $\Delta \tau = 0$ represents a transparent universe.}
    \label{fig:delta_tau_plot}
\end{figure}

table~\ref{tab:delta_tau} and figure~\ref{fig:delta_tau_plot} present the differential optical depth $\Delta \tau$ reconstructed via GPR and the FKM methods. Both model-independent methodologies yield results consistent with a transparent universe ($\Delta \tau = 0$) within a $2\sigma$ confidence level across all redshift bins. Generally, GPR provides tighter constraints due to its global smoothness prior, whereas FKM exhibits larger uncertainties reflecting its sensitivity to local data fluctuations. The significant uncertainty in the highest redshift bin ($z_{\text{mid}}=1.907$) is driven by the sparsity and extrapolation of CC data beyond $z \approx 1.97$.

Notably, several redshift bins exhibit negative central values for $\Delta \tau$. Since standard photon absorption or scattering mechanisms fundamentally require $\Delta \tau \ge 0$, these negative excursions argue strongly against cosmic opacity as the primary cause of any potential CDDR deviation. Given that the overall results are consistent with a transparent universe within the error bars, these mild negative values lack statistical significance. Rather than pointing to explicit residual systematics or calibration errors, they are most likely driven by random statistical fluctuations in the observational data.

To further quantify the overall cosmic transparency and test for any cumulative redshift-dependent photon attenuation, we evaluate the average of the opacity derivative. Motivated by the widely used phenomenological model $\tau(z) = 2\epsilon z$ \cite{Avgoustidis_2009} (which corresponds to $\eta(z) = (1+z)^\epsilon$), the opacity implies a constant derivative $d\tau/dz$. Thus, we calculate the global average of the opacity derivative:
\begin{equation}
    \left\langle \frac{d\tau}{dz} \right\rangle = \frac{\sum_j w_j (\Delta\tau_j / \Delta z_j)}{\sum_j w_j}, \quad \text{where} \quad w_j = \left(\frac{\Delta z_j}{\sigma_{\Delta\tau_j}}\right)^2.
\end{equation}
Using this robust statistical estimator, we obtain $\langle d\tau/dz \rangle = 0.0409 \pm 0.1024$ for the GPR method and $0.0730 \pm 0.1607$ for the FKM method. Both opacity derivatives are consistent with zero well within the $1\sigma$ confidence level. In summary, we find no statistically significant evidence for cosmic opacity. The consistency between the GPR and FKM reconstructions reinforces the robustness of the CDDR in the late-time universe. Future surveys with improved precision and extended redshift coverage will be essential to disentangle these statistical fluctuations from potential new physics or subtle systematic effects.

\subsection{Constraints on Potential New Physics Affecting Cosmic Transparency}
Although the analyses in the previous subsection support a transparent universe consistent with the standard CDDR, the residual observational uncertainties can still be utilized to place stringent bounds on exotic physical models. As described in section~\ref{sec:intro}, photon number conservation is one of the fundamental assumptions underlying the CDDR. Deviation from $\eta(z)=1$ can be interpreted as evidence for exotic physical processes that attenuate the photon flux. Physically, the ADD is derived from geometric measurements and remains unaffected by photon loss. Conversely, the LD is inferred from the observed flux ($F \propto D_L^{-2}$). If photons are absorbed or converted into dark sector candidates, such as Axion-Like Particles (ALPs) or Mini-Charged Particles (MCPs), the flux is reduced by a survival probability $\mathcal{P}(z)$. This attenuation ($F_{\text{obs}} = \mathcal{P}(z) F_{\text{true}}$) artificially inflates the observed LD to $D_{L, \text{obs}} = D_{L, \text{true}} / \sqrt{\mathcal{P}(z)}$. Following established cosmic opacity frameworks \cite{Avgoustidis_2009,Tiwari_2017,PhysRevD.92.123539,buenabad2022constraintsaxionscosmicdistance}, this effect directly modifies the distance duality parameter as: $\eta(z) = {D_{L, \text{obs}}}/({D_A(1+z)^2})= \mathcal{P}(z)^{-1/2}$.

To constrain these physical models, we perform a direct forward-modeling analysis using the full combined datasets. Instead of relying on the intermediate reconstructed $\eta(z)$, we assume a flat $\Lambda$CDM background cosmology modified by the opacity effects. The theoretical LD is given by:
\begin{equation}
    D_L^{\text{th}}(z) = \frac{D_L^{\Lambda\text{CDM}}(z, \Omega_m, H_0)}{\sqrt{\mathcal{P}(z, \mathbf{p})}},
\end{equation}
where $\mathbf{p}$ represents the opacity model parameters. We define the total $\chi^2$ as the sum of contributions from SNIa, BAO, and CC data. Then, we perform a grid search analysis over the cosmological parameter space ($\Omega_m$) and the opacity parameters ($P/L$ for ALPs or $\kappa$ for MCPs). To ensure robust constraints independent of the Hubble tension, we profile over the Hubble constant $H_0$ as a nuisance parameter by minimizing the $\chi^2$ at each grid point. Furthermore, the supernova absolute magnitude $M_B$ is analytically marginalized in the SNIa likelihood calculation \cite{Conley_2010}.

\subsubsection{Axion-Like Particles (ALPs)}

Axion-like particles are generic predictions of many extensions of the Standard Model, such as string theory compactifications \cite{Svrcek_2006}. In the presence of extragalactic magnetic fields $B$, photon-ALP oscillations $\gamma \xleftrightarrow{B} \phi$ induce a frequency-independent dimming of distant sources. The average probability that a photon emitted by a supernova at redshift $z$ is observed by us after traversing the magnetic fields in the intergalactic medium which is written as \cite{Burrage_2009,Davis_2009,Schelpe_2010}:
\begin{equation}
    \mathcal{P}(z)=A+(1-A)\exp\left(-\frac{3}{2}D_\text{C}(z)\frac{P}{L}\right),
\end{equation}
where $D_\text{C}(z)=c\int_0^z\frac{dz'}{H(z')}$ is the comoving distance to the source, and $A$ is the ALP flux parameter\footnote{The ALP flux parameter is defined by the initial photon and ALP fluxes ($I_\gamma, I_\phi$) at the source as $A = \frac{2}{3} (1 + I_\phi/I_\gamma)$. It characterizes three distinct physical regimes: (i) $A=2/3$ corresponds to the case where the source emits only photons ($I_\phi=0$), leading to standard cosmic dimming; (ii) $A=1$ occurs when the initial fluxes are already in equilibrium ($I_\phi/I_\gamma=1/2$), resulting in no net change in photon number; and (iii) $A>1$ represents an ALP-rich source ($I_\phi/I_\gamma > 1/2$), which can lead to anomalous brightening of the observed source as ALPs convert back into photons.} (here taken as $A=2/3$ for initially unpolarized light from supernova), $P$ is the conversion probability within a single magnetic domain of size $L$.

\begin{figure}[h]
    \centering
    \includegraphics[width=0.49\textwidth]{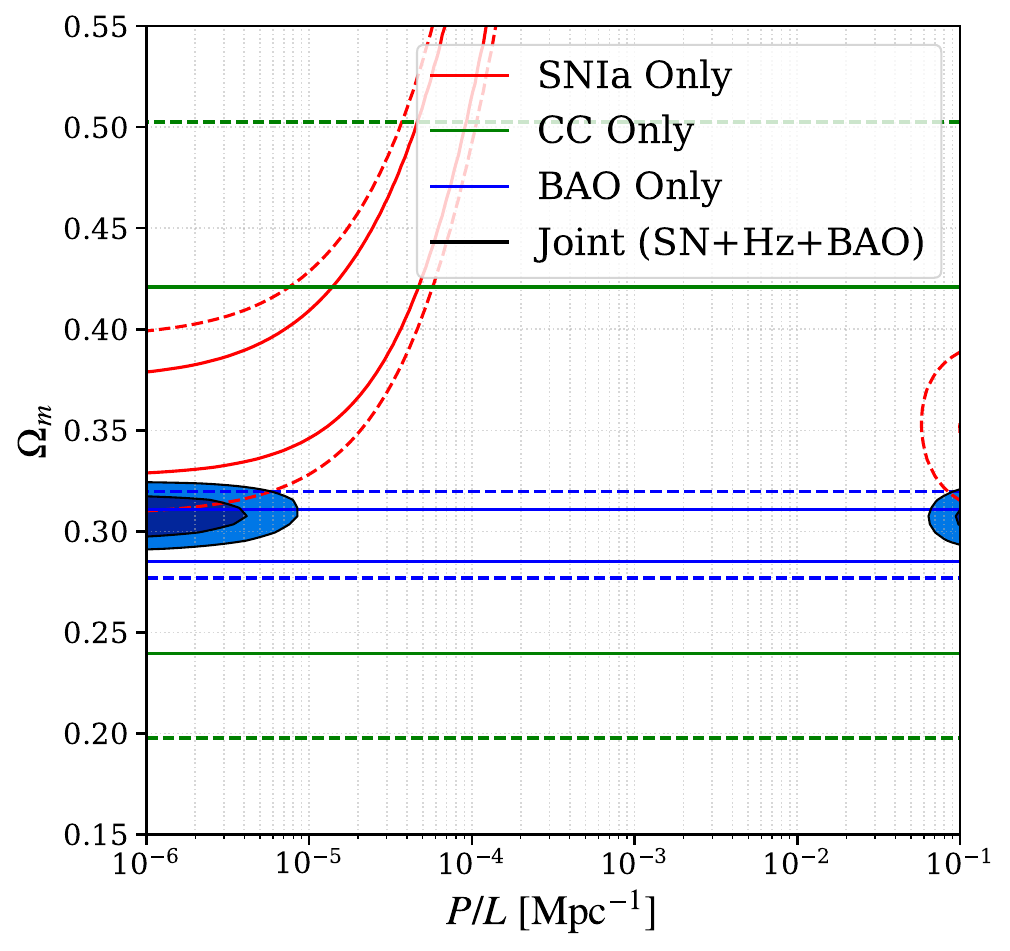}
    \includegraphics[width=0.49\textwidth]{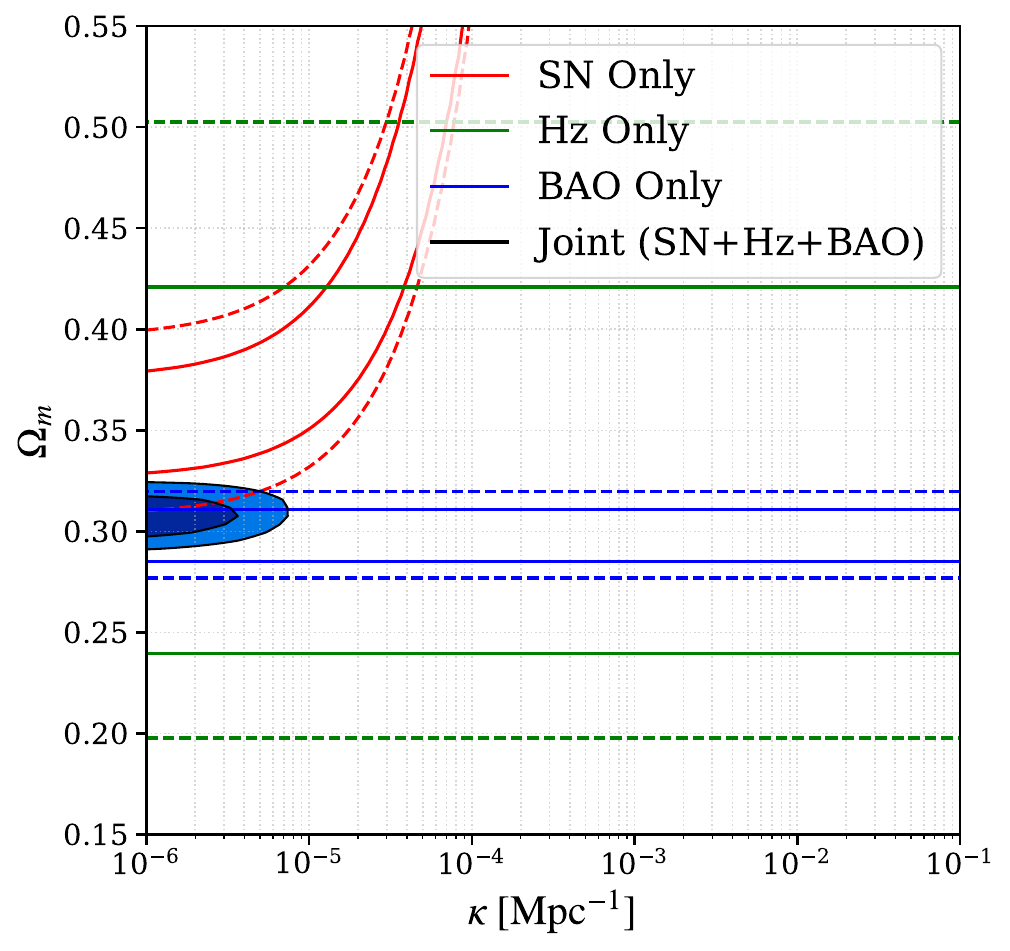}
    \caption{Left panel: Constraints on the coherent Axion-Like Particle (ALP) model. Shaded regions show the 68\% and 95\% confidence levels on the $P/L - \Omega_m$ plane, assuming a mixing constant $A=2/3$. Right panel: Constraints on the Mini-Charged Particle (MCP) model. Shaded regions indicate the 68\% and 95\% confidence levels on the $\Omega_m - \kappa$ plane.}
    \label{fig:alp_mcp_constraints}
\end{figure}

To account for cosmic expansion explicitly, we use the redshift-evolving form (accounting for the redshift evolution of the magnetic field and photon frequency) \cite{nair_cosmic_2012}:
\begin{equation}
    \mathcal{P}(z) = A + (1-A)\exp\left[-\frac{P}{L}\frac{H(z)-H_{0}}{\Omega_{m}H^2_{0}}c\right].\label{eq:alptranverse}
\end{equation}
We then perform a grid scan over the parameter space of $\Omega_m \in[0.15, 0.55]$ and $P/L \in [10^{-6}, 10^{-1}]~\mathrm{Mpc}^{-1}$ (sampled in logarithmic scale). This specific prior range is chosen empirically; we initially conducted a grid search within a broader parameter space and gradually narrowed it down for the final sampling. The resulting constraints on the $P/L - \Omega_m$ plane, obtained after profiling over $H_0$, are shown in the left panel of figure~\ref{fig:alp_mcp_constraints}. In this parameter space, the small and large $P/L$ regions (blue contours) physically correspond to the weak-mixing and strong-mixing regimes, respectively. As observed from the individual probe contours, relying solely on SNIa data leads to a significant parameter degeneracy in the low-mixing region. However, the inclusion of robust geometric distance indicators from BAO and CC breaks this degeneracy. The joint analysis reveals that the effective conversion parameter is tightly constrained, yielding a 95\% confidence-level excluded interval of $P/L \in[6.43 \times 10^{-6},\, 7.05 \times 10^{-2}]~\mathrm{Mpc}^{-1}$.

To translate these constraints into physical bounds on ALP parameters, we define the dimensionless variables $B_\text{nG}/M_{10}$ for the magnetic field $B$ and the coupling scale $M$ (common expression used is the coupling constant $g_{\phi\gamma}$, they are reciprocals of each other) as:
\begin{equation}
    B_\text{nG}=\frac{B}{1\text{nG}}, \quad M_{10}=\frac{M}{10^{10}\text{GeV}}, \label{eq:BM}
\end{equation}
where $M = 1/g_{\phi\gamma}$ is the coupling energy scale. For a photon propagating through a constant, coherent magnetic field domain of size $L$, the probability of converting into an ALP is given by \cite{PhysRevD.37.1237}:
\begin{equation}
    P(L) = \sin^2(2\theta) \sin^2\left(\frac{\Delta(L)}{\cos 2\theta}\right). \label{eq:P_coherent}
\end{equation}
This conversion probability consists of two fundamental components: the first part, $\sin^2(2\theta)$, represents the mixing strength (amplitude) between the photon and the ALP, while the second part describes the oscillation phase as a function of the propagation distance. The relevant mixing parameters are defined as:
\begin{equation}
    \Delta(L) = \frac{m_{\text{eff}}^2 L}{4 \omega_\gamma}, \quad \tan 2\theta = \frac{2B \omega_\gamma}{M m_{\text{eff}}^2}. \label{eq:probability}
\end{equation}
Here, $\omega_\gamma$ is the photon energy, $B$ is the strength of the transverse magnetic field, and $m_\phi$ represents the rest mass of the ALP. The term $m_{\text{eff}}^2 = |m_\phi^2 - \omega_p^2|$ is the effective mass squared difference that governs the oscillation phase, where $\omega_p^2 = 4\pi \alpha n_e / m_e$ denotes the plasma frequency of the medium, which acts as an effective mass for the photons. In this study, we restrict our attention to the very light ALP regime ($m_\phi \ll \omega_p$), where the ALP mass becomes negligible and the mass difference is dominated by the plasma frequency, such that $m_{\text{eff}}^2 \simeq \omega_p^2$. Furthermore, in the weak mixing regime where the coupling scale $M$ is sufficiently large, the mixing angle $\theta$ is infinitesimal, allowing for the approximation $\tan 2\theta \simeq \sin 2\theta$.

\begin{figure}[h]
    \centering
    \includegraphics[width=0.49\textwidth]{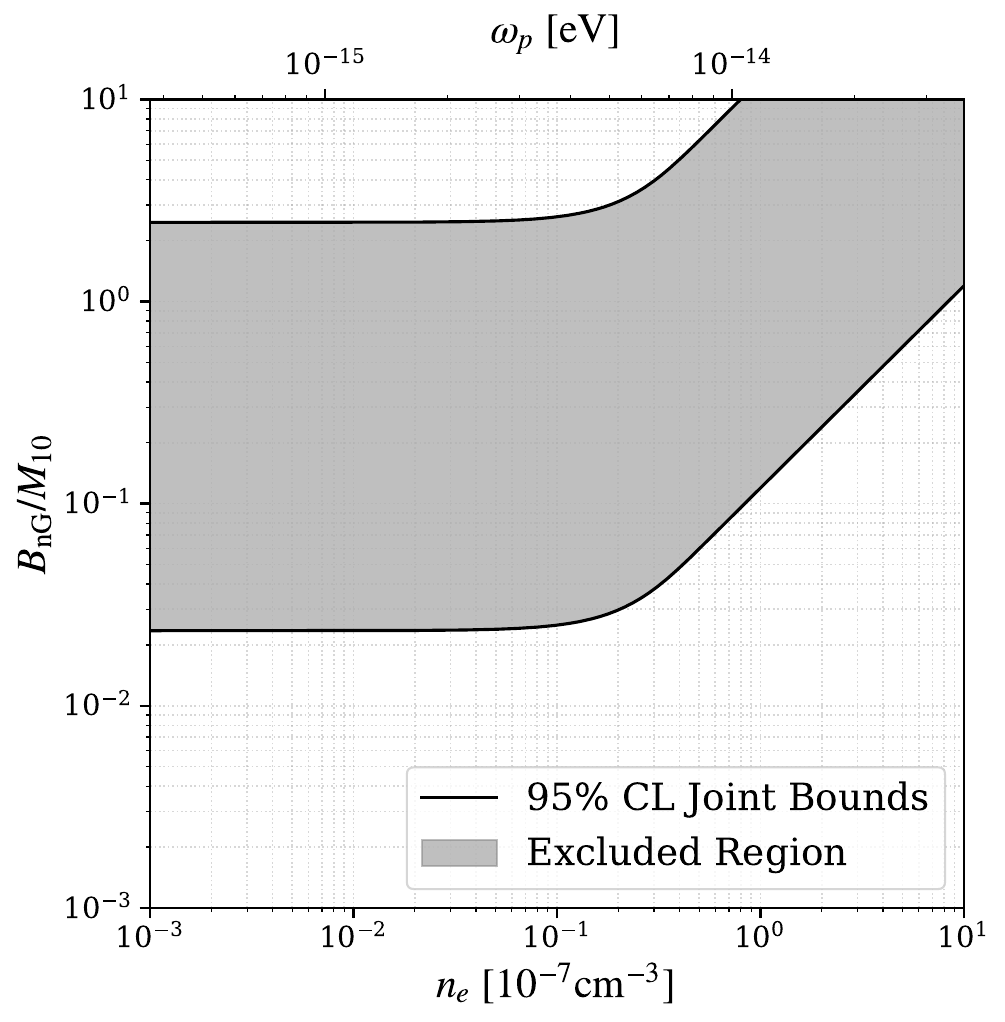}
    \includegraphics[width=0.49\textwidth]{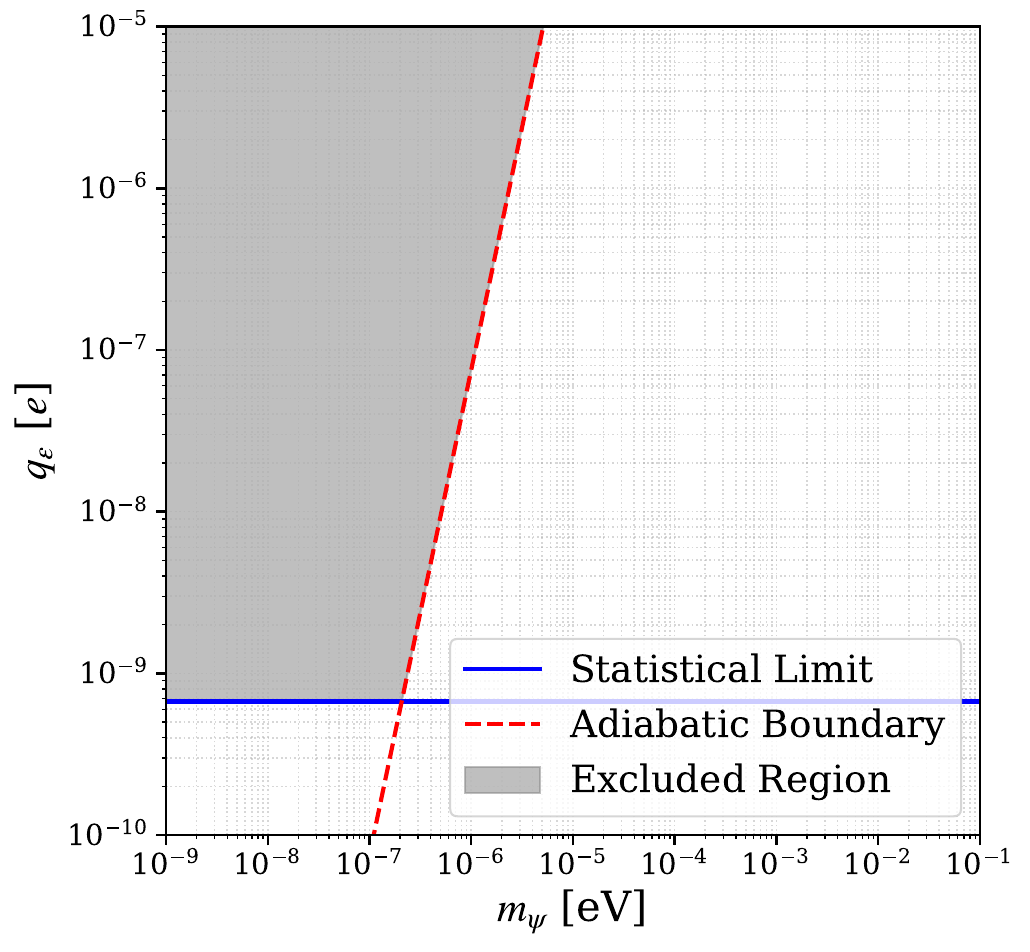}
    \caption{Left panel: Constraints on the environmental parameters (product of ALP coupling times magnetic field $B_\text{nG}/M_{10}$ vs. average electron density $n_e$ or corresponding plasma frequency $\omega_p$) required for effective photon-ALP conversion, grey region is the excluded region. Right panel: Exclusion regions in the MCP mass ($m_\psi$) versus electric charge ($q_\epsilon$) plane derived from the opacity test. The blue line represents the upper limit obtained through the data, the red dashed line represents the upper limit under the adiabatic condition, and the gray area is the excluded region.}
    \label{fig:physics_constraints}
\end{figure}

By combining eqs.~(\ref{eq:BM})--(\ref{eq:probability}) with the statistically excluded interval $P/L \in[6.43 \times 10^{-6},\, 7.05 \times 10^{-2}]~\mathrm{Mpc}^{-1}$, we map our constraints onto the $(B_{\mathrm{nG}}/M_{10}, n_e)$ plane (grey band, left panel of figure~\ref{fig:physics_constraints}). At low densities ($n_e \lesssim 10^{-8} \text{ cm}^{-3}$), the conversion is in the coherent regime where the $n_e$ dependence analytically cancels out, producing a flat band. Here, the lower and upper edges correspond to coupling scales of $M \approx 6.0 \times 10^{11}~\mathrm{GeV}$ and $M \approx 5.7 \times 10^{9}~\mathrm{GeV}$, respectively. At higher densities ($n_e \gtrsim 10^{-7} \text{ cm}^{-3}$), the effective plasma mass suppresses the mixing (incoherent regime), bending the band upward as stronger couplings (smaller $M$) are required to produce an equivalent dimming signal.

Physically, the lower edge ($M \approx 6.0 \times 10^{11}~\mathrm{GeV}$) defines the sensitivity threshold of current cosmological data; weaker couplings produce dimming too faint to be distinguished from the standard $\Lambda$CDM model. Conversely, the upper edge ($M \approx 5.7 \times 10^{9}~\mathrm{GeV}$) marks a saturation threshold. In this extreme strong-coupling regime, the photon survival probability drops rapidly to a constant $\mathcal{P}(z) = 2/3$. Because the SNIa absolute magnitude $M_B$ is marginalized as a free nuisance parameter in our fit, this redshift-independent dimming is perfectly degenerate with a shift in $M_B$. Consequently, while intermediate couplings are excluded, the extreme strong-coupling region evades detection as the ALP effects become indistinguishable from the intrinsic properties of standard candles.

\subsubsection{Mini-Charged Particles (MCPs)}
An alternative opacity source arises from hypothetical mini-charged particles (MCPs) with a fractional electric charge $q_\epsilon \ll e$ \cite{HOLDOM1986196}. Astrophysical and cosmological constraints on MCPs have been established through various channels, including SN1987A cooling \cite{Chang_2018,PhysRevLett.133.251004}, modified stellar evolution \cite{Vogel_2014,PhysRevD.109.083011}, galactic magnetic fields \cite{Stebbins_2019}, and the Cosmic Microwave Background (CMB) \cite{Burrage_200902}. While recent geomagnetic studies have achieved exquisite local sensitivity \cite{arza_geomagnetic_2026}, cosmological distance measurements offer a unique global perspective. This approach provides a complementary, large-scale verification of photon conservation over Gpc scales, independent of local environment assumptions.

In this model, photons can decay into MCP pairs in background magnetic fields via the process $\gamma \xrightarrow{B} \psi + \bar{\psi}$, leading to an exponential extinction law $\mathcal{P}(z) = \exp(-\kappa D_C(z))$ \cite{Ahlers_2009}. Mathematically, this survival probability is recovered from the ALP case in eq.~(\ref{eq:alptranverse}) in the limit $A \to 0$ by identifying the extinction coefficient as $\kappa \equiv 3P/(2L)$. In the deep non-perturbative regime, $\kappa$ is expressed as:
\begin{equation}
    \kappa = \frac{e^{8/3}}{4\Gamma(\frac{1}{6})\Gamma(\frac{13}{6})}
    \left(\frac{2B^{2}q_{\epsilon}^8}{3\omega_\gamma}\right)^{1/3} f,
\end{equation}
where $\Gamma$ denotes the Gamma function, and $f$ is a numerical factor of order unity accounting for the MCP spin and photon polarization (specifically, $f=1$ and $2/3$ for the parallel and perpendicular polarizations of a Dirac fermion, respectively). This expression is valid subject to the adiabatic condition $3\omega_\gamma e q_{\epsilon} B / (2 m_{\psi}^3) \ll 1$ \cite{Ahlers_2007}. 

In our analysis, we fix the environmental parameters to $B=1~\mathrm{nG}$ and $\omega_\gamma=1.5~\mathrm{eV}$ to execute a grid-based likelihood scan over $\Omega_m$ and $\kappa$. The resulting constraints, shown in the right panel of figure~\ref{fig:alp_mcp_constraints}, allow us to place a 95\% C.L. upper limit on the phenomenological extinction coefficient of $\kappa \lesssim 5.09 \times 10^{-6}~\mathrm{Mpc}^{-1}$. Translating this into the particle physics plane ($m_\psi, q_\epsilon$) in figure~\ref{fig:physics_constraints}, we exclude a significant parameter space. For the assumed magnetic field strength, our data disfavors MCPs with mass $m_\psi \lesssim 2.07 \times 10^{-7}~\mathrm{eV}$ and charge $q_\epsilon \gtrsim 6.69 \times 10^{-10}e$ (95\% C.L.), where the intersection of the constraint contours marks the regime where the adiabatic approximation breaks down.

\section{Discussion and Conclusion}

In this work, we have performed a comprehensive, model-independent test of the Cosmic Distance Duality Relation (CDDR) and cosmic transparency by combining the latest Baryon Acoustic Oscillations (BAO) data from DESI DR2 with Type Ia Supernovae from Pantheon+. To ensure robustness against methodological systematics, we employed two complementary redshift-matching techniques: Gaussian Process Regression (GPR) for capturing global trends and the Free-Knots Method (FKM) for assessing local variations. Our analysis leads to the following key conclusions:

\begin{itemize}
    \item \emph{Robustness of the CDDR:} 
    Using both GPR and FKM reconstruction methods, we first perform direct null tests on the CDDR and demonstrate that the null hypothesis ($\eta(z)=1$) holds within statistical uncertainties. Specifically, the relation remains consistent with observational data when utilizing the Planck 2018 (P18) calibration or when leaving the nuisance parameters free (Noprior). For instance, under the linear parameterization $\eta(z) = 1 + \eta_1 z$, the GPR analysis yields $\eta_1 = 0.0227^{+0.0272}_{-0.0265}$, which is compatible with the standard expectation of $\eta_1=0$ at the $1\sigma$ level. This result reinforces the Etherington reciprocity theorem as a robust property of the late-time universe described by the standard cosmological model.

    \item \emph{Calibration Sensitivity and the Hubble Tension:} 
    The analysis demonstrates the degeneracy between the CDDR parameter $\eta(z)$ and the absolute distance calibration, governed by $M_B$ and $r_d$. While the results under the Planck 2018 prior support the reciprocity theorem, the adoption of the local distance ladder prior (R22) introduces a modest statistical discrepancy (up to $1.8\sigma$ in FKM). Conversely, assuming the validity of the CDDR ($\eta=1$) leads to an internally calibrated supernova absolute magnitude of $M_B \approx -19.44$ mag. This value aligns with Inverse Distance Ladder estimates but exhibits a difference from the SH0ES calibration. These deviations and the Hubble tension may share a common origin, consistent with findings in related studies exploring calibration inconsistencies \cite{10.1093/mnras/stab1588,renzi_resilience_2022}. Theoretically, this sensitivity reinforces that the CDDR is intrinsically intertwined with the cosmic expansion history and spatial curvature $\Omega_k$ \cite{Di_Valentino_2019,PhysRevD.103.L041301}. A genuine violation of the CDDR could also potentially mimic the signatures of a non-flat universe or other physical phenomena \cite{the_euclid_theory_working_group_cosmology_2013}. Consequently, the CDDR serves as a fundamental consistency check for the standard cosmological framework, underscoring the ongoing challenges in reconciling early- and late-universe distance anchors within our current metric description of the universe.

    \item \emph{Constraints on Cosmic Opacity and Dark Sector Physics:} 
    Investigating the transparency of the universe, the reconstructed differential optical depth $\Delta \tau$ is largely consistent with zero, and the observed fluctuations are compatible with statistical noise. This is further supported by the average of the opacity derivative, which is compatible with zero ($\langle d\tau/dz \rangle \approx 0$). Based on these results, stringent constraints are placed on dark sector physics. For Axion-Like Particles (ALPs), an excluded interval for the photon-ALP coupling scale is identified as $M \in[5.7 \times 10^9, \,\, 6.0 \times 10^{11}]$~GeV (95\% C.L.). For Mini-Charged Particles (MCPs), the region characterized by mass $m_\psi \lesssim 2.07 \times 10^{-7}$~eV and electric charge $q_\epsilon \gtrsim 6.69 \times 10^{-10}e$ is similarly excluded (95\% C.L.). These results provide a robust, model-independent verification of photon conservation over cosmological scales.
\end{itemize}

In summary, current observations from DESI DR2 and Pantheon+ support the validity of CDDR. However, the non-negligible impact of calibration priors highlights that the CDDR test is linked to the Hubble tension. As we move into the era of ultra-precision cosmology, disentangling these degeneracies will be crucial for distinguishing between subtle systematic effects and genuine signatures of new physics.

\section*{Acknowledgments}
We sincerely thank the anonymous referee for the valuable comments and suggestions that improved our manuscript. We are also grateful to Pushty Shrimankar for helpful email correspondence. This work is supported by the National SKA Program of China (Grants Nos. 2022SKA0110200 and 2022SKA0110203), the National Key R\&D Intergovernmental Cooperation Program of China (2024YFA1611500), the Regional Collaborative Innovation Project of Xinjiang Uyghur Autonomous Region (2022E01013), and the National Natural Science Foundation of China (12173078). This research has also made use of the following open-source packages: \texttt{matplotlib} \cite{2007CSE.....9...90H}, \texttt{numpy} \cite{2020Natur.585..357H} and \texttt{scipy} \cite{2020NatMe..17..261V}. Our code and MCMC chain can be found in \url{https://github.com/irosphis/CDDR_Test}.


\bibliographystyle{JHEP}
\bibliography{biblio.bib}

@ARTICLE{1933PMag...15..761E,
       author = {{Etherington}, I.~M.~H.},
        title = "{On the Definition of Distance in General Relativity.}",
      journal = {Philosophical Magazine},
         year = 1933,
        month = jan,
       volume = {15},
       number = {18},
        pages = {761},
       adsurl = {https://ui.adsabs.harvard.edu/abs/1933PMag...15..761E}
}

@INPROCEEDINGS{1971grc..conf..104E,
       author = {{Ellis}, G.~F.~R.},
        title = "{Relativistic Cosmology}",
    booktitle = {General Relativity and Cosmology},
         year = 1971,
       editor = {{Sachs}, R.~K.},
        month = jan,
        pages = {104-182},
       adsurl = {https://ui.adsabs.harvard.edu/abs/1971grc..conf..104E}
}

@article{desicollaboration2025desidr2resultsi,
  title = {DESI DR2 results. I. Baryon acoustic oscillations from the Lyman alpha forest},
  author = {Abdul Karim, M. and Aguilar, J. and Ahlen, S. and Allende Prieto, C. and Alves, O. and Anand, A. and Andrade, U. and Armengaud, E. and Aviles, A. and Bailey, S. and Bault, A. and Behera, J. and BenZvi, S. and Bianchi, D. and Blake, C. and Brodzeller, A. and Brooks, D. and Buckley-Geer, E. and Burtin, E. and Calderon, R. and Canning, R. and Carnero Rosell, A. and Carrilho, P. and Casas, L. and Castander, F. J. and Cereskaite, R. and Charles, M. and Chaussidon, E. and Chaves-Montero, J. and Chebat, D. and Claybaugh, T. and Cole, S. and Cooper, A. P. and Cuceu, A. and Dawson, K. S. and de Belsunce, R. and de la Macorra, A. and de Mattia, A. and Deiosso, N. and Della Costa, J. and Dey, A. and Dey, B. and Ding, Z. and Doel, P. and Edelstein, J. and Eisenstein, D. J. and Elbers, W. and Fagrelius, P. and Fanning, K. and Ferraro, S. and Font-Ribera, A. and Forero-Romero, J. E. and Garcia-Quintero, C. and Garrison, L. H. and Gazta\~naga, E. and Gil-Mar\'{\i}n, H. and Gontcho, S. Gontcho A. and Gonzalez-Morales, A. X. and Gordon, C. and Green, D. and Gutierrez, G. and Guy, J. and Hahn, C. and Herbold, M. and Herrera-Alcantar, H. K. and Ho, M. and Ho, M.-F. and Honscheid, K. and Howlett, C. and Huterer, D. and Ishak, M. and Juneau, S. and Kara\ifmmode \mbox{\c{c}}\else \c{c}\fi{}ayl��, N. G. and Kehoe, R. and Kent, S. and Kirkby, D. and Kisner, T. and Kitaura, F.-S. and Koposov, S. E. and Kremin, A. and Lahav, O. and Lamman, C. and Landriau, M. and Lang, D. and Lasker, J. and Le Goff, J. M. and Le Guillou, L. and Leauthaud, A. and Levi, M. E. and Li, Q. and Li, T. S. and Lodha, K. and Lokken, M. and Magneville, C. and Manera, M. and Martini, P. and Matthewson, W. L. and McDonald, P. and Meisner, A. and Mena-Fern\'andez, J. and Miquel, R. and Moustakas, J. and Mu\~noz-Guti\'errez, A. and Mu\~noz-Santos, D. and Myers, A. D. and Newman, J. A. and Niz, G. and Noriega, H. E. and Paillas, E. and Palanque-Delabrouille, N. and Pan, J. and Percival, W. J. and P\'erez-R\`afols, I. and Pieri, M. M. and Poppett, C. and Prada, F. and Rabinowitz, D. and Raichoor, A. and Ram\'{\i}rez-P\'erez, C. and Rashkovetskyi, M. and Ravoux, C. and Rich, J. and Rockosi, C. and Ross, A. J. and Rossi, G. and Ruhlmann-Kleider, V. and Sanchez, E. and Sanders, N. and Satyavolu, S. and Schlegel, D. and Schubnell, M. and Seo, H. and Shafieloo, A. and Sharples, R. and Silber, J. and Sinigaglia, F. and Sprayberry, D. and Tan, T. and Tarl\'e, G. and Taylor, P. and Turner, W. and Valdes, F. and Vargas-Maga\~na, M. and Walther, M. and Weaver, B. A. and Wolfson, M. and Y\`eche, C. and Zarrouk, P. and Zhou, R. and Zou, H.},
  collaboration = {DESI Collaboration},
  journal = {Phys. Rev. D},
  volume = {112},
  issue = {8},
  pages = {083514},
  numpages = {28},
  year = {2025},
  month = {Oct},
  publisher = {American Physical Society},
  doi = {10.1103/2wwn-xjm5},
  url = {https://link.aps.org/doi/10.1103/2wwn-xjm5}
}

@article{desicollaboration2025desidr2resultsii,
  title = {DESI DR2 results. II. Measurements of baryon acoustic oscillations and cosmological constraints},
  author = {Abdul Karim, M. and Aguilar, J. and Ahlen, S. and Alam, S. and Allen, L. and Prieto, C. Allende and Alves, O. and Anand, A. and Andrade, U. and Armengaud, E. and Aviles, A. and Bailey, S. and Baltay, C. and Bansal, P. and Bault, A. and Behera, J. and BenZvi, S. and Bianchi, D. and Blake, C. and Brieden, S. and Brodzeller, A. and Brooks, D. and Buckley-Geer, E. and Burtin, E. and Calderon, R. and Canning, R. and Rosell, A. Carnero and Carrilho, P. and Casas, L. and Castander, F. J. and Charles, M. and Chaussidon, E. and Chaves-Montero, J. and Chebat, D. and Chen, X. and Claybaugh, T. and Cole, S. and Cooper, A. P. and Cuceu, A. and Dawson, K. S. and de la Macorra, A. and de Mattia, A. and Deiosso, N. and Della Costa, J. and Demina, R. and Dey, A. and Dey, B. and Ding, Z. and Doel, P. and Edelstein, J. and Eisenstein, D. J. and Elbers, W. and Fagrelius, P. and Fanning, K. and Fern\'andez-Garc\'{\i}a, E. and Ferraro, S. and Font-Ribera, A. and Forero-Romero, J. E. and Frenk, C. S. and Garcia-Quintero, C. and Garrison, L. H. and Gazta\~naga, E. and Gil-Mar\'{\i}n, H. and Gontcho, S. Gontcho A. and Gonzalez, D. and Gonzalez-Morales, A. X. and Gordon, C. and Green, D. and Gutierrez, G. and Guy, J. and Hadzhiyska, B. and Hahn, C. and He, S. and Herbold, M. and Herrera-Alcantar, H. K. and Ho, M.-F. and Honscheid, K. and Howlett, C. and Huterer, D. and Ishak, M. and Juneau, S. and Kamble, N. V. and Kara\ifmmode \mbox{\c{c}}\else \c{c}\fi{}ayl��, N. G. and Kehoe, R. and Kent, S. and Kim, A. G. and Kirkby, D. and Kisner, T. and Koposov, S. E. and Kremin, A. and Krolewski, A. and Lahav, O. and Lamman, C. and Landriau, M. and Lang, D. and Lasker, J. and Le Goff, J. M. and Le Guillou, L. and Leauthaud, A. and Levi, M. E. and Li, Q. and Li, T. S. and Lodha, K. and Lokken, M. and Lozano-Rodr\'{\i}guez, F. and Magneville, C. and Manera, M. and Martini, P. and Matthewson, W. L. and Meisner, A. and Mena-Fern\'andez, J. and Menegas, A. and Mergulh\~ao, T. and Miquel, R. and Moustakas, J. and Mu\~noz-Guti\'errez, A. and Mu\~noz-Santos, D. and Myers, A. D. and Nadathur, S. and Naidoo, K. and Napolitano, L. and Newman, J. A. and Niz, G. and Noriega, H. E. and Paillas, E. and Palanque-Delabrouille, N. and Pan, J. and Peacock, J. A. and Ibanez, M. P. and Percival, W. J. and P\'erez-Fern\'andez, A. and P\'erez-R\`afols, I. and Pieri, M. M. and Poppett, C. and Prada, F. and Rabinowitz, D. and Raichoor, A. and Ram\'{\i}rez-P\'erez, C. and Rashkovetskyi, M. and Ravoux, C. and Rich, J. and Rocher, A. and Rockosi, C. and Rohlf, J. and Rom\'an-Herrera, J. O. and Ross, A. J. and Rossi, G. and Ruggeri, R. and Ruhlmann-Kleider, V. and Samushia, L. and Sanchez, E. and Sanders, N. and Schlegel, D. and Schubnell, M. and Seo, H. and Shafieloo, A. and Sharples, R. and Silber, J. and Sinigaglia, F. and Sprayberry, D. and Tan, T. and Tarl\'e, G. and Taylor, P. and Turner, W. and Ure\~na-L\'opez, L. A. and Vaisakh, R. and Valdes, F. and Valogiannis, G. and Vargas-Maga\~na, M. and Verde, L. and Walther, M. and Weaver, B. A. and Weinberg, D. H. and White, M. and Wolfson, M. and Y\`eche, C. and Yu, J. and Zaborowski, E. A. and Zarrouk, P. and Zhai, Z. and Zhang, H. and Zhao, C. and Zhao, G. B. and Zhou, R. and Zou, H.},
  collaboration = {DESI Collaboration},
  journal = {Phys. Rev. D},
  volume = {112},
  issue = {8},
  pages = {083515},
  numpages = {40},
  year = {2025},
  month = {Oct},
  publisher = {American Physical Society},
  doi = {10.1103/tr6y-kpc6},
  url = {https://link.aps.org/doi/10.1103/tr6y-kpc6}
}

@article{Conley_2010,
   title={SUPERNOVA CONSTRAINTS AND SYSTEMATIC UNCERTAINTIES FROM THE FIRST THREE YEARS OF THE SUPERNOVA LEGACY SURVEY},
   volume={192},
   ISSN={1538-4365},
   url={http://dx.doi.org/10.1088/0067-0049/192/1/1},
   DOI={10.1088/0067-0049/192/1/1},
   number={1},
   journal={The Astrophysical Journal Supplement Series},
   publisher={American Astronomical Society},
   author={Conley, A. and Guy, J. and Sullivan, M. and Regnault, N. and Astier, P. and Balland, C. and Basa, S. and Carlberg, R. G. and Fouchez, D. and Hardin, D. and Hook, I. M. and Howell, D. A. and Pain, R. and Palanque-Delabrouille, N. and Perrett, K. M. and Pritchet, C. J. and Rich, J. and Ruhlmann-Kleider, V. and Balam, D. and Baumont, S. and Ellis, R. S. and Fabbro, S. and Fakhouri, H. K. and Fourmanoit, N. and González-Gaitán, S. and Graham, M. L. and Hudson, M. J. and Hsiao, E. and Kronborg, T. and Lidman, C. and Mourao, A. M. and Neill, J. D. and Perlmutter, S. and Ripoche, P. and Suzuki, N. and Walker, E. S.},
   year={2010},
   month=dec, pages={1} }

@article{Uzan_2004,
   title={Distance duality relation from x-ray and Sunyaev-Zel’dovich observationsof clusters},
   volume={70},
   ISSN={1550-2368},
   url={http://dx.doi.org/10.1103/PhysRevD.70.083533},
   DOI={10.1103/physrevd.70.083533},
   number={8},
   journal={Physical Review D},
   publisher={American Physical Society (APS)},
   author={Uzan, Jean-Philippe and Aghanim, Nabila and Mellier, Yannick},
   year={2004},
   month=oct }

@article{Santana_2017,
   title={How does light move in a generic metric-affine background?},
   volume={95},
   ISSN={2470-0029},
   url={http://dx.doi.org/10.1103/PhysRevD.95.061501},
   DOI={10.1103/physrevd.95.061501},
   number={6},
   journal={Physical Review D},
   publisher={American Physical Society (APS)},
   author={Santana, Lucas T. and Calvão, Maurício O. and Reis, Ribamar R. R. and Siffert, Beatriz B.},
   year={2017},
   month=mar }

@article{buenabad2022constraintsaxionscosmicdistance,
	title = {Constraints on axions from cosmic distance measurements},
	volume = {2022},
	issn = {1029-8479},
	url = {https://link.springer.com/10.1007/JHEP02(2022)103},
	doi = {10.1007/JHEP02(2022)103},
	language = {en},
	number = {2},
	urldate = {2025-05-26},
	journal = {Journal of High Energy Physics},
	author = {Buen-Abad, Manuel A. and Fan, JiJi and Sun, Chen},
	month = feb,
	year = {2022},
	pages = {103},
}

@article{Tiwari_2017,
   title={Constraining axionlike particles using the distance-duality relation},
   volume={95},
   ISSN={2470-0029},
   url={http://dx.doi.org/10.1103/PhysRevD.95.023005},
   DOI={10.1103/physrevd.95.023005},
   number={2},
   journal={Physical Review D},
   publisher={American Physical Society (APS)},
   author={Tiwari, Prabhakar},
   year={2017},
   month=jan }

@article{Azevedo_2021,
   title={Distance-duality in theories with a nonminimal coupling to gravity},
   volume={104},
   ISSN={2470-0029},
   url={http://dx.doi.org/10.1103/PhysRevD.104.084079},
   DOI={10.1103/physrevd.104.084079},
   number={8},
   journal={Physical Review D},
   publisher={American Physical Society (APS)},
   author={Azevedo, R. P. L. and Avelino, P. P.},
   year={2021},
   month=oct }

@article{Cao_2016,
   title={Testing the gas mass density profile of galaxy clusters with distance duality relation},
   volume={457},
   ISSN={1365-2966},
   url={http://dx.doi.org/10.1093/mnras/stv2999},
   DOI={10.1093/mnras/stv2999},
   number={1},
   journal={Monthly Notices of the Royal Astronomical Society},
   publisher={Oxford University Press (OUP)},
   author={Cao, Shuo and Biesiada, Marek and Zheng, Xiaogang and Zhu, Zong-Hong},
   year={2016},
   month=jan, pages={281–287} }

@article{More_2009,
    doi = {10.1088/0004-637X/696/2/1727},
    url = {https://dx.doi.org/10.1088/0004-637X/696/2/1727},
    year = {2009},
    month = {apr},
    publisher = {The American Astronomical Society},
    volume = {696},
    number = {2},
    pages = {1727},
    author = {More, Surhud and Bovy, Jo and Hogg, David W.},
    title = {COSMIC TRANSPARENCY: A TEST WITH THE BARYON ACOUSTIC FEATURE AND TYPE Ia SUPERNOVAE},
    journal = {The Astrophysical Journal}
}

@article{Remya_Nair_2012,
    doi = {10.1088/1475-7516/2012/12/028},
    url = {https://dx.doi.org/10.1088/1475-7516/2012/12/028},
    year = {2012},
    month = {dec},
    publisher = {},
    volume = {2012},
    number = {12},
    pages = {028},
    author = {Remya Nair and Sanjay Jhingan and Deepak Jain},
    title = {Cosmic distance duality and cosmic transparency},
    journal = {Journal of Cosmology and Astroparticle Physics}
}

@article{Brout_2022,
   title={The Pantheon+ Analysis: Cosmological Constraints},
   volume={938},
   ISSN={1538-4357},
   url={http://dx.doi.org/10.3847/1538-4357/ac8e04},
   DOI={10.3847/1538-4357/ac8e04},
   number={2},
   journal={The Astrophysical Journal},
   publisher={American Astronomical Society},
   author={Brout, Dillon and Scolnic, Dan and Popovic, Brodie and Riess, Adam G. and Carr, Anthony and Zuntz, Joe and Kessler, Rick and Davis, Tamara M. and Hinton, Samuel and Jones, David and Kenworthy, W. D’Arcy and Peterson, Erik R. and Said, Khaled and Taylor, Georgie and Ali, Noor and Armstrong, Patrick and Charvu, Pranav and Dwomoh, Arianna and Meldorf, Cole and Palmese, Antonella and Qu, Helen and Rose, Benjamin M. and Sanchez, Bruno and Stubbs, Christopher W. and Vincenzi, Maria and Wood, Charlotte M. and Brown, Peter J. and Chen, Rebecca and Chambers, Ken and Coulter, David A. and Dai, Mi and Dimitriadis, Georgios and Filippenko, Alexei V. and Foley, Ryan J. and Jha, Saurabh W. and Kelsey, Lisa and Kirshner, Robert P. and Möller, Anais and Muir, Jessie and Nadathur, Seshadri and Pan, Yen-Chen and Rest, Armin and Rojas-Bravo, Cesar and Sako, Masao and Siebert, Matthew R. and Smith, Mat and Stahl, Benjamin E. and Wiseman, Phil},
   year={2022},
   month=oct, pages={110}
}

@article{xu_model-independent_2022,
	title = {Model-independent {Test} for the {Cosmic} {Distance}–{Duality} {Relation} with {Pantheon} and {eBOSS} {DR16} {Quasar} {Sample}},
	volume = {939},
	issn = {0004-637X, 1538-4357},
	url = {https://iopscience.iop.org/article/10.3847/1538-4357/ac9793},
	doi = {10.3847/1538-4357/ac9793},
	language = {en},
	number = {2},
	urldate = {2025-04-04},
	journal = {The Astrophysical Journal},
	author = {Xu, Bing and Wang, Zhenzhen and Zhang, Kaituo and Huang, Qihong and Zhang, Jianjian},
	month = nov,
	year = {2022},
	pages = {115},
}

@article{PhysRevD.69.101305,
  title = {Cosmic distance-duality as a probe of exotic physics and acceleration},
  author = {Bassett, Bruce A. and Kunz, Martin},
  journal = {Phys. Rev. D},
  volume = {69},
  issue = {10},
  pages = {101305},
  numpages = {5},
  year = {2004},
  month = {May},
  publisher = {American Physical Society},
  doi = {10.1103/PhysRevD.69.101305},
  url = {https://link.aps.org/doi/10.1103/PhysRevD.69.101305}
}

@article{DeBernardis:2006ii,
    author = "De Bernardis, Francesco and Giusarma, Elena and Melchiorri, Alessandro",
    title = "{Constraints on dark energy and distance duality from Sunyaev Zel'dovich effect and Chandra X-ray measurements}",
    eprint = "gr-qc/0606029",
    archivePrefix = "arXiv",
    doi = "10.1142/S0218271806008486",
    journal = "Int. J. Mod. Phys. D",
    volume = "15",
    pages = "759--766",
    year = "2006"
}

@article{Holanda_2010,
   title={TESTING THE DISTANCE–DUALITY RELATION WITH GALAXY CLUSTERS AND TYPE Ia SUPERNOVAE},
   volume={722},
   ISSN={2041-8213},
   url={http://dx.doi.org/10.1088/2041-8205/722/2/L233},
   DOI={10.1088/2041-8205/722/2/l233},
   number={2},
   journal={The Astrophysical Journal},
   publisher={American Astronomical Society},
   author={Holanda, R. F. L. and Lima, J. A. S. and Ribeiro, M. B.},
   year={2010},
   month=oct, pages={L233–L237} 
}

@ARTICLE{deleo2025distinguishingdistancedualitybreaking,
       author = {{De Leo}, Chiara and {Martinelli}, Matteo and {D'Agostino}, Rocco and {Gianfagna}, Giulia and {Martins}, C.~J.~A.~P.},
        title = "{Distinguishing distance duality breaking models using electromagnetic and gravitational waves measurements}",
      journal = {Journal of Cosmology and Astroparticle Physics},
         year = 2025,
        month = nov,
       volume = {2025},
       number = {11},
          eid = {001},
        pages = {001},
          doi = {10.1088/1475-7516/2025/11/001},
archivePrefix = {arXiv},
       eprint = {2505.13613},
 primaryClass = {astro-ph.CO},
       adsurl = {https://ui.adsabs.harvard.edu/abs/2025JCAP...11..001D}
}

@article{PhysRevD.103.103513,
  title = {Machine learning forecasts of the cosmic distance duality relation with strongly lensed gravitational wave events},
  author = {Arjona, Rub\'en and Lin, Hai-Nan and Nesseris, Savvas and Tang, Li},
  journal = {Phys. Rev. D},
  volume = {103},
  issue = {10},
  pages = {103513},
  numpages = {15},
  year = {2021},
  month = {May},
  publisher = {American Physical Society},
  doi = {10.1103/PhysRevD.103.103513},
  url = {https://link.aps.org/doi/10.1103/PhysRevD.103.103513}
}

@article{Rana_2017,
doi = {10.1088/1475-7516/2017/07/010},
url = {https://dx.doi.org/10.1088/1475-7516/2017/07/010},
year = {2017},
month = {jul},
publisher = {},
volume = {2017},
number = {07},
pages = {010},
author = {Rana, Akshay and Jain, Deepak and Mahajan, Shobhit and Mukherjee, Amitabha and Holanda, R.F.L.},
title = {Probing the cosmic distance duality relation using time delay lenses},
journal = {Journal of Cosmology and Astroparticle Physics},
}

@article{Corasaniti_2006,
   title={The impact of cosmic dust on supernova cosmology},
   volume={372},
   ISSN={1365-2966},
   url={http://dx.doi.org/10.1111/j.1365-2966.2006.10825.x},
   DOI={10.1111/j.1365-2966.2006.10825.x},
   number={1},
   journal={Monthly Notices of the Royal Astronomical Society},
   publisher={Oxford University Press (OUP)},
   author={Corasaniti, P. S.},
   year={2006},
   month=oct, pages={191–198} }

@article{Qi_2025,
   title={Testing the Cosmic Distance Duality Relation Using Strong Gravitational Lensing Time Delays and Type Ia Supernovae},
   volume={979},
   ISSN={1538-4357},
   url={http://dx.doi.org/10.3847/1538-4357/ad9de4},
   DOI={10.3847/1538-4357/ad9de4},
   number={1},
   journal={The Astrophysical Journal},
   publisher={American Astronomical Society},
   author={Qi, Jing-Zhao and Jiang, Yi-Fan and Hou, Wan-Ting and Zhang, Xin},
   year={2025},
   month=jan, pages={2} }

@ARTICLE{tonghua2023recentobservationstellinglight,
       author = {{Liu}, Tonghua and {Cao}, Shuo and {Ma}, Shuai and {Liu}, Yuting and {Zheng}, Chenfa and {Wang}, Jieci},
        title = "{What are recent observations telling us in light of improved tests of distance duality relation?}",
      journal = {Physics Letters B},
         year = 2023,
        month = mar,
       volume = {838},
          eid = {137687},
        pages = {137687},
          doi = {10.1016/j.physletb.2023.137687},
archivePrefix = {arXiv},
       eprint = {2301.02997},
 primaryClass = {astro-ph.CO},
       adsurl = {https://ui.adsabs.harvard.edu/abs/2023PhLB..83837687L}
}

@article{teixeira_implications_2025,
  title = {Implications of distance duality violation for the ${H}_{0}$ tension and evolving dark energy},
  author = {Teixeira, Elsa M. and Giar\`e, William and Hogg, Natalie B. and Montandon, Thomas and Poudou, Ad\`ele and Poulin, Vivian},
  journal = {Phys. Rev. D},
  volume = {112},
  issue = {2},
  pages = {023515},
  numpages = {19},
  year = {2025},
  month = {Jul},
  publisher = {American Physical Society},
  doi = {10.1103/zzmp-rxrh},
  url = {https://link.aps.org/doi/10.1103/zzmp-rxrh}
}

@ARTICLE{alfano_cosmic_2025,
       author = {{Alfano}, Anna Chiara and {Luongo}, Orlando},
        title = "{Cosmic distance duality after DESI 2024 data release and dark energy evolution}",
      journal = {Physics of the Dark Universe},
         year = 2026,
        month = feb,
       volume = {51},
          eid = {102205},
        pages = {102205},
          doi = {10.1016/j.dark.2025.102205},
archivePrefix = {arXiv},
       eprint = {2501.15233},
 primaryClass = {astro-ph.CO},
       adsurl = {https://ui.adsabs.harvard.edu/abs/2026PDU....5102205A}
}

@ARTICLE{favale2024quantification2dvs3d,
       author = {{Favale}, Arianna and {G{\'o}mez-Valent}, Adri{\`a} and {Migliaccio}, Marina},
        title = "{Quantification of 2D vs 3D BAO tension using SNIa as a redshift interpolator and test of the Etherington relation}",
      journal = {Physics Letters B},
         year = 2024,
        month = nov,
       volume = {858},
          eid = {139027},
        pages = {139027},
          doi = {10.1016/j.physletb.2024.139027},
archivePrefix = {arXiv},
       eprint = {2405.12142},
 primaryClass = {astro-ph.CO},
       adsurl = {https://ui.adsabs.harvard.edu/abs/2024PhLB..85839027F}
}

@article{PhysRevD.99.063507,
  title = {Testing the Etherington distance duality relation at higher redshifts: Combined radio quasar and gravitational wave data},
  author = {Qi, Jing-Zhao and Cao, Shuo and Zheng, Chenfa and Pan, Yu and Li, Zejun and Li, Jin and Liu, Tonghua},
  journal = {Phys. Rev. D},
  volume = {99},
  issue = {6},
  pages = {063507},
  numpages = {9},
  year = {2019},
  month = {Mar},
  publisher = {American Physical Society},
  doi = {10.1103/PhysRevD.99.063507},
  url = {https://link.aps.org/doi/10.1103/PhysRevD.99.063507}
}

@article{Bora_2021,
   title={A test of cosmic distance duality relation using SPT-SZ  galaxy clusters, Type Ia supernovae, and cosmic chronometers},
   volume={2021},
   ISSN={1475-7516},
   url={http://dx.doi.org/10.1088/1475-7516/2021/06/052},
   DOI={10.1088/1475-7516/2021/06/052},
   number={06},
   journal={Journal of Cosmology and Astroparticle Physics},
   publisher={IOP Publishing},
   author={Bora, Kamal and Desai, Shantanu},
   year={2021},
   month=jun, pages={052} }

@article{huang_opacity-free_2024,
    title = {An opacity-free method of testing the cosmic distance duality relation using strongly lensed gravitational wave signals},
    journal = {Physics of the Dark Universe},
    volume = {47},
    pages = {101810},
    year = {2025},
    issn = {2212-6864},
    doi = {https://doi.org/10.1016/j.dark.2025.101810},
    url = {https://www.sciencedirect.com/science/article/pii/S2212686425000056},
    author = {Shun-Jia Huang and En-Kun Li and Jian-dong Zhang and Xian Chen and Zucheng Gao and Xin-yi Lin and Yi-Ming Hu},
}

@article{Meng_2012,
   title={MORPHOLOGY OF GALAXY CLUSTERS: A COSMOLOGICAL MODEL-INDEPENDENT TEST OF THE COSMIC DISTANCE-DUALITY RELATION},
   volume={745},
   ISSN={1538-4357},
   url={http://dx.doi.org/10.1088/0004-637X/745/1/98},
   DOI={10.1088/0004-637x/745/1/98},
   number={1},
   journal={The Astrophysical Journal},
   publisher={American Astronomical Society},
   author={Meng, Xiao-Lei and Zhang, Tong-Jie and Zhan, Hu and Wang, Xin},
   year={2012},
   month=jan, pages={98} }

@article{Liao_2019,
   title={The Cosmic Distance Duality Relation with Strong Lensing and Gravitational Waves: An Opacity-free Test},
   volume={885},
   ISSN={1538-4357},
   url={http://dx.doi.org/10.3847/1538-4357/ab4819},
   DOI={10.3847/1538-4357/ab4819},
   number={1},
   journal={The Astrophysical Journal},
   publisher={American Astronomical Society},
   author={Liao, Kai},
   year={2019},
   month=oct, pages={70} }

@article{PhysRevD.99.083523,
  title = {Testing the cosmic distance-duality relation from future gravitational wave standard sirens},
  author = {Fu, Xiangyun and Zhou, Lu and Chen, Jun},
  journal = {Phys. Rev. D},
  volume = {99},
  issue = {8},
  pages = {083523},
  numpages = {9},
  year = {2019},
  month = {Apr},
  publisher = {American Physical Society},
  doi = {10.1103/PhysRevD.99.083523},
  url = {https://link.aps.org/doi/10.1103/PhysRevD.99.083523}
}

@article{yang2019constraintscosmicdistanceduality,
title = {Constraints on the cosmic distance duality relation with simulated data of gravitational waves from the Einstein Telescope},
journal = {Astroparticle Physics},
volume = {108},
pages = {57-62},
year = {2019},
issn = {0927-6505},
doi = {https://doi.org/10.1016/j.astropartphys.2019.01.005},
url = {https://www.sciencedirect.com/science/article/pii/S0927650518302536},
author = {Tao Yang and R.F.L. Holanda and Bin Hu},
}

@article{cao2011distancedualityrelationtemperature,
       author = {{Cao}, Shuo and {Zhu}, ZongHong},
        title = "{The distance duality relation and the temperature profile of galaxy clusters}",
      journal = {Science China Physics, Mechanics, and Astronomy},
         year = 2011,
        month = dec,
       volume = {54},
       number = {12},
        pages = {2260-2264},
          doi = {10.1007/s11433-011-4559-7},
archivePrefix = {arXiv},
       eprint = {1102.2750},
 primaryClass = {astro-ph.CO},
       adsurl = {https://ui.adsabs.harvard.edu/abs/2011SCPMA..54.2260C}
}

@article{Wang_2022,
doi = {10.3847/1538-4357/ac3755},
url = {https://dx.doi.org/10.3847/1538-4357/ac3755},
year = {2022},
month = {jan},
publisher = {The American Astronomical Society},
volume = {924},
number = {2},
pages = {97},
author = {Wang, F. Y. and Hu, J. P. and Zhang, G. Q. and Dai, Z. G.},
title = {Standardized Long Gamma-Ray Bursts as a Cosmic Distance Indicator},
journal = {The Astrophysical Journal}
}

@article{Zheng_2020,
doi = {10.3847/1538-4357/ab7995},
url = {https://dx.doi.org/10.3847/1538-4357/ab7995},
year = {2020},
month = {apr},
publisher = {The American Astronomical Society},
volume = {892},
number = {2},
pages = {103},
author = {Zheng, Xiaogang and Liao, Kai and Biesiada, Marek and Cao, Shuo and Liu, Tong-Hua and Zhu, Zong-Hong},
title = {Multiple Measurements of Quasars Acting as Standard Probes: Exploring the Cosmic Distance Duality Relation at Higher Redshift},
journal = {The Astrophysical Journal},
}

@ARTICLE{keil_probing_2025,
       author = {{Keil}, Felicitas and {Nesseris}, Savvas and {Tutusaus}, Isaac and {Blanchard}, Alain},
        title = "{Probing the distance duality relation with machine learning and recent data}",
      journal = {Journal of Cosmology and Astroparticle Physics},
         year = 2026,
        month = jan,
       volume = {2026},
       number = {1},
          eid = {022},
        pages = {022},
          doi = {10.1088/1475-7516/2026/01/022},
archivePrefix = {arXiv},
       eprint = {2504.01750},
 primaryClass = {astro-ph.CO},
       adsurl = {https://ui.adsabs.harvard.edu/abs/2026JCAP...01..022K}
}

@article{tang_deep_2023,
	title = {Deep learning method in testing the cosmic distance duality relation},
	volume = {47},
	issn = {1674-1137, 2058-6132},
	url = {http://arxiv.org/abs/2210.04228},
	doi = {10.1088/1674-1137/ac945b},
	language = {en},
	number = {1},
	urldate = {2024-09-30},
	journal = {Chinese Physics C},
	author = {Tang, Li and Lin, Hai-Nan and Liu, Liang},
	month = jan,
	year = {2023},
	note = {arXiv:2210.04228 [astro-ph, physics:gr-qc]},
	pages = {015101},
}

@article{wu_null_2023,
	title = {Null test for cosmic curvature using {Gaussian} process*},
	volume = {47},
	issn = {1674-1137, 2058-6132},
	url = {https://iopscience.iop.org/article/10.1088/1674-1137/acc647},
	doi = {10.1088/1674-1137/acc647},
	language = {en},
	number = {5},
	urldate = {2025-03-26},
	journal = {Chinese Physics C},
	author = {Wu, Peng-Ju and Qi, Jing-Zhao and Zhang, Xin},
	month = may,
	year = {2023},
	pages = {055106},
}

@article{bengaly_null_2022,
	title = {A null test of the {Cosmological} {Principle} with {BAO} measurements},
	volume = {35},
	issn = {22126864},
	url = {https://linkinghub.elsevier.com/retrieve/pii/S2212686422000103},
	doi = {10.1016/j.dark.2022.100966},
	language = {en},
	urldate = {2025-03-11},
	journal = {Physics of the Dark Universe},
	author = {Bengaly, Carlos},
	month = mar,
	year = {2022},
	pages = {100966},
}

@ARTICLE{gao_null_2025,
       author = {{Gao}, Shengqing and {Gao}, Qing and {Gong}, Yungui and {Lu}, Xuchen},
        title = "{Null tests with Gaussian process}",
      journal = {Science China Physics, Mechanics, and Astronomy},
         year = 2025,
        month = jun,
       volume = {68},
       number = {8},
          eid = {280408},
        pages = {280408},
          doi = {10.1007/s11433-025-2682-1},
archivePrefix = {arXiv},
       eprint = {2503.15943},
 primaryClass = {astro-ph.CO},
       adsurl = {https://ui.adsabs.harvard.edu/abs/2025SCPMA..6880408G}
}

@article{ma_statistical_2018,
	title = {Statistical {Test} of {Distance}–{Duality} {Relation} with {Type} {Ia} {Supernovae} and {Baryon} {Acoustic} {Oscillations}},
	volume = {861},
	issn = {0004-637X, 1538-4357},
	url = {https://iopscience.iop.org/article/10.3847/1538-4357/aac88f},
	doi = {10.3847/1538-4357/aac88f},
	language = {en},
	number = {2},
	urldate = {2025-04-02},
	journal = {The Astrophysical Journal},
	author = {Ma, Cong and Corasaniti, Pier-Stefano},
	month = jul,
	year = {2018},
	pages = {124},
}

@article{Holanda_2016,
doi = {10.1088/1475-7516/2016/02/054},
url = {https://dx.doi.org/10.1088/1475-7516/2016/02/054},
year = {2016},
month = {feb},
publisher = {},
volume = {2016},
number = {02},
pages = {054},
author = {Holanda, R.F.L. and Busti, V.C. and Alcaniz, J.S.},
title = {Probing the cosmic distance duality with strong gravitational lensing and supernovae Ia data},
journal = {Journal of Cosmology and Astroparticle Physics},
}

@ARTICLE{yang_testing_2024,
       author = {{Yang}, Fan and {Fu}, Xiangyun and {Xu}, Bing and {Zhang}, Kaituo and {Huang}, Yang and {Yang}, Ying},
        title = "{Testing the cosmic distance duality relation using Type Ia supernovae and radio quasars through model-independent methods}",
      journal = {Chinese Physics C},
         year = 2025,
        month = oct,
       volume = {49},
       number = {10},
          eid = {105108},
        pages = {105108},
          doi = {10.1088/1674-1137/ade4a3},
archivePrefix = {arXiv},
       eprint = {2407.05559},
 primaryClass = {astro-ph.CO},
       adsurl = {https://ui.adsabs.harvard.edu/abs/2025ChPhC..49j5108Y}
}

@article{PhysRevD.90.124064,
  title = {Breaking of the equivalence principle in the electromagnetic sector and its cosmological signatures},
  author = {Hees, Aur\'elien and Minazzoli, Olivier and Larena, Julien},
  journal = {Phys. Rev. D},
  volume = {90},
  issue = {12},
  pages = {124064},
  numpages = {14},
  year = {2014},
  month = {Dec},
  publisher = {American Physical Society},
  doi = {10.1103/PhysRevD.90.124064},
  url = {https://link.aps.org/doi/10.1103/PhysRevD.90.124064}
}

@article{Liao_2016,
doi = {10.3847/0004-637X/822/2/74},
url = {https://dx.doi.org/10.3847/0004-637X/822/2/74},
year = {2016},
month = {may},
publisher = {The American Astronomical Society},
volume = {822},
number = {2},
pages = {74},
author = {Liao, Kai and Li, Zhengxiang and Cao, Shuo and Biesiada, Marek and Zheng, Xiaogang and Zhu, Zong-Hong},
title = {THE DISTANCE DUALITY RELATION FROM STRONG GRAVITATIONAL LENSING},
journal = {The Astrophysical Journal},
}

@article{wang_testing_2024,
	title = {Testing the cosmic distance duality relation with {Type} {Ia} supernova and transverse {BAO} measurements},
	volume = {84},
	issn = {1434-6052},
	url = {http://arxiv.org/abs/2407.12250},
	doi = {10.1140/epjc/s10052-024-13049-1},
	language = {en},
	number = {7},
	urldate = {2024-09-20},
	journal = {The European Physical Journal C},
	author = {Wang, Min and Fu, Xiangyun and Xu, Bing and Huang, Yang and Yang, Ying and Lu, Zhenyan},
	month = jul,
	year = {2024},
	note = {arXiv:2407.12250 [astro-ph, physics:gr-qc]},
	pages = {702},
}

@article{Li_2011,
   title={COSMOLOGICAL-MODEL-INDEPENDENT TESTS FOR THE DISTANCE–DUALITY RELATION FROM GALAXY CLUSTERS AND TYPE Ia SUPERNOVA},
   volume={729},
   ISSN={2041-8213},
   url={http://dx.doi.org/10.1088/2041-8205/729/1/L14},
   DOI={10.1088/2041-8205/729/1/l14},
   number={1},
   journal={The Astrophysical Journal},
   publisher={American Astronomical Society},
   author={Li, Zhengxiang and Wu, Puxun and Yu, Hongwei},
   year={2011},
   month=feb, pages={L14} }

@article{Zhang_2023,
   title={Kernel Selection for Gaussian Process in Cosmology: With Approximate Bayesian Computation Rejection and Nested Sampling},
   volume={266},
   ISSN={1538-4365},
   url={http://dx.doi.org/10.3847/1538-4365/accb92},
   DOI={10.3847/1538-4365/accb92},
   number={2},
   journal={The Astrophysical Journal Supplement Series},
   publisher={American Astronomical Society},
   author={Zhang, Hao and Wang, Yu-Chen and Zhang, Tong-Jie and Zhang, Tingting},
   year={2023},
   month=may, pages={27} }

@article{Favale_2023,
  title     = {Cosmic chronometers to calibrate the ladders and measure the curvature of the Universe. A model-independent study},
  volume    = {523},
  issn      = {1365-2966},
  url       = {http://dx.doi.org/10.1093/mnras/stad1621},
  doi       = {10.1093/mnras/stad1621},
  number    = {3},
  journal   = {Monthly Notices of the Royal Astronomical Society},
  publisher = {Oxford University Press (OUP)},
  author    = {Favale, Arianna and Gómez-Valent, Adrià and Migliaccio, Marina},
  year      = {2023},
  month     = jun,
  pages     = {3406–3422}
}

@article{Riess_2022,
   title={A Comprehensive Measurement of the Local Value of the Hubble Constant with 1 km s-1 Mpc-1 Uncertainty from the Hubble Space Telescope and the SH0ES Team},
   volume={934},
   ISSN={2041-8213},
   url={http://dx.doi.org/10.3847/2041-8213/ac5c5b},
   DOI={10.3847/2041-8213/ac5c5b},
   number={1},
   journal={The Astrophysical Journal Letters},
   publisher={American Astronomical Society},
   author={Riess, Adam G. and Yuan, Wenlong and Macri, Lucas M. and Scolnic, Dan and Brout, Dillon and Casertano, Stefano and Jones, David O. and Murakami, Yukei and Anand, Gagandeep S. and Breuval, Louise and Brink, Thomas G. and Filippenko, Alexei V. and Hoffmann, Samantha and Jha, Saurabh W. and D’arcy Kenworthy, W. and Mackenty, John and Stahl, Benjamin E. and Zheng, WeiKang},
   year={2022},
   month=jul, pages={L7} }

@article{Dinda_2023,
   title={Model independent bounds on type Ia supernova absolute peak magnitude},
   volume={107},
   ISSN={2470-0029},
   url={http://dx.doi.org/10.1103/PhysRevD.107.063513},
   DOI={10.1103/physrevd.107.063513},
   number={6},
   journal={Physical Review D},
   publisher={American Physical Society (APS)},
   author={Dinda, Bikash R. and Banerjee, Narayan},
   year={2023},
   month=mar }

@article{planck2018,
   title={Planck2018 results: VI. Cosmological parameters},
   volume={641},
   ISSN={1432-0746},
   url={http://dx.doi.org/10.1051/0004-6361/201833910},
   DOI={10.1051/0004-6361/201833910},
   journal={Astronomy \& Astrophysics},
   publisher={EDP Sciences},
   author={Aghanim, N. and Akrami, Y. and Ashdown, M. and Aumont, J. and Baccigalupi, C. and Ballardini, M. and Banday, A. J. and Barreiro, R. B. and Bartolo, N. and Basak, S. and Battye, R. and Benabed, K. and Bernard, J.-P. and Bersanelli, M. and Bielewicz, P. and Bock, J. J. and Bond, J. R. and Borrill, J. and Bouchet, F. R. and Boulanger, F. and Bucher, M. and Burigana, C. and Butler, R. C. and Calabrese, E. and Cardoso, J.-F. and Carron, J. and Challinor, A. and Chiang, H. C. and Chluba, J. and Colombo, L. P. L. and Combet, C. and Contreras, D. and Crill, B. P. and Cuttaia, F. and de Bernardis, P. and de Zotti, G. and Delabrouille, J. and Delouis, J.-M. and Di Valentino, E. and Diego, J. M. and Doré, O. and Douspis, M. and Ducout, A. and Dupac, X. and Dusini, S. and Efstathiou, G. and Elsner, F. and Enßlin, T. A. and Eriksen, H. K. and Fantaye, Y. and Farhang, M. and Fergusson, J. and Fernandez-Cobos, R. and Finelli, F. and Forastieri, F. and Frailis, M. and Fraisse, A. A. and Franceschi, E. and Frolov, A. and Galeotta, S. and Galli, S. and Ganga, K. and Génova-Santos, R. T. and Gerbino, M. and Ghosh, T. and González-Nuevo, J. and Górski, K. M. and Gratton, S. and Gruppuso, A. and Gudmundsson, J. E. and Hamann, J. and Handley, W. and Hansen, F. K. and Herranz, D. and Hildebrandt, S. R. and Hivon, E. and Huang, Z. and Jaffe, A. H. and Jones, W. C. and Karakci, A. and Keihänen, E. and Keskitalo, R. and Kiiveri, K. and Kim, J. and Kisner, T. S. and Knox, L. and Krachmalnicoff, N. and Kunz, M. and Kurki-Suonio, H. and Lagache, G. and Lamarre, J.-M. and Lasenby, A. and Lattanzi, M. and Lawrence, C. R. and Le Jeune, M. and Lemos, P. and Lesgourgues, J. and Levrier, F. and Lewis, A. and Liguori, M. and Lilje, P. B. and Lilley, M. and Lindholm, V. and López-Caniego, M. and Lubin, P. M. and Ma, Y.-Z. and Macías-Pérez, J. F. and Maggio, G. and Maino, D. and Mandolesi, N. and Mangilli, A. and Marcos-Caballero, A. and Maris, M. and Martin, P. G. and Martinelli, M. and Martínez-González, E. and Matarrese, S. and Mauri, N. and McEwen, J. D. and Meinhold, P. R. and Melchiorri, A. and Mennella, A. and Migliaccio, M. and Millea, M. and Mitra, S. and Miville-Deschênes, M.-A. and Molinari, D. and Montier, L. and Morgante, G. and Moss, A. and Natoli, P. and Nørgaard-Nielsen, H. U. and Pagano, L. and Paoletti, D. and Partridge, B. and Patanchon, G. and Peiris, H. V. and Perrotta, F. and Pettorino, V. and Piacentini, F. and Polastri, L. and Polenta, G. and Puget, J.-L. and Rachen, J. P. and Reinecke, M. and Remazeilles, M. and Renzi, A. and Rocha, G. and Rosset, C. and Roudier, G. and Rubiño-Martín, J. A. and Ruiz-Granados, B. and Salvati, L. and Sandri, M. and Savelainen, M. and Scott, D. and Shellard, E. P. S. and Sirignano, C. and Sirri, G. and Spencer, L. D. and Sunyaev, R. and Suur-Uski, A.-S. and Tauber, J. A. and Tavagnacco, D. and Tenti, M. and Toffolatti, L. and Tomasi, M. and Trombetti, T. and Valenziano, L. and Valiviita, J. and Van Tent, B. and Vibert, L. and Vielva, P. and Villa, F. and Vittorio, N. and Wandelt, B. D. and Wehus, I. K. and White, M. and White, S. D. M. and Zacchei, A. and Zonca, A.},
   year={2020},
   month=sep, pages={A6} }

@article{Scolnic_2022,
  title     = {The Pantheon+ Analysis: The Full Data Set and Light-curve Release},
  volume    = {938},
  issn      = {1538-4357},
  url       = {http://dx.doi.org/10.3847/1538-4357/ac8b7a},
  doi       = {10.3847/1538-4357/ac8b7a},
  number    = {2},
  journal   = {The Astrophysical Journal},
  publisher = {American Astronomical Society},
  author    = {Scolnic, Dan and Brout, Dillon and Carr, Anthony and Riess, Adam G. and Davis, Tamara M. and Dwomoh, Arianna and Jones, David O. and Ali, Noor and Charvu, Pranav and Chen, Rebecca and Peterson, Erik R. and Popovic, Brodie and Rose, Benjamin M. and Wood, Charlotte M. and Brown, Peter J. and Chambers, Ken and Coulter, David A. and Dettman, Kyle G. and Dimitriadis, Georgios and Filippenko, Alexei V. and Foley, Ryan J. and Jha, Saurabh W. and Kilpatrick, Charles D. and Kirshner, Robert P. and Pan, Yen-Chen and Rest, Armin and Rojas-Bravo, Cesar and Siebert, Matthew R. and Stahl, Benjamin E. and Zheng, WeiKang},
  year      = {2022},
  month     = oct,
  pages     = {113}
}

@article{emcee,
   author = {{Foreman-Mackey}, D. and {Hogg}, D.~W. and {Lang}, D. and {Goodman}, J.},
    title = {emcee: The MCMC Hammer},
  journal = {PASP},
     year = 2013,
   volume = 125,
    pages = {306-312},
   eprint = {1202.3665},
      doi = {10.1086/670067}
}

@article{moresco_setting_2020,
	title = {Setting the Stage for Cosmic Chronometers. {II}. Impact of Stellar Population Synthesis Models Systematics and Full Covariance Matrix},
	volume = {898},
	issn = {0004-637X, 1538-4357},
	url = {https://iopscience.iop.org/article/10.3847/1538-4357/ab9eb0},
	doi = {10.3847/1538-4357/ab9eb0},
	pages = {82},
	number = {1},
	journal = {The Astrophysical Journal},
	shortjournal = {{ApJ}},
	author = {Moresco, Michele and Jimenez, Raul and Verde, Licia and Cimatti, Andrea and Pozzetti, Lucia},
	urldate = {2026-01-13},
	date = {2020-07-01},
	langid = {english},
    year = {2020},
}

@ARTICLE{moresco_unveiling_2022,
       author = {{Moresco}, Michele and {Amati}, Lorenzo and {Amendola}, Luca and {Birrer}, Simon and {Blakeslee}, John P. and {Cantiello}, Michele and {Cimatti}, Andrea and {Darling}, Jeremy and {Della Valle}, Massimo and {Fishbach}, Maya and {Grillo}, Claudio and {Hamaus}, Nico and {Holz}, Daniel and {Izzo}, Luca and {Jimenez}, Raul and {Lusso}, Elisabeta and {Meneghetti}, Massimo and {Piedipalumbo}, Ester and {Pisani}, Alice and {Pourtsidou}, Alkistis and {Pozzetti}, Lucia and {Quartin}, Miguel and {Risaliti}, Guido and {Rosati}, Piero and {Verde}, Licia},
        title = "{Unveiling the Universe with emerging cosmological probes}",
      journal = {Living Reviews in Relativity},
         year = 2022,
        month = dec,
       volume = {25},
       number = {1},
          eid = {6},
        pages = {6},
          doi = {10.1007/s41114-022-00040-z},
archivePrefix = {arXiv},
       eprint = {2201.07241},
 primaryClass = {astro-ph.CO},
       adsurl = {https://ui.adsabs.harvard.edu/abs/2022LRR....25....6M}
}

@article{nair_cosmic_2012,
	title = {Cosmic distance duality and cosmic transparency},
	volume = {2012},
	issn = {1475-7516},
	url = {https://iopscience.iop.org/article/10.1088/1475-7516/2012/12/028},
	doi = {10.1088/1475-7516/2012/12/028},
	language = {en},
	number = {12},
	urldate = {2025-04-15},
	journal = {Journal of Cosmology and Astroparticle Physics},
	author = {Nair, Remya and Jhingan, Sanjay and Jain, Deepak},
	month = dec,
	year = {2012},
	pages = {028--028},
}

@article{Avgoustidis_2009,
   title={Consistency among distance measurements: transparency, BAO scale and accelerated expansion},
   volume={2009},
   ISSN={1475-7516},
   url={http://dx.doi.org/10.1088/1475-7516/2009/06/012},
   DOI={10.1088/1475-7516/2009/06/012},
   number={06},
   journal={Journal of Cosmology and Astroparticle Physics},
   publisher={IOP Publishing},
   author={Avgoustidis, Anastasios and Verde, Licia and Jimenez, Raul},
   year={2009},
   month=jun, pages={012–012} }

@article{the_euclid_theory_working_group_cosmology_2013,
	title = {Cosmology and {Fundamental} {Physics} with the {Euclid} {Satellite}},
	volume = {16},
	issn = {2367-3613, 1433-8351},
	url = {http://link.springer.com/10.12942/lrr-2013-6},
	doi = {10.12942/lrr-2013-6},
	language = {en},
	number = {1},
	urldate = {2025-12-04},
	journal = {Living Reviews in Relativity},
	author = {{The Euclid Theory Working Group} and Amendola, Luca and Appleby, Stephen and Bacon, David and Baker, Tessa and Baldi, Marco and Bartolo, Nicola and Blanchard, Alain and Bonvin, Camille and Borgani, Stefano and Branchini, Enzo and Burrage, Clare and Camera, Stefano and Carbone, Carmelita and Casarini, Luciano and Cropper, Mark and De Rham, Claudia and Di Porto, Cinzia and Ealet, Anne and Ferreira, Pedro G. and Finelli, Fabio and García-Bellido, Juan and Giannantonio, Tommaso and Guzzo, Luigi and Heavens, Alan and Heisenberg, Lavinia and Heymans, Catherine and Hoekstra, Henk and Hollenstein, Lukas and Holmes, Rory and Horst, Ole and Jahnke, Knud and Kitching, Thomas D. and Koivisto, Tomi and Kunz, Martin and La Vacca, Giuseppe and March, Marisa and Majerotto, Elisabetta and Markovic, Katarina and Marsh, David and Marulli, Federico and Massey, Richard and Mellier, Yannick and Mota, David F. and Nunes, Nelson J. and Percival, Will and Pettorino, Valeria and Porciani, Cristiano and Quercellini, Claudia and Read, Justin and Rinaldi, Massimiliano and Sapone, Domenico and Scaramella, Roberto and Skordis, Constantinos and Simpson, Fergus and Taylor, Andy and Thomas, Shaun and Trotta, Roberto and Verde, Licia and Vernizzi, Filippo and Vollmer, Adrian and Wang, Yun and Weller, Jochen and Zlosnik, Tom},
	month = dec,
	year = {2013},
	pages = {6},
}

@article{Svrcek_2006,
   title={Axions in string theory},
   volume={2006},
   ISSN={1029-8479},
   url={http://dx.doi.org/10.1088/1126-6708/2006/06/051},
   DOI={10.1088/1126-6708/2006/06/051},
   number={06},
   journal={Journal of High Energy Physics},
   publisher={Springer Science and Business Media LLC},
   author={Svrcek, Peter and Witten, Edward},
   year={2006},
   month=jun, pages={051–051} }

@article{renzi_resilience_2022,
	title = {The resilience of the {Etherington}–{Hubble} relation},
	volume = {513},
	copyright = {https://creativecommons.org/licenses/by/4.0/},
	issn = {0035-8711, 1365-2966},
	url = {https://academic.oup.com/mnras/article/513/3/4004/6567878},
	doi = {10.1093/mnras/stac1030},
	language = {en},
	number = {3},
	urldate = {2026-01-20},
	journal = {Monthly Notices of the Royal Astronomical Society},
	author = {Renzi, Fabrizio and Hogg, Natalie B and Giarè, William},
	month = may,
	year = {2022},
	pages = {4004--4014},
}

@article{10.1093/mnras/stab1588,
    author = {Efstathiou, George},
    title = {To H0 or not to H0?},
    journal = {Monthly Notices of the Royal Astronomical Society},
    volume = {505},
    number = {3},
    pages = {3866-3872},
    year = {2021},
    month = {06},
    doi = {10.1093/mnras/stab1588},
    url = {https://doi.org/10.1093/mnras/stab1588}
}

@inbook{Rasmussen2004,
author="Rasmussen, Carl Edward",
editor="Bousquet, Olivier
and von Luxburg, Ulrike
and R{\"a}tsch, Gunnar",
title="Gaussian Processes in Machine Learning",
bookTitle="Advanced Lectures on Machine Learning: ML Summer Schools 2003, Canberra, Australia, February 2 - 14, 2003, T{\"u}bingen, Germany, August 4 - 16, 2003, Revised Lectures",
year="2004",
publisher="Springer Berlin Heidelberg",
address="Berlin, Heidelberg",
pages="63--71",
isbn="978-3-540-28650-9",
doi="10.1007/978-3-540-28650-9_4",
url="https://doi.org/10.1007/978-3-540-28650-9_4",
}

@article{Sanchez_2024,
doi = {10.3847/1538-4357/ad739a},
url = {https://doi.org/10.3847/1538-4357/ad739a},
year = {2024},
month = {oct},
publisher = {The American Astronomical Society},
volume = {975},
number = {1},
pages = {5},
author = {Sánchez, B. O. and Brout, D. and Vincenzi, M. and Sako, M. and Herner, K. and Kessler, R. and Davis, T. M. and Scolnic, D. and Acevedo, M. and Lee, J. and Möller, A. and Qu, H. and Kelsey, L. and Wiseman, P. and Armstrong, P. and Rose, B. and Camilleri, R. and Chen, R. and Galbany, L. and Kovacs, E. and Lidman, C. and Popovic, B. and Smith, M. and Shah, P. and Sullivan, M. and Toy, M. and Abbott, T. M. C. and Aguena, M. and Allam, S. and Alves, O. and Annis, J. and Asorey, J. and Avila, S. and Bacon, D. and Brooks, D. and Burke, D. L. and Carnero Rosell, A. and Carollo, D. and Carretero, J. and da Costa, L. N. and Castander, F. J. and Desai, S. and Diehl, H. T. and Duarte, J. and Everett, S. and Ferrero, I. and Flaugher, B. and Frieman, J. and García-Bellido, J. and Gatti, M. and Gaztanaga, E. and Giannini, G. and Glazebrook, K. and González-Gaitán, S. and Gruendl, R. A. and Gutierrez, G. and Hinton, S. R. and Hollowood, D. L. and Honscheid, K. and James, D. J. and Kuehn, K. and Lahav, O. and Lee, S. and Lewis, G. F. and Lin, H. and Marshall, J. L. and Mena-Fernández, J. and Miquel, R. and Myles, J. and Nichol, R. C. and Ogando, R. L. C. and Palmese, A. and Pereira, M. E. S. and Pieres, A. and Plazas Malagón, A. A. and Porredon, A. and Romer, A. K. and Sanchez, E. and Sanchez Cid, D. and Sevilla-Noarbe, I. and Suchyta, E. and Swanson, M. E. C. and Tarle, G. and Tucker, B. E. and Tucker, D. L. and Vikram, V. and Walker, A. R. and Weaverdyck, N. and (DES Collaboration)},
title = {The Dark Energy Survey Supernova Program: Light Curves and 5 Yr Data Release},
journal = {The Astrophysical Journal},
}

@article{PhysRevD.103.L041301,
  title = {Curvature tension: Evidence for a closed universe},
  author = {Handley, Will},
  journal = {Phys. Rev. D},
  volume = {103},
  issue = {4},
  pages = {L041301},
  numpages = {7},
  year = {2021},
  month = {Feb},
  publisher = {American Physical Society},
  doi = {10.1103/PhysRevD.103.L041301},
  url = {https://link.aps.org/doi/10.1103/PhysRevD.103.L041301}
}

@article{Di_Valentino_2019,
   title={Planck evidence for a closed Universe and a possible crisis for cosmology},
   volume={4},
   ISSN={2397-3366},
   url={http://dx.doi.org/10.1038/s41550-019-0906-9},
   DOI={10.1038/s41550-019-0906-9},
   number={2},
   journal={Nature Astronomy},
   publisher={Springer Science and Business Media LLC},
   author={Di Valentino, Eleonora and Melchiorri, Alessandro and Silk, Joseph},
   year={2019},
   month=nov, pages={196–203} }

@ARTICLE{rubin2025unionunitycosmology2000,
       author = {{Rubin}, David and {Aldering}, Greg and {Betoule}, Marc and {Fruchter}, Andy and {Huang}, Xiaosheng and {Kim}, Alex G. and {Lidman}, Chris and {Linder}, Eric and {Perlmutter}, Saul and {Ruiz-Lapuente}, Pilar and {Suzuki}, Nao},
        title = "{Union through UNITY: Cosmology with 2000 SNe Using a Unified Bayesian Framework}",
      journal = {The Astrophysical Journal},
         year = 2025,
        month = jun,
       volume = {986},
       number = {2},
          eid = {231},
        pages = {231},
          doi = {10.3847/1538-4357/adc0a5},
archivePrefix = {arXiv},
       eprint = {2311.12098},
 primaryClass = {astro-ph.CO},
       adsurl = {https://ui.adsabs.harvard.edu/abs/2025ApJ...986..231R}
}

@article{Liang_2013,
   title={A consistent test of the distance–duality relation with galaxy clusters and Type Ia Supernovae},
   volume={436},
   ISSN={0035-8711},
   url={http://dx.doi.org/10.1093/mnras/stt1589},
   DOI={10.1093/mnras/stt1589},
   number={2},
   journal={Monthly Notices of the Royal Astronomical Society},
   publisher={Oxford University Press (OUP)},
   author={Liang, Nan and Li, Zhengxiang and Wu, Puxun and Cao, Shuo and Liao, Kai and Zhu, Zong-Hong},
   year={2013},
   month=oct, pages={1017–1022} }

@article{HOLDOM1986196,
title = {Two U(1)'s and $\epsilon$ charge shifts},
journal = {Physics Letters B},
volume = {166},
number = {2},
pages = {196-198},
year = {1986},
issn = {0370-2693},
doi = {https://doi.org/10.1016/0370-2693(86)91377-8},
url = {https://www.sciencedirect.com/science/article/pii/0370269386913778},
author = {Bob Holdom}
}

@article{Ahlers_2009,
   title={Hubble diagram as a probe of minicharged particles},
   volume={80},
   ISSN={1550-2368},
   url={http://dx.doi.org/10.1103/PhysRevD.80.023513},
   DOI={10.1103/physrevd.80.023513},
   number={2},
   journal={Physical Review D},
   publisher={American Physical Society (APS)},
   author={Ahlers, Markus},
   year={2009},
   month=jul }

@article{Zhang_2014,
doi = {10.1088/1674-4527/14/10/002},
url = {https://doi.org/10.1088/1674-4527/14/10/002},
year = {2014},
month = {oct},
publisher = {},
volume = {14},
number = {10},
pages = {1221},
author = "Zhang, Cong and Zhang, Han and Yuan, Qiang and Liu, Siming and Zhang, Tong-Jie and Sun, Yan-Chun",
title = {Four new observational H(z) data from luminous red galaxies in the Sloan Digital Sky Survey data release seven},
journal = {Research in Astronomy and Astrophysics},
}

@article{Jimenez_2003,
doi = {10.1086/376595},
url = {https://doi.org/10.1086/376595},
year = {2003},
month = {aug},
publisher = {},
volume = {593},
number = {2},
pages = {622},
author = {Jimenez, Raul and Verde, Licia and Treu, Tommaso and Stern, Daniel},
title = {Constraints on the Equation of State of Dark Energy and the Hubble Constant from Stellar Ages and the Cosmic Microwave Background*},
journal = {The Astrophysical Journal},
}

@article{PhysRevD.71.123001,
  title = {Constraints on the redshift dependence of the dark energy potential},
  author = {Simon, Joan and Verde, Licia and Jimenez, Raul},
  journal = {Phys. Rev. D},
  volume = {71},
  issue = {12},
  pages = {123001},
  numpages = {18},
  year = {2005},
  month = {Jun},
  publisher = {American Physical Society},
  doi = {10.1103/PhysRevD.71.123001},
  url = {https://link.aps.org/doi/10.1103/PhysRevD.71.123001}
}

@ARTICLE{Moresco_2012,
       author = {{Moresco}, M. and {Cimatti}, A. and {Jimenez}, R. and {Pozzetti}, L. and {Zamorani}, G. and {Bolzonella}, M. and {Dunlop}, J. and {Lamareille}, F. and {Mignoli}, M. and {Pearce}, H. and {Rosati}, P. and {Stern}, D. and {Verde}, L. and {Zucca}, E. and {Carollo}, C.~M. and {Contini}, T. and {Kneib}, J.-P. and {Le F{\`e}vre}, O. and {Lilly}, S.~J. and {Mainieri}, V. and {Renzini}, A. and {Scodeggio}, M. and {Balestra}, I. and {Gobat}, R. and {McLure}, R. and {Bardelli}, S. and {Bongiorno}, A. and {Caputi}, K. and {Cucciati}, O. and {de la Torre}, S. and {de Ravel}, L. and {Franzetti}, P. and {Garilli}, B. and {Iovino}, A. and {Kampczyk}, P. and {Knobel}, C. and {Kova{\v{c}}}, K. and {Le Borgne}, J.-F. and {Le Brun}, V. and {Maier}, C. and {Pell{\'o}}, R. and {Peng}, Y. and {Perez-Montero}, E. and {Presotto}, V. and {Silverman}, J.~D. and {Tanaka}, M. and {Tasca}, L.~A.~M. and {Tresse}, L. and {Vergani}, D. and {Almaini}, O. and {Barnes}, L. and {Bordoloi}, R. and {Bradshaw}, E. and {Cappi}, A. and {Chuter}, R. and {Cirasuolo}, M. and {Coppa}, G. and {Diener}, C. and {Foucaud}, S. and {Hartley}, W. and {Kamionkowski}, M. and {Koekemoer}, A.~M. and {L{\'o}pez-Sanjuan}, C. and {McCracken}, H.~J. and {Nair}, P. and {Oesch}, P. and {Stanford}, A. and {Welikala}, N.},
        title = "{Improved constraints on the expansion rate of the Universe up to z \raisebox{-0.5ex}\textasciitilde 1.1 from the spectroscopic evolution of cosmic chronometers}",
      journal = {Journal of Cosmology and Astroparticle Physics},
         year = 2012,
        month = aug,
       volume = {2012},
       number = {8},
          eid = {006},
        pages = {006},
          doi = {10.1088/1475-7516/2012/08/006},
archivePrefix = {arXiv},
       eprint = {1201.3609},
 primaryClass = {astro-ph.CO},
       adsurl = {https://ui.adsabs.harvard.edu/abs/2012JCAP...08..006M}
}

@article{Daniel_Stern_2010,
   title={Cosmic chronometers: constraining the equation of state of dark energy. I: H(z) measurements},
   volume={2010},
   ISSN={1475-7516},
   url={http://dx.doi.org/10.1088/1475-7516/2010/02/008},
   DOI={10.1088/1475-7516/2010/02/008},
   number={02},
   journal={Journal of Cosmology and Astroparticle Physics},
   publisher={IOP Publishing},
   author={Daniel Stern and Raul Jimenez and Licia Verde and Marc Kamionkowski and S. Adam Stanford},
   year={2010},
   month=feb, pages={008–008} }

@article{Moresco_2015,
    author = {{Moresco}, M.},
    title = "{Raising the bar: new constraints on the Hubble parameter with cosmic chronometers at z \raisebox{-0.5ex}\textasciitilde 2.}",
    journal = {Monthly Notices of the Royal Astronomical Society: Letters},
     year = 2015,
    month = jun,
   volume = {450},
    pages = {L16-L20},
      doi = {10.1093/mnrasl/slv037},
    archivePrefix = {arXiv},
   eprint = {1503.01116},
    primaryClass = {astro-ph.CO},
   adsurl = {https://ui.adsabs.harvard.edu/abs/2015MNRAS.450L..16M}
}

@article{Moresco_2016,
doi = {10.1088/1475-7516/2016/05/014},
url = {https://doi.org/10.1088/1475-7516/2016/05/014},
year = {2016},
month = {may},
publisher = {},
volume = {2016},
number = {05},
pages = {014},
author = {Moresco, Michele and Pozzetti, Lucia and Cimatti, Andrea and Jimenez, Raul and Maraston, Claudia and Verde, Licia and Thomas, Daniel and Citro, Annalisa and Tojeiro, Rita and Wilkinson, David},
title = {A 6\% measurement of the Hubble parameter at $z \sim 0.45$: direct evidence of the epoch of cosmic re-acceleration},
journal = {Journal of Cosmology and Astroparticle Physics},
}

@article{Lewis:2019xzd,
   author = "Lewis, Antony",
   title = "{GetDist: a Python package for analysing Monte Carlo samples}",
   eprint = "1910.13970",
   archivePrefix = "arXiv",
   primaryClass = "astro-ph.IM",
   doi = "10.1088/1475-7516/2025/08/025",
   journal = "JCAP",
   volume = "08",
   pages = "025",
   year = "2025"
}

@article{PhysRevD.92.123539,
  title = {Is the Universe transparent?},
  author = {Liao, Kai and Avgoustidis, A. and Li, Zhengxiang},
  journal = {Phys. Rev. D},
  volume = {92},
  issue = {12},
  pages = {123539},
  numpages = {9},
  year = {2015},
  month = {Dec},
  publisher = {American Physical Society},
  doi = {10.1103/PhysRevD.92.123539},
  url = {https://link.aps.org/doi/10.1103/PhysRevD.92.123539}
}

@article{Burrage_2009,
   title={Detecting chameleons: The astronomical polarization produced by chameleonlike scalar fields},
   volume={79},
   ISSN={1550-2368},
   url={http://dx.doi.org/10.1103/PhysRevD.79.044028},
   DOI={10.1103/physrevd.79.044028},
   number={4},
   journal={Physical Review D},
   publisher={American Physical Society (APS)},
   author={Burrage, Clare and Davis, Anne-Christine and Shaw, Douglas J.},
   year={2009},
   month=feb }

@article{Davis_2009,
   title={Effect of a chameleon scalar field on the cosmic microwave background},
   volume={80},
   ISSN={1550-2368},
   url={http://dx.doi.org/10.1103/PhysRevD.80.064016},
   DOI={10.1103/physrevd.80.064016},
   number={6},
   journal={Physical Review D},
   publisher={American Physical Society (APS)},
   author={Davis, Anne-Christine and Schelpe, Camilla A. O. and Shaw, Douglas J.},
   year={2009},
   month=sep }

@article{Schelpe_2010,
   title={Chameleon-photon mixing in a primordial magnetic field},
   volume={82},
   ISSN={1550-2368},
   url={http://dx.doi.org/10.1103/PhysRevD.82.044033},
   DOI={10.1103/physrevd.82.044033},
   number={4},
   journal={Physical Review D},
   publisher={American Physical Society (APS)},
   author={Schelpe, Camilla A. O.},
   year={2010},
   month=aug }

@article{PhysRevD.37.1237,
  title = {Mixing of the photon with low-mass particles},
  author = {Raffelt, Georg and Stodolsky, Leo},
  journal = {Phys. Rev. D},
  volume = {37},
  issue = {5},
  pages = {1237--1249},
  numpages = {0},
  year = {1988},
  month = {Mar},
  publisher = {American Physical Society},
  doi = {10.1103/PhysRevD.37.1237},
  url = {https://link.aps.org/doi/10.1103/PhysRevD.37.1237}
}

@article{Ahlers_2007,
   title={Light from the hidden sector: Experimental signatures of paraphotons},
   volume={76},
   ISSN={1550-2368},
   url={http://dx.doi.org/10.1103/PhysRevD.76.115005},
   DOI={10.1103/physrevd.76.115005},
   number={11},
   journal={Physical Review D},
   publisher={American Physical Society (APS)},
   author={Ahlers, M. and Gies, H. and Jaeckel, J. and Redondo, J. and Ringwald, A.},
   year={2007},
   month=dec }

@article{PhysRevLett.133.251004,
  title = {Self-Interacting Dark Sectors in Supernovae Can Behave as a Relativistic Fluid},
  author = {Fiorillo, Damiano F. G. and Vitagliano, Edoardo},
  journal = {Phys. Rev. Lett.},
  volume = {133},
  issue = {25},
  pages = {251004},
  numpages = {8},
  year = {2024},
  month = {Dec},
  publisher = {American Physical Society},
  doi = {10.1103/PhysRevLett.133.251004},
  url = {https://link.aps.org/doi/10.1103/PhysRevLett.133.251004}
}

@article{Chang_2018,
   title={Supernova 1987A constraints on sub-GeV dark sectors, millicharged particles, the QCD axion, and an axion-like particle},
   volume={2018},
   ISSN={1029-8479},
   url={http://dx.doi.org/10.1007/JHEP09(2018)051},
   DOI={10.1007/jhep09(2018)051},
   number={9},
   journal={Journal of High Energy Physics},
   publisher={Springer Science and Business Media LLC},
   author={Chang, Jae Hyeok and Essig, Rouven and McDermott, Samuel D.},
   year={2018},
   month=sep }

@article{Vogel_2014,
doi = {10.1088/1475-7516/2014/02/029},
url = {https://doi.org/10.1088/1475-7516/2014/02/029},
year = {2014},
month = {feb},
publisher = {},
volume = {2014},
number = {02},
pages = {029},
author = {Vogel, Hendrik and Redondo, Javier},
title = {Dark radiation constraints on minicharged particles in models with a hidden photon},
journal = {Journal of Cosmology and Astroparticle Physics},
}

@article{PhysRevD.109.083011,
  title = {New bounds on light millicharged particles from the tip of the red-giant branch},
  author = {Fung, Audrey and Heeba, Saniya and Liu, Qinrui and Muralidharan, Varun and Schutz, Katelin and Vincent, Aaron C.},
  journal = {Phys. Rev. D},
  volume = {109},
  issue = {8},
  pages = {083011},
  numpages = {12},
  year = {2024},
  month = {Apr},
  publisher = {American Physical Society},
  doi = {10.1103/PhysRevD.109.083011},
  url = {https://link.aps.org/doi/10.1103/PhysRevD.109.083011}
}

@article{Stebbins_2019,
   title={New limits on charged dark matter from large-scale coherent magnetic fields},
   volume={2019},
   ISSN={1475-7516},
   url={http://dx.doi.org/10.1088/1475-7516/2019/12/003},
   DOI={10.1088/1475-7516/2019/12/003},
   number={12},
   journal={Journal of Cosmology and Astroparticle Physics},
   publisher={IOP Publishing},
   author={Stebbins, Albert and Krnjaic, Gordan},
   year={2019},
   month=dec, pages={003–003} }

@article{Burrage_200902,
   title={Late time CMB anisotropies constrain mini-charged particles},
   volume={2009},
   ISSN={1475-7516},
   url={http://dx.doi.org/10.1088/1475-7516/2009/11/002},
   DOI={10.1088/1475-7516/2009/11/002},
   number={11},
   journal={Journal of Cosmology and Astroparticle Physics},
   publisher={IOP Publishing},
   author={Burrage, C and Jaeckel, J and Redondo, J and Ringwald, A},
   year={2009},
   month=nov, pages={002–002} }

@ARTICLE{arza_geomagnetic_2026,
       author = {{Arza}, Ariel and {Gong}, Yuanlin and {Shu}, Jing and {Wu}, Lei and {Yuan}, Qiang and {Zhu}, Bin},
        title = "{Geomagnetic Constraints on Millicharged Dark Matter}",
      journal = {Physical Review Letters},
         year = 2026,
        month = jan,
       volume = {136},
       number = {4},
          eid = {041001},
        pages = {041001},
          doi = {10.1103/8xqd-dbrz},
archivePrefix = {arXiv},
       eprint = {2501.14949},
 primaryClass = {hep-ph},
       adsurl = {https://ui.adsabs.harvard.edu/abs/2026PhRvL.136d1001A}
}

@ARTICLE{2007CSE.....9...90H,
       author = {{Hunter}, John D.},
        title = "{Matplotlib: A 2D Graphics Environment}",
      journal = {Computing in Science and Engineering},
         year = 2007,
        month = jan,
       volume = {9},
       number = {3},
        pages = {90-95},
          doi = {10.1109/MCSE.2007.55},
       adsurl = {https://ui.adsabs.harvard.edu/abs/2007CSE.....9...90H}
}

@ARTICLE{2020Natur.585..357H,
       author = {{Harris}, Charles R. and {Millman}, K. Jarrod and {van der Walt}, St{\'e}fan J. and {Gommers}, Ralf and {Virtanen}, Pauli and {Cournapeau}, David and {Wieser}, Eric and {Taylor}, Julian and {Berg}, Sebastian and {Smith}, Nathaniel J. and {Kern}, Robert and {Picus}, Matti and {Hoyer}, Stephan and {van Kerkwijk}, Marten H. and {Brett}, Matthew and {Haldane}, Allan and {del R{\'\i}o}, Jaime Fern{\'a}ndez and {Wiebe}, Mark and {Peterson}, Pearu and {G{\'e}rard-Marchant}, Pierre and {Sheppard}, Kevin and {Reddy}, Tyler and {Weckesser}, Warren and {Abbasi}, Hameer and {Gohlke}, Christoph and {Oliphant}, Travis E.},
        title = "{Array programming with NumPy}",
      journal = {Nature},
         year = 2020,
        month = sep,
       volume = {585},
       number = {7825},
        pages = {357-362},
          doi = {10.1038/s41586-020-2649-2},
archivePrefix = {arXiv},
       eprint = {2006.10256},
 primaryClass = {cs.MS},
       adsurl = {https://ui.adsabs.harvard.edu/abs/2020Natur.585..357H}
}

@ARTICLE{2020NatMe..17..261V,
       author = {{Virtanen}, Pauli and {Gommers}, Ralf and {Oliphant}, Travis E. and {Haberland}, Matt and {Reddy}, Tyler and {Cournapeau}, David and {Burovski}, Evgeni and {Peterson}, Pearu and {Weckesser}, Warren and {Bright}, Jonathan and {van der Walt}, St{\'e}fan J. and {Brett}, Matthew and {Wilson}, Joshua and {Millman}, K. Jarrod and {Mayorov}, Nikolay and {Nelson}, Andrew R.~J. and {Jones}, Eric and {Kern}, Robert and {Larson}, Eric and {Carey}, C.~J. and {Polat}, {\.I}lhan and {Feng}, Yu and {Moore}, Eric W. and {VanderPlas}, Jake and {Laxalde}, Denis and {Perktold}, Josef and {Cimrman}, Robert and {Henriksen}, Ian and {Quintero}, E.~A. and {Harris}, Charles R. and {Archibald}, Anne M. and {Ribeiro}, Ant{\^o}nio H. and {Pedregosa}, Fabian and {van Mulbregt}, Paul and {SciPy 1.  0 Contributors}},
        title = "{SciPy 1.0: fundamental algorithms for scientific computing in Python}",
      journal = {Nature Medicine},
         year = 2020,
        month = feb,
       volume = {17},
        pages = {261-272},
          doi = {10.1038/s41592-019-0686-2},
archivePrefix = {arXiv},
       eprint = {1907.10121},
 primaryClass = {cs.MS},
       adsurl = {https://ui.adsabs.harvard.edu/abs/2020NatMe..17..261V}
}







\appendix
\section{Kernel Selection and Hyperparameter Posterior Distribution in Gaussian Process Regression}\label{appendix:a}

\begin{figure}[htbp]
    \centering
    {\includegraphics[width=0.495\textwidth]{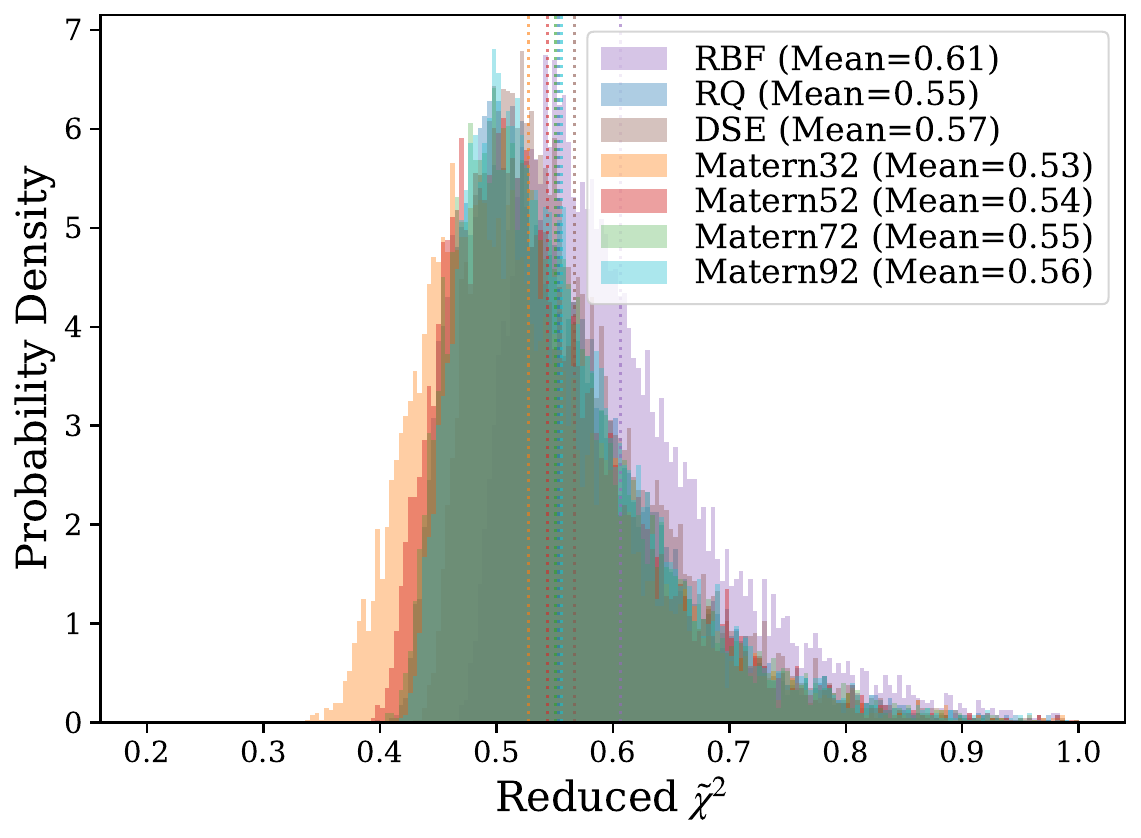}}\hfill
    {\includegraphics[width=0.495\textwidth]{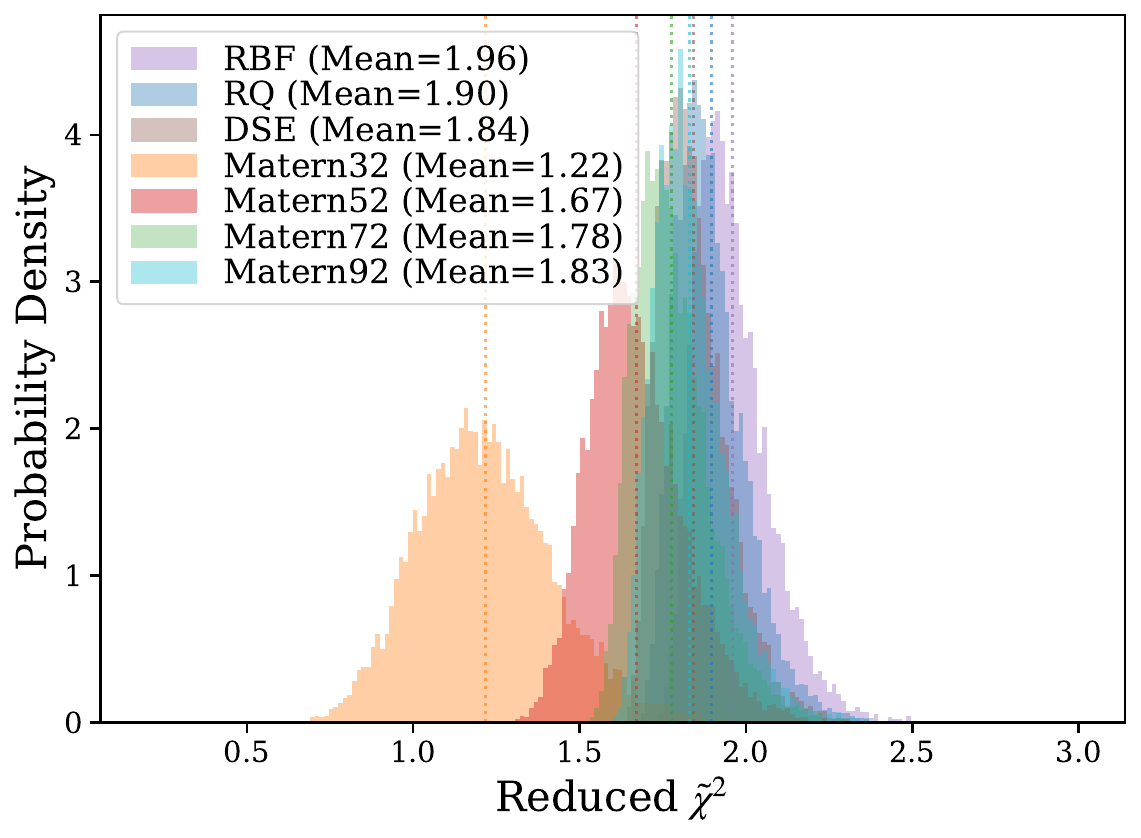}}
    \caption{
    Posterior distributions of the reduced $\tilde{\chi}^2$ for different covariance kernels.
    Left: $H(z)$ reconstruction.
    Right: $m_B(z)$ reconstruction based on compressed supernova data.}
    \label{fig:gp_chi2}
\end{figure}

\begin{figure}[htbp]
    \centering
    {\includegraphics[width=0.48\textwidth]{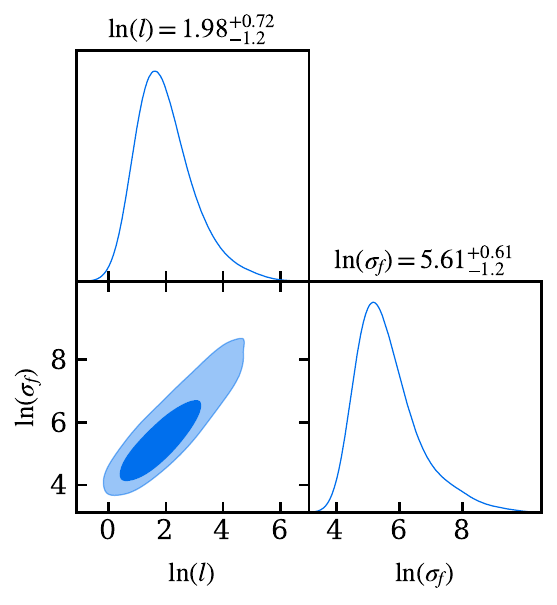}}\hfill
    {\includegraphics[width=0.48\textwidth]{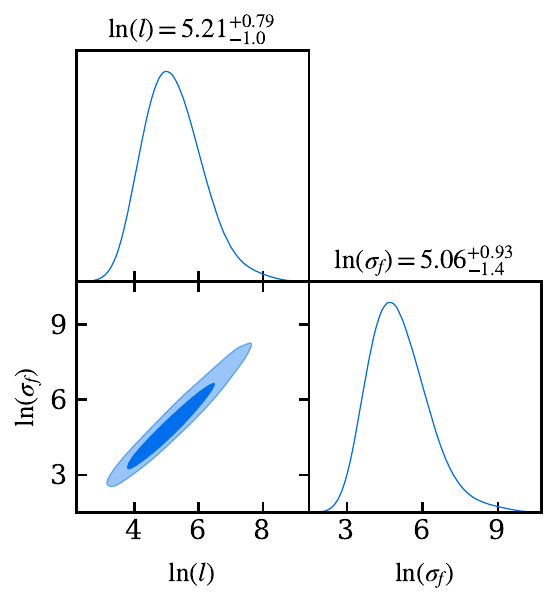}}
    \caption{
    Posterior distributions of the GP hyperparameters inferred from Bayesian sampling (using Matérn 3/2 kernel). 
    Left: $H(z)$ reconstruction.
    Right: $m_B(z)$ reconstruction.}
    \label{fig:gp_hyperparameters}
\end{figure}

To examine the robustness of our GPR analysis with respect to the choice of covariance kernel, we consider several commonly used kernels, including the squared exponential (SE), rational quadratic (RQ), double squared exponential (DSE), and Matérn kernels with different smoothness parameters (3/2, 5/2, 7/2 and 9/2).
For each kernel, we reconstruct the expansion history $H(z)$ and evaluate the goodness-of-fit using the reduced chi-square statistic, following Ref.~\cite{Favale_2023}. Specifically, for a given observational dataset $\mathbf{y}_{\text{obs}}$ and its corresponding full covariance matrix $\mathbf{C}_{\text{data}}$, we first optimize the hyperparameters of the GP for a specific kernel. The GP then provides a multivariate Gaussian posterior distribution at the observed redshifts, characterized by a predictive mean vector $\boldsymbol{\mu}_{\text{GP}}$ and a reconstruction covariance matrix $\boldsymbol{\Sigma}_{\text{GP}}$.

figure~\ref{fig:gp_chi2} shows the posterior distributions of the reduced $\tilde{\chi}^2$ values obtained from Monte Carlo realizations of the reconstructed functions. These distributions are generated by drawing $N = 10^4$ random function realizations $\mathbf{y}^{(\mu)}_{\text{GP}}$ (where $\mu = 1, \dots, N$) from this posterior distribution $\mathcal{N}(\boldsymbol{\mu}_{\text{GP}}, \boldsymbol{\Sigma}_{\text{GP}})$. This Monte Carlo sampling fully accounts for both the observational data covariance and the GP reconstruction uncertainty. For each realization, we calculate the $\chi^2$ statistic against the original observational data using the actual data covariance matrix:
\begin{equation}
\chi^2_{(\mu)} = \left( \mathbf{y}^{(\mu)}_{\text{GP}} - \mathbf{y}_{\text{obs}} \right)^T \mathbf{C}_{\text{data}}^{-1} \left( \mathbf{y}^{(\mu)}_{\text{GP}} - \mathbf{y}_{\text{obs}} \right).
\end{equation}
Finally, we compute the reduced chi-square for each realization, defined as $\tilde{\chi}^2_{(\mu)} = \chi^2_{(\mu)} / N_{\text{dof}}$, where $N_{\text{dof}}$ is the number of degrees of freedom (the number of data points minus the number of kernel hyperparameters). We then calculate the average $\tilde{\chi}^2$ from these $10^4$ realizations to quantify the goodness-of-fit for each kernel.

For the $H(z)$ reconstruction, all kernels lead to comparable $\tilde{\chi}^2$ distributions and provide statistically acceptable fits. To quantify relative differences, we adopt the kernel comparison method of Ref.~\cite{Favale_2023}. The probability that kernel $K_i$ yields a smaller reduced chi-square than kernel $K_j$ is defined as:
\begin{equation}
P\!\left(\tilde{\chi}^2_{K_i} < \tilde{\chi}^2_{K_j}\right)
= \frac{1}{1 + P_j / P_i},
\end{equation}
where $P_i$ and $P_j$ are the corresponding statistical weights.
These weights are estimated from Monte Carlo realizations via
\begin{equation}
\frac{P_j}{P_i} \simeq \frac{N_j}{N_i},
\end{equation}
with $N_i$ denoting the number of realizations for which $\tilde{\chi}^2_{K_i} < \tilde{\chi}^2_{K_j}$.

Using the SE kernel as a reference, the relative statistical weights for the $H(z)$ reconstruction are summarized in table~\ref{tab:kernel_weights_hz}.
Among the kernels considered, the Matérn $3/2$ kernel shows the smallest average reduced $\tilde{\chi}^2$ and probability $P(\tilde{\chi}^2_{\text{SE}} < \tilde{\chi}^2_k)$, although the differences with respect to other kernels are mild.

\begin{table}[ht]
\centering
\caption{Relative statistical weights of different kernels for the $H(z)$ reconstruction, taking the SE kernel as reference. The bolded one represents the best result.}
\label{tab:kernel_weights_hz}
\begin{tabular}{ccc}
\hline
Kernel & $P_k / P_{\text{SE}}$ & $P(\tilde{\chi}^2_{\text{SE}} < \tilde{\chi}^2_k)$ \\
\hline
RQ        & 2.232 & 0.309 \\
DSE       & 1.839 & 0.352 \\
\textbf{Matérn 3/2} & \textbf{3.050} & \textbf{0.247} \\
Matérn 5/2 & 2.554 & 0.281 \\
Matérn 7/2 & 2.331 & 0.300 \\
Matérn 9/2 & 2.131 & 0.319 \\
\hline
\end{tabular}
\end{table}

For the reconstruction of the supernova absolute magnitude $m_B(z)$, the situation is qualitatively different. Owing to the large number of supernova data points, performing a full kernel exploration within GPR becomes computationally expensive. To make the analysis feasible, we adopt a compressed-point approach similar to that of Ref.~\cite{xu_model-independent_2022}, in which the information of the full supernova sample is summarized into 36 effective data points along with their corresponding compressed covariance matrix. The goodness-of-fit evaluation for $m_B(z)$ is then efficiently performed on this compressed dataset.
The resulting relative statistical weights are reported in table~\ref{tab:kernel_weights_mb}. In this case, the Matérn $3/2$ kernel is clearly favored over smoother kernels.

\begin{table}[ht]
\centering
\caption{Relative statistical weights of different kernels for the $m_B(z)$ reconstruction, with the SE kernel as reference. The bolded one represents the best result.}
\label{tab:kernel_weights_mb}
\begin{tabular}{ccc}
\hline
Kernel & $P_k / P_{\text{SE}}$ & $P(\tilde{\chi}^2_{\text{SE}} < \tilde{\chi}^2_k)$ \\
\hline
RQ        & 2.054 & 0.327 \\
DSE       & 3.968 & 0.201 \\
\textbf{Matérn 3/2} & \textbf{525.316} & \textbf{0.002} \\
Matérn 5/2 & 17.215 & 0.055 \\
Matérn 7/2 & 7.475 & 0.118 \\
Matérn 9/2 & 4.429 & 0.184 \\
\hline
\end{tabular}
\end{table}

Based on the above kernel comparison for both $H(z)$ and $m_B(z)$ reconstructions, we adopt the Matérn $3/2$ kernel as the fiducial choice throughout this work. All results presented in the main text are obtained using this kernel. Finally, we emphasize that throughout this work we adopt a fully Bayesian GPR framework, in which the hyperparameters of the covariance kernels are inferred via posterior sampling rather than fixed at maximum-likelihood values. As illustrated in figure~\ref{fig:gp_hyperparameters}, this approach allows the uncertainty in the GP hyperparameters to be consistently propagated into the reconstructed functions and subsequent cosmological analyses.

\end{document}